%% file: GGI_aps.tex
\documentclass[aps, prd, floatfix, notitlepage, nofootinbib, superscriptaddress, onecolumn]{revtex4-1}

\usepackage{latexsym}
\usepackage{amsmath}
\usepackage{amssymb}
\usepackage{amsfonts}
\usepackage{textcomp}
\usepackage{color}
\usepackage{multirow}
\usepackage{epigraph}

\usepackage[mathscr,scaled=1.15]{urwchancal}
\DeclareFontFamily{OT1}{pzc}{}
\DeclareFontShape{OT1}{pzc}{m}{it}%
{<-> s * [1.15] pzcmi7t}{}
\DeclareMathAlphabet{\mathpzc}{OT1}{pzc}{m}{it}

\usepackage{color}
\usepackage{supertabular}
\usepackage{placeins}
\usepackage{epsfig}
\usepackage{graphicx}

\definecolor{purple}{rgb}{0.5,0,0.5}
\definecolor{blue}{rgb}{0.0,0,0.9}
\definecolor{prdblue}{rgb}{0.133,0.118,0.498}
\usepackage[colorlinks=true, pdfstartview=FitV, linkcolor=prdblue, citecolor= prdblue, urlcolor=prdblue]{hyperref}

\usepackage{xspace}

\newcommand{\mevnospace}{\ensuremath{{\mathrm{\,Me\kern -0.1em V}}}}
\newcommand{\gevnospace}{\ensuremath{{\mathrm{\,Ge\kern -0.1em V}}}}
\newcommand{\tevnospace}{\ensuremath{{\mathrm{\,Te\kern -0.1em V}}}}
\newcommand{\mev}{\mevnospace\xspace}

\def\be{\begin{equation}}
\def\ee{\end{equation}}
\def\bpm{\begin{pmatrix}}
\def\epm{\end{pmatrix}}
\def\bsm{\begin{bmatrix}}
\def\esm{\end{bmatrix}}
\def\bvm{\begin{vmatrix}}
\def\evm{\end{vmatrix}}
\def\bVM{\begin{Vmatrix}}
\def\eVM{\end{Vmatrix}}
\def\bea{\begin{eqnarray}}
\def\eea{\end{eqnarray}}

\makeatletter
\def\l@subsection#1#2{}
\def\l@subsubsection#1#2{}
\makeatother

\begin{document}

\title{GGI Lectures on Exotic Hadrons}

\date{\today}

\author{Luciano Maiani}
\email{luciano.maiani@cern.ch}
\affiliation{CERN, 1211 Geneva 23, Switzerland}
\author{Alessandro Pilloni}
\email{alessandro.pilloni@unime.it}
\affiliation{Dipartimento MIFT,
Universit\`a degli Studi di Messina, Viale Ferdinando Stagno d'Alcontres 31, I-98166 Messina, Italy}
\affiliation{INFN Sezione di Catania, Via Santa Sofia 64, I-95123 Catania, Italy}

\begin{abstract}
It is well known that M.~Gell-Mann, introducing quarks in 1964 to describe the known mesons and baryons, hinted at the existence of  further $qq\bar q\bar q$ mesons (tetraquarks) and  $qqqq\bar q$ baryons (pentaquarks). In 1977, R. Jaffe proposed a model of the lightest scalar mesons as {\it diquark-antidiquark} pairs and A.~de~Rujula, H.~Georgi and S.~Glashow coined the term {\it hadron molecules}, to describe possible hadrons made by meson-antimeson pairs bound by the familiar nuclear forces, also an overall tetraquark system. The two alternative pictures have been employed to interpret the unexpected hadron discovered by Belle in 2003, the $X(3872)$, confirmed by BaBar and seen in many other High Energy experiments. Since then, a wealth of Exotic Hadrons have been discovered, mesons and baryons that cannot be described by the classical Gell-Mann, $q\bar q$ and $qqq$, configurations, opening a new chapter of Hadron Spectroscopy. \\
\vspace{-.0cm}\\
{\it Galileo Galilei Institute, Firenze (Italy), February 21-25, 2022}
\end{abstract}

\maketitle

\setcounter{tocdepth}{1}
\tableofcontents
\newpage
\section{Introduction}
\label{sec:intro}
\input{chapts/Overview}

\newpage
\section{Masses of mesons and baryons in the Constituent Quark Model}
\label{sec:CQM}

\input{chapts/CQM}

\newpage
\section{Light and Heavy Tetraquarks}
\label{sec:tetra}

\input{chapts/Tetraqs}

\newpage
\section{Tetraquarks and the Eightfold Way. Di-$J/\psi$ resonances}
\label{sec:8fold}
\input{chapts/8fold}

\newpage
\section{$X(3872)$ and its missing partners}
\label{sec:partners}

\input{chapts/missingpart}

\newpage
\section{Tetraquarks in the Born-Oppenheimer approximation}
\label{sec:bornoppe}
\input{chapts/bornoppe}

\newpage
\label{sec:ninfty}
\section{Multiquark  states in $N$ colours, in the $N\to\infty$  limit}
\input{chapts/N_infty}

\newpage
\section{Weinberg's  criterion applied to $X(3872)$ and $T^+_{cc}(3875)$}
\label{sec:weinberg}
\input{chapts/WeinbergCrit}

\newpage
 
\bibliographystyle{apsrev4-2.bst}
\bibliography{quattro}
\end{document}

%% file: chapts/Overview.tex

\epigraph{Baryons can now be constructed from quarks by using the combinations $qqq$, $qqqq\bar q$, etc., while mesons are constructed out of $q\bar q$, $qq\bar q\bar q$, etc.}{M. Gell-Mann, 1964~\cite{GellMann:1964nj}.}

Quantum corrections, however, make so that any hadron will be a superposition of states with  arbitrary numbers of $q\bar q$ pairs, if meson,  and  $qqq+q\bar q$ pairs, if baryon. Can we find a meaningful way to count the number of quarks confined inside a hadron? 

A first possibility relies on the fact that heavy quark pairs are difficult to be created or destroyed by QCD forces inside hadrons. As a consequence, hadrons decaying with one $J/\psi$ (or  $\Upsilon$) in the final state are good candidates to contain a leading, or {\it valence},  $c\bar c$ (or $b\bar b$) pair. Since we have a good theoretical control on the spectrum of pure $c\bar c$ (charmonia) and $b \bar b$ (bottomonia) states, any hidden charm or bottom state not fitting in the theoretical spectrum is a good candidate for being an exotic, multiquark state. 
\begin{figure}[b]
 \begin{center}
   \includegraphics[width=15.0truecm]{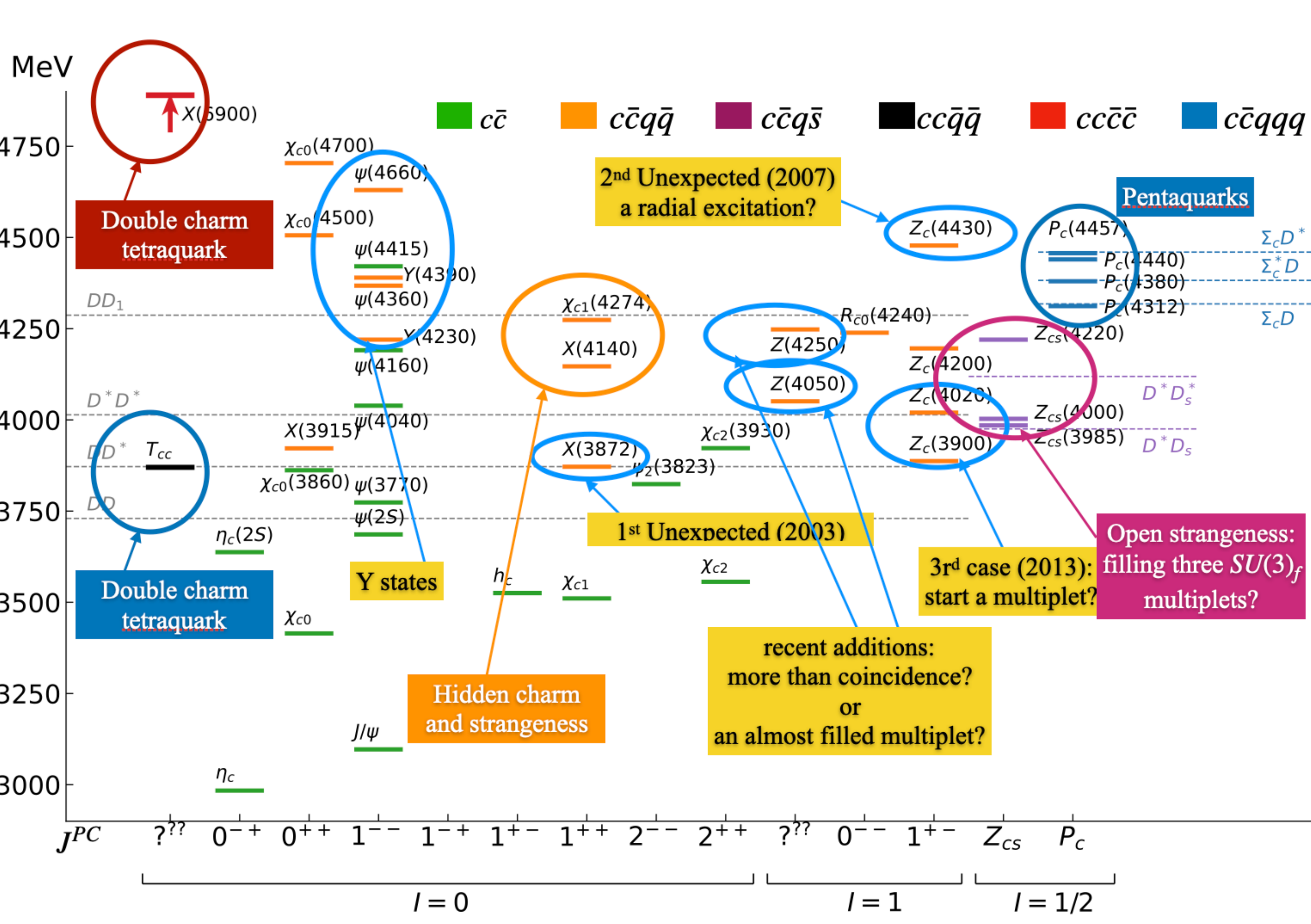}
  \caption{\footnotesize{Anticipated and unexpected charmonia. In green the observed ordinary charmonium lines. Compilation and figure adapted from~\cite{JPAC:2021rxu}.}}
\label{unexch}
\end{center}
\end{figure}
These {\it unanticipated} charmonia or bottomonia are the $X$ and $Y$ mesons discovered in the last two decades, the first examples being the $X(3872)$ meson discovered by Belle in 2003 and the $Y(4260)$ meson, discovered by BaBar in 2005.

Another definitely unambiguous class of exotic states are hidden charm or bottom states which are also {\it electrically charged}. Charged charmonia/botttomonia,  unanticipated by default, are indicated as $Z$ particles, the first example being $Z(4430)$, discovered by Belle in 2008 and confirmed by LHCb in 2014. The pentaquarks observed by LHCb in 2015 and 2019 complete the list. Fig.~\ref{unexch} gives an updated compilation  of the observed, predicted and unanticipated charmonia prepared for the present Lecture Notes (older compilations can be found in~\cite{Olsen:2015zcy,Esposito:2016noz,Brambilla:2019esw}).

 A short characterization of $X,~Y$ and $Z$ particles goes as follows.
\begin{itemize}
\item $X$, e.g. $X(3872)$: neutral, typically seen to decay into $\psi+{\rm pions}$,  positive parity, $J^{PC}=0^{++},~ 1^{+\pm}, ~2^{++}$;
\item $Y$, e.g. $Y(4260)$: neutral, seen in $e^+e^-$ annihilation (possibly with extra Initial State Radiation, that lowers the nominal center-of-mass energy of the collision): 
$e^+ e^- \to Y$, therefore $J^{PC}=1^{--}$;
\item $Z$, e.g.  $Z(4430)$: isospin triplet (charged and neutral), typically positive parity, manifest four-valence quarks, mostly seen to decay into $\psi+\pi $, some into $h_c(1P) +\pi$ and into open charm mesons, valence quarks $c\bar c u \bar d$; 
\item $Z_b$ observed by Belle:  $Z_b(10610),~Z_b(10650)$, decaying into $\Upsilon(nS)\,\pi$, $h_b(nP)\,\pi$ and open bottom mesons, valence quarks $b\bar b u \bar d$;
\item $X$ resonances with hidden charm and hidden strangeness have been seen by LHCb, e.g. $X(4140)\to \psi+\phi$,  $J^{PC}=1^{++}$
\item Hidden charm--open strangeness ($c\bar c u \bar s$)  seen in 2021 by BES III: $Z_{cs}^+(3985)\to \psi+K^+$ and by LHCb: $Z_{cs}
^+(4003)\to \psi+K^+$.
\end{itemize}

In addition:
\begin{itemize}
\item four-charm tetraquarks have been seen as di-$J/\psi$ resonances by LHCb, e.g. $X(6900)\to \psi+\psi\to 2(\mu^+\mu^-)$
\item Pentaquark resonances have been observed by LHCb~\cite{LHCb:2019kea} in the decay channel $J/\psi+p$, valence quarks ($cuud\bar c$).
\end{itemize}
In the present Lectures, I will concentrate on $X$, $Z$ and doubly charmed tetraquarks, presenting present studies about their mass spectrum and other properties.

\emph{\bf{Generating multiquark hadrons in QCD.}}
\begin{figure}[b]
 \begin{center}
   \includegraphics[width=9.0truecm]{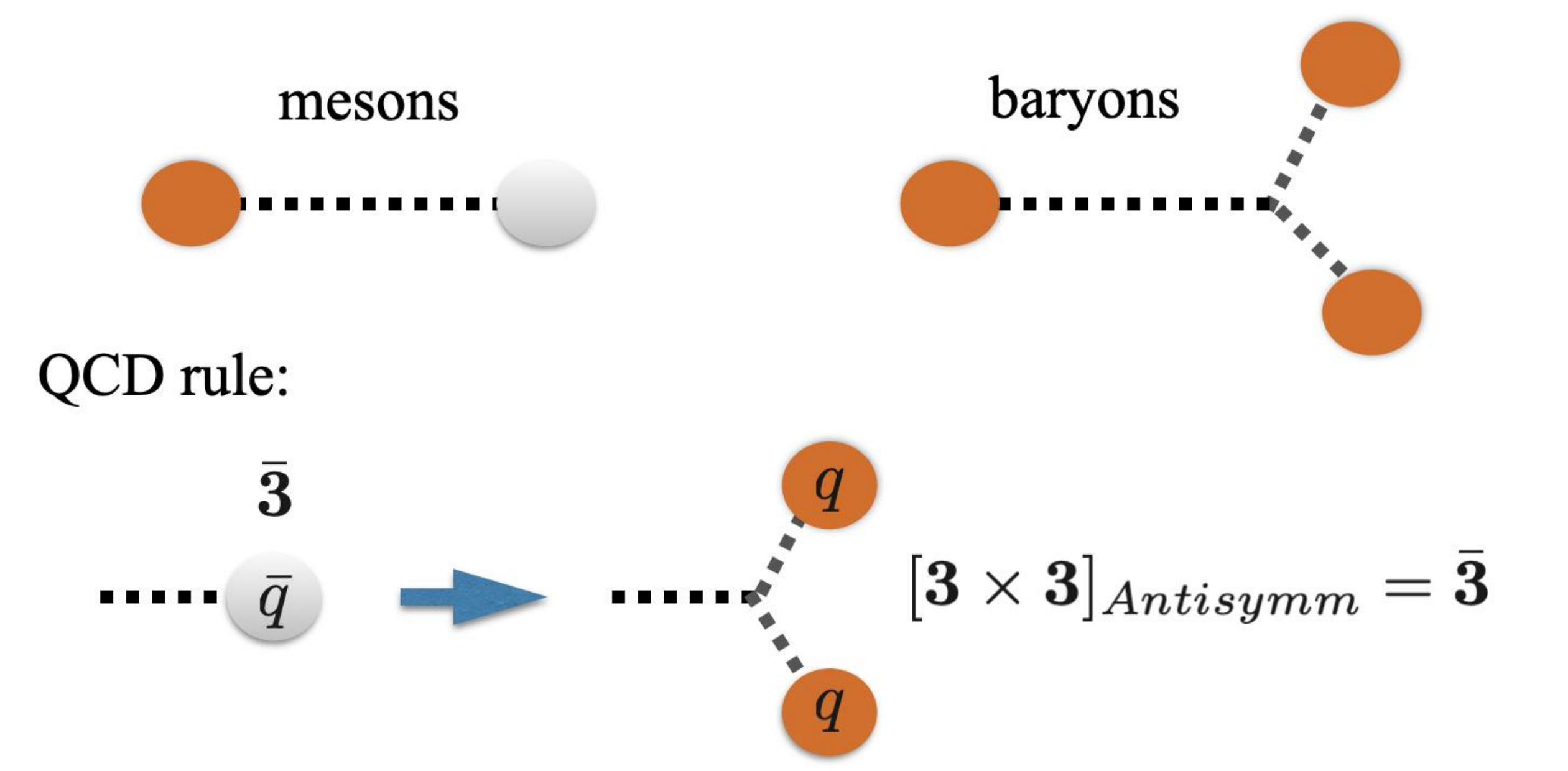}
  \caption{\footnotesize{Color strings in ordinary mesons and baryons.}}
\label{strings}
\end{center}
\end{figure}
A reasonable description of QCD forces acting inside mesons and baryons can be made in terms of oriented strings originating from quarks and ending into antiquarks. The end of QCD strings carries a representation ${\bf 3}$ of $SU(3)_{c}$, so that strings originating from the three quarks in the baryon can join in one point in a $SU(3)_c$-invariant way,\footnote{Three representations ${\bf 3}$ are combined in color singlet by the antisymmetric, invariant tensor $\epsilon^{ijk}$, see Lecture~\ref{sec:CQM}.} see baryons in Fig.~\ref{strings}.

\begin{figure}[t]
 \begin{center}
   \includegraphics[width=10.0truecm]{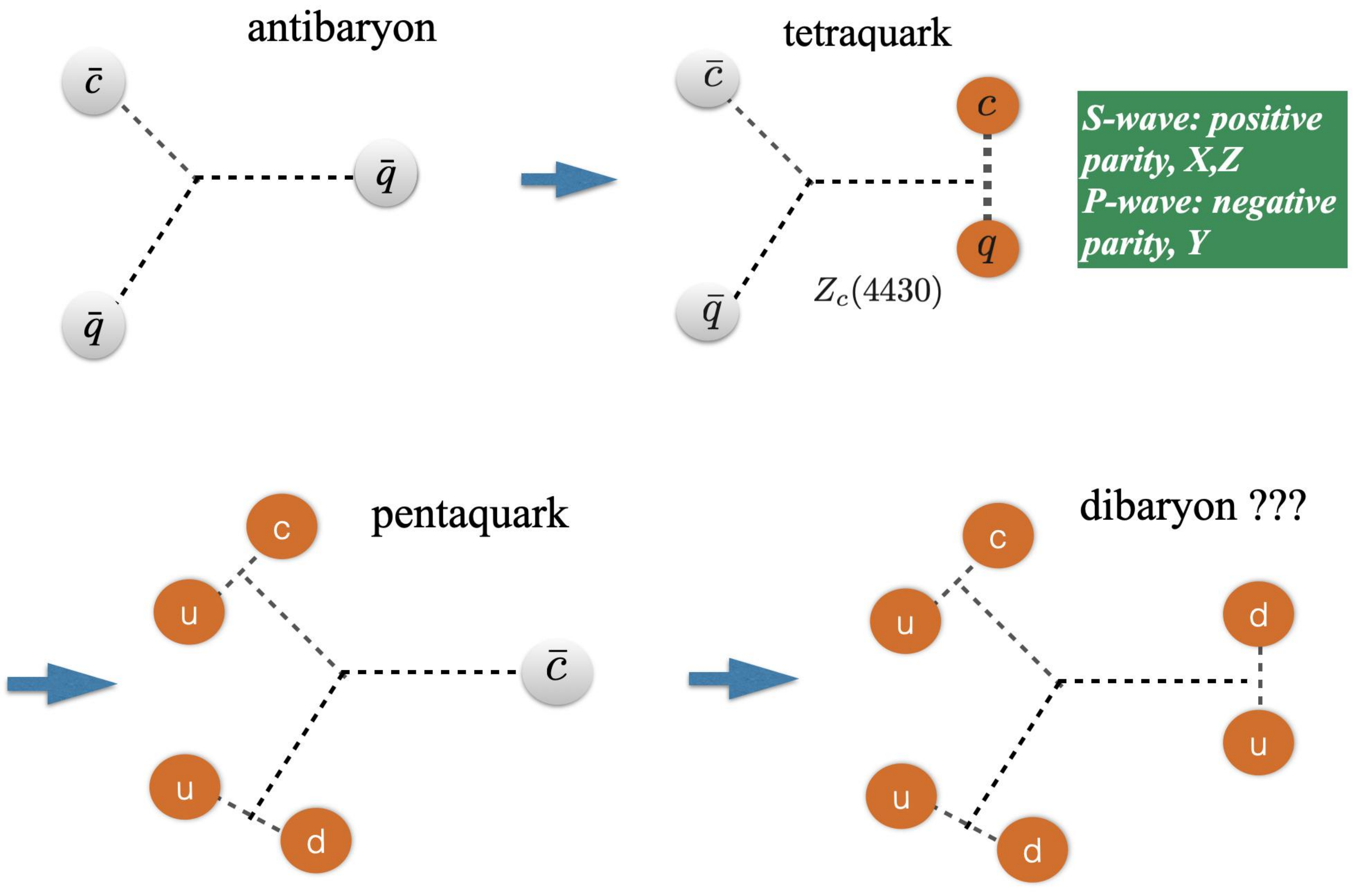}
  \caption{\footnotesize{Color strings in multiquark states.}}
\label{multiq}
\end{center}
\end{figure}

New QCD singlets  can be obtained from meson or baryon configurations  by  substituting an antiquark with a color ${\bf{\bar 3}}$  diquark, where the strings originating from the two quarks are attached to the string that was ending with the antiquark to form an invariant three-string junction, Fig.~\ref{multiq}.
The new configuration should correspond to a bound state, similarly to the original one. In particular, this does apply in the limit where the quarks in the diquark become heavy, thereby fusing into a single  ${\bf{\bar 3}}$ source equivalent to the original antiquark source, except possibly for spin-spin interactions that vanish for infinite quark mass (Lectures~\ref{sec:CQM} and~\ref{sec:bornoppe}).
 In this way, starting from the antibaryon all the known exotics, $X,~Y,~Z$, pentaquarks and more (the dibaryon) are reproduced, each corresponding to a new series of the complete hadron spectrum.
 
 The $X$ and $Z$ correspond to tetraquark ground states in $S$-wave (positive parity), the $Y$ to negative parity $P$-waves.
   Pentaquarks have negative parity in the ground $S$-wave and positive parity in $P$-wave. Dibaryons close the spectroscopic series, with positive parity in the ground, $S$-wave, state.

\emph{\bf{Molecules and Hadrocharmonia.}}
In $B$ meson decays, a $D^*\bar D$  pair with low relative momentum coud form a {\it hadron molecule}, a shallow bound state made possible by attractive one-pion exchange forces, similar to a deuteron formed from a low-momentum $pn$  pair. As proposed in~\cite{Tornqvist:1993ng}, $X(3872)$ would be composed by four quarks arranged in two color singlet mesons bound by forces similar to those that bind atomic nuclei. The closeness of $X(3872)$  mass to the $D^*\bar D$ threshold would be a natural consequence of this picture.

\begin{figure}[t]
 \begin{center}
   \includegraphics[width=9.0truecm]{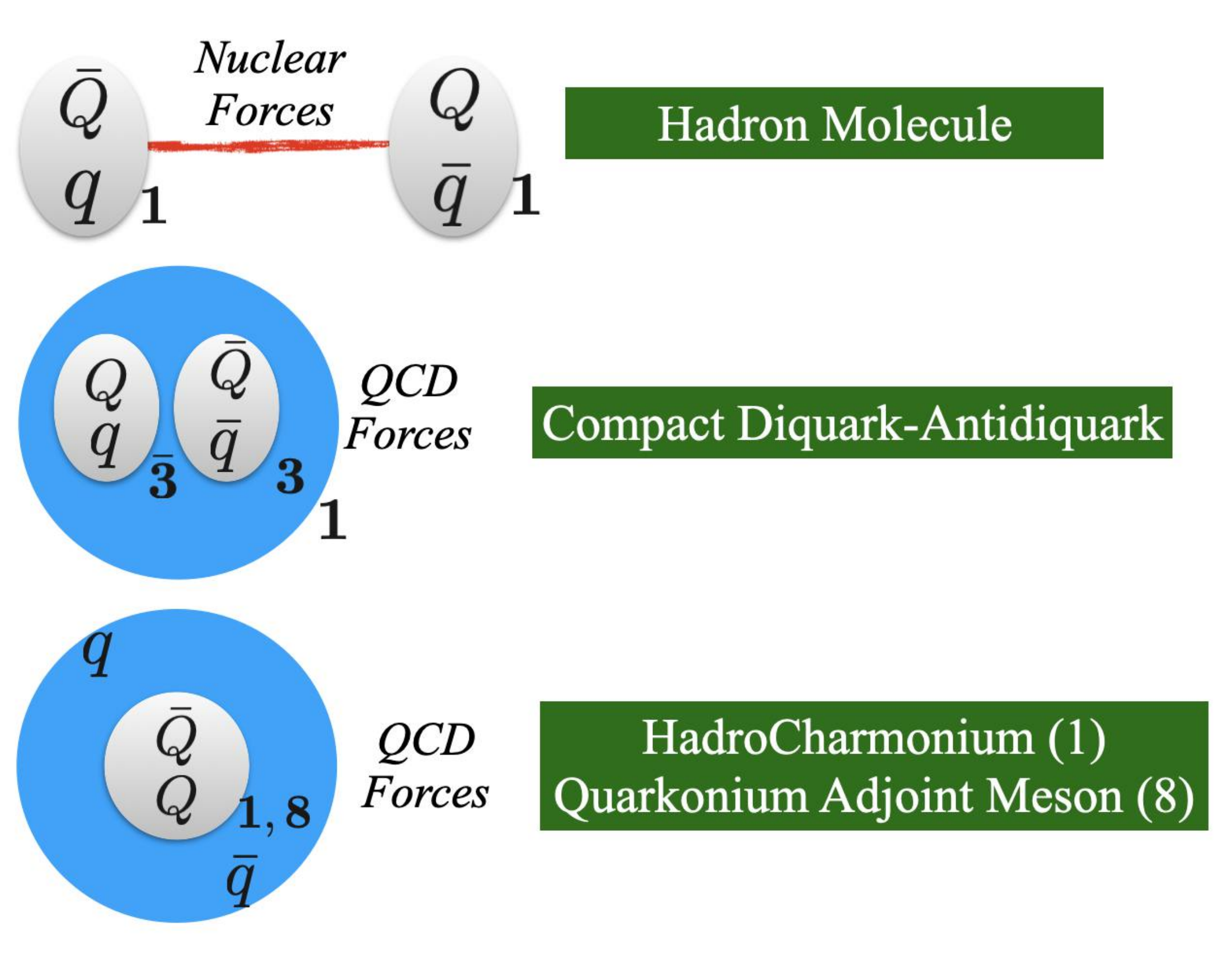}
 \caption{\footnotesize{Schematic representation of hadron molecules (see~\cite{Guo:2017jvc}), compact tetraquarks (see~\cite{Maiani:2004vq}), hadrocharmonium and quarkonium adjoint mesons (see~\cite{Dubynskiy:2008mq,Braaten:2014qka}).}}
\label{nocons}
\end{center}
\end{figure}

Another possibility, suggested by the fact that several exotic states seem to decay into a single channel with an ordinary $c\bar c$ state, is the hadrocharmonium hypothesis~\cite{Dubynskiy:2008mq} where some exotic hadrons are supposed to be made by a $c\bar c$ meson core, surrounded by a cloud of  light quark and antiquark pairs. Similarly, adjoint hadrocharmonia with the $c\bar c$ core in color octet have been proposed~\cite{Braaten:2014qka}. These possibilities are illustrated in Fig.~\ref{nocons}.

There is no consensus yet. In these lectures, focused on the compact tetraquark picture, we shall examine several facts which may eventually lead to a clarification and to the choice of the correct theory of exotic hadrons.

%% file: chapts/CQM.tex

Starting from a brief description of the way we arrived to the Standard Theory, I illustrate in this Lecture the Constituent Quark Model for meson and baryon spectrum. The model was first intruduced by Sakharov and Zeldovich~\cite{Zeldovich:1967rt}. It was formulated in QCD in~\cite{DeRujula:1976zlg} and applied to tetraquarks in~\cite{Maiani:2004vq} (see also~\cite{Georgi:1982jb,Ali:2019roi}).

\emph{\bf{First attempts.}}
Fermi and Yang, in 1949, proposed nucleons to be an elementary doublet: $F=(p, n)$ and mesons to be composite nucleon-antinucleon states
 \be
  \pi= \left(F\bar F\right)\notag
 \ee
 The natural symmetry of the Fermi-Yang model is {\it  Isospin}, the group $SU(2)$ of unitary $2 \times  2$ complex matrices with unit determinant, acting on the basic nucleon doublet

 \bea
\left(\begin{array}{c}p \\ n \end{array}\right) \to U\, \left(\begin{array}{c}p \\ n \end{array}\right), \quad U\in SU(2)
\eea
 To include strange mesons, in 1956 Sakata introduced the $\Lambda$ baryon as a new constituent, carrying negative unit of strangeness: $S= (p, n, \Lambda)$,  with:
 \be
 \pi,~K,\dots=(S\bar S),~\text{Baryons}=\left(SS\bar S\right)\label{sakmes}
 \ee
  The  basic symmetry of the Sakata model was the extension of isotopic spin to $SU(3)$, the unitary, $3\times 3$ transformations of the Sakata triplet:
 \bea
\left(\begin{array}{c} p \\ n \\ \Lambda \end{array}\right) \to U\, \left(\begin{array}{c}p \\ n\\ \Lambda \end{array}\right), \quad U\in SU(3) \label{su3}
\eea

The Sakata model reproduced well the quantum numbers of the pseudoscalar mesons observed at the time (pions and Kaons).
However, a clear prediction of the model was that {\it there must exist baryons with strangeness $S=+1$}, which, unfortunately, is a wrong prediction: no such particles have been seen until today!

\subsection{{The {Eightfold Way} (Gell-Mann and Ne'eman, 1962).}} 

Gell-Mann and Ne'eman reject the idea that there may be hadrons which are elementary and others that are composite. Treating all hadrons on the same footing (a principle called {\it Nuclear Democracy}), they keep Sakata's $SU(3)$ as the basic symmetry and try to find, in an abstract way, representations of the group (i.e. particle multiplets) that fit the observed quantum numbers of mesons and baryons.\footnote{In his  imaginative style, Gell-Mann says that the procedure is similar to the recipe of the French cuisine to cook a pheasant by putting it inside two veal steaks. When it is cooked, you eat the pheasant and throw the steaks away. In their case, the steaks were the Sakata constituents, needed to identify the symmetry, and the pheasant the $SU(3)$ symmetry, that you keep after throwing away the constituents. Gell-Mann and Zweig succeeded later to find the good constituents to go with the correct symmetry.}

\begin{figure}[b]
 \begin{center}
   \includegraphics[width=5.6truecm]{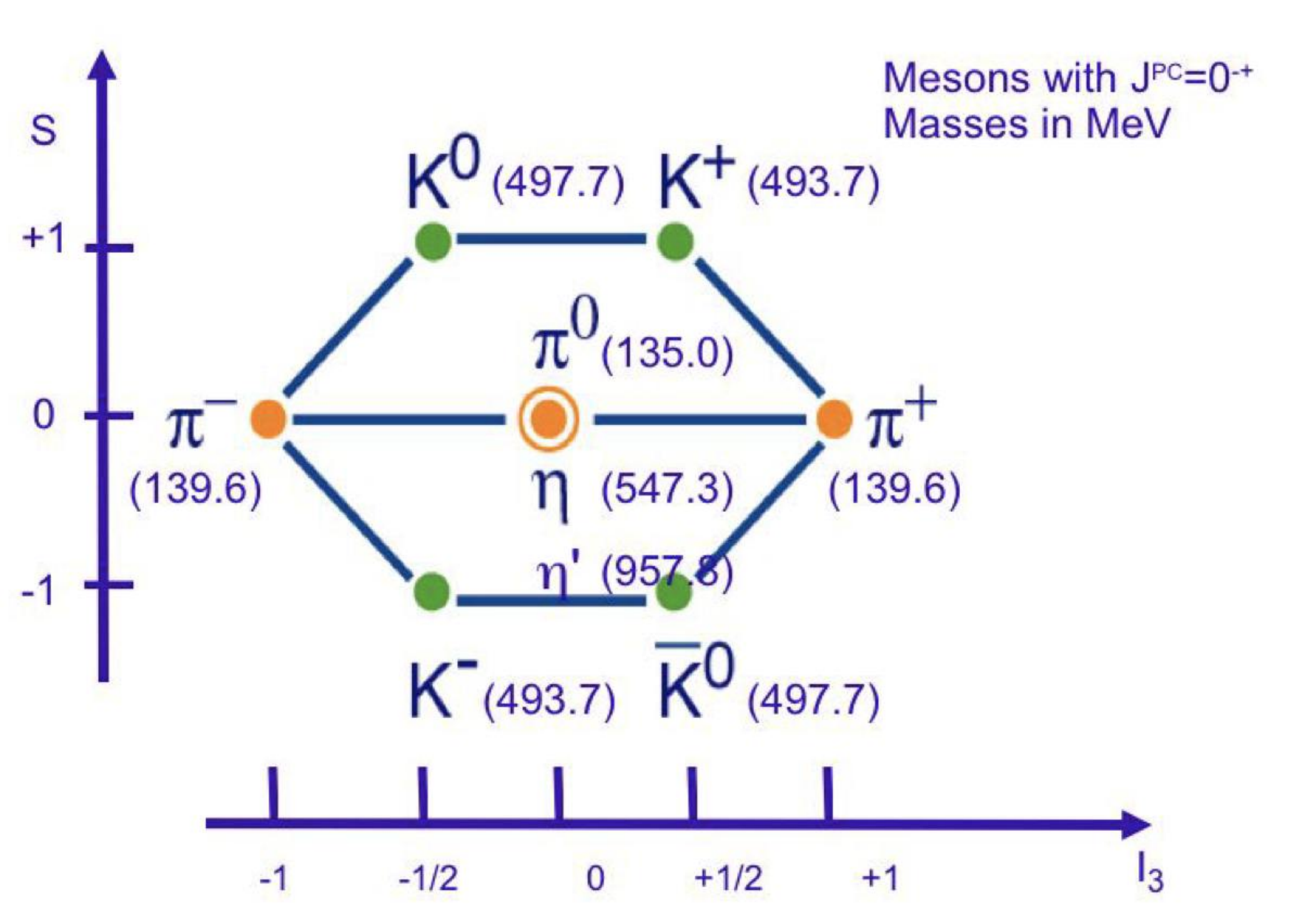} \hspace{2cm}
    \includegraphics[width=5.6truecm]{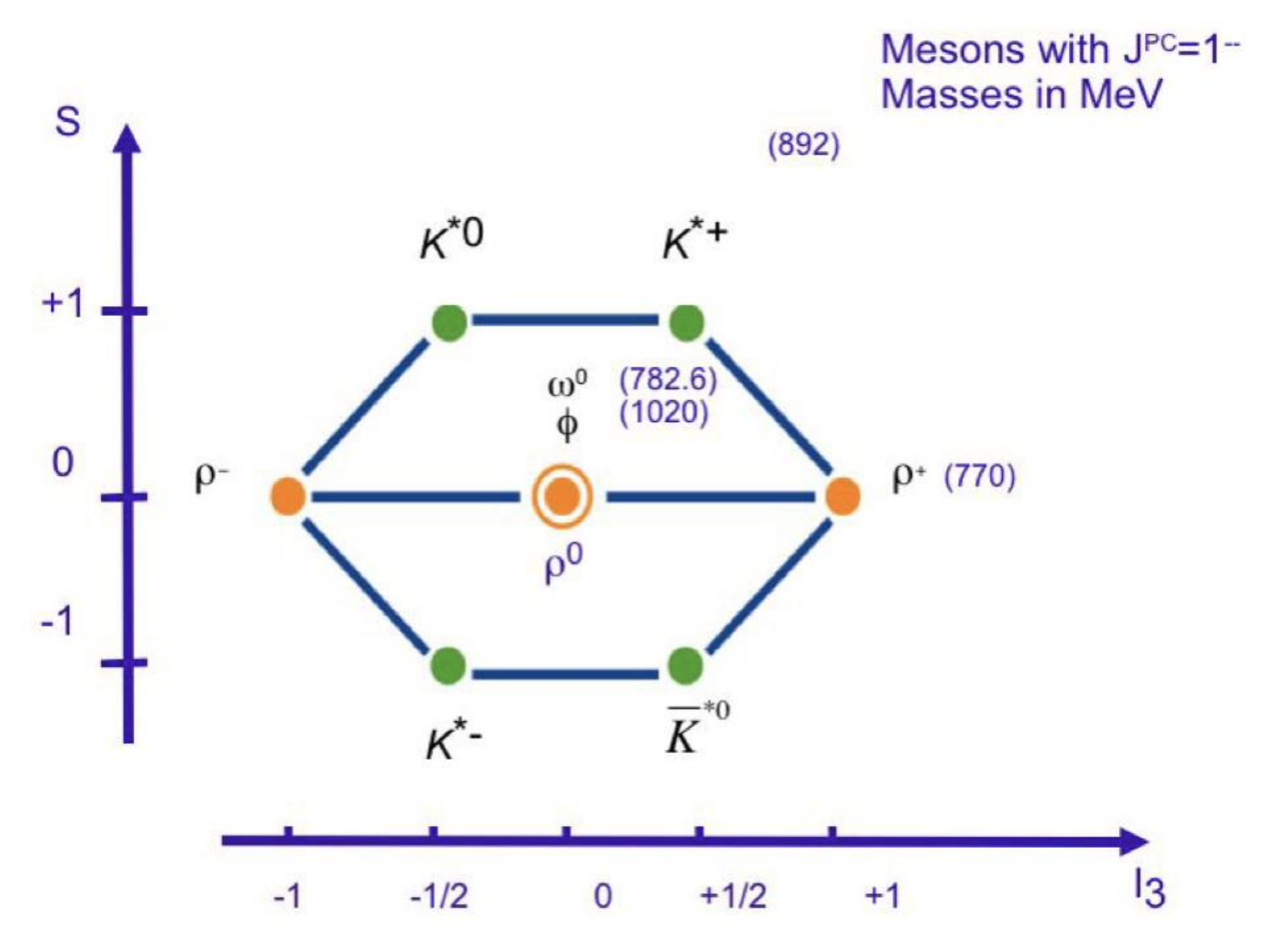}
 \caption{\footnotesize{The Eightfold Way: (left) the octet of pseudoscalar mesons; (right) the octet of vector mesons.}}
\label{8way}
\end{center}
\end{figure}

For mesons, the assignement in~\eqref{sakmes} corresponds to the eight-dimensional,  octet, and one-dimensional, singlet, representations, Fig.~\ref{8way}. 

\begin{figure}[htb!]
 \begin{center}
   \includegraphics[width=5.6truecm]{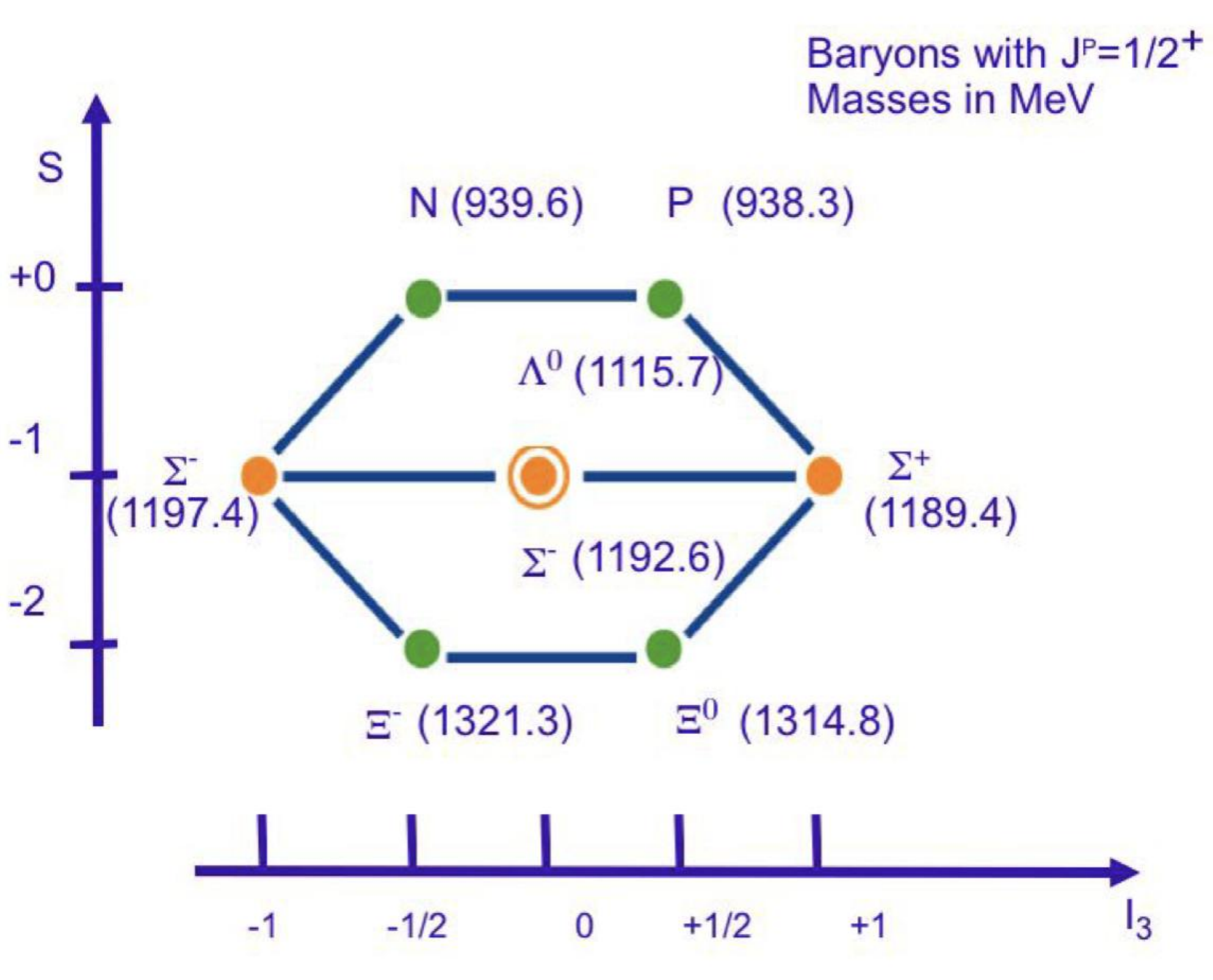} \hspace{2cm}
   \includegraphics[width=5.6truecm]{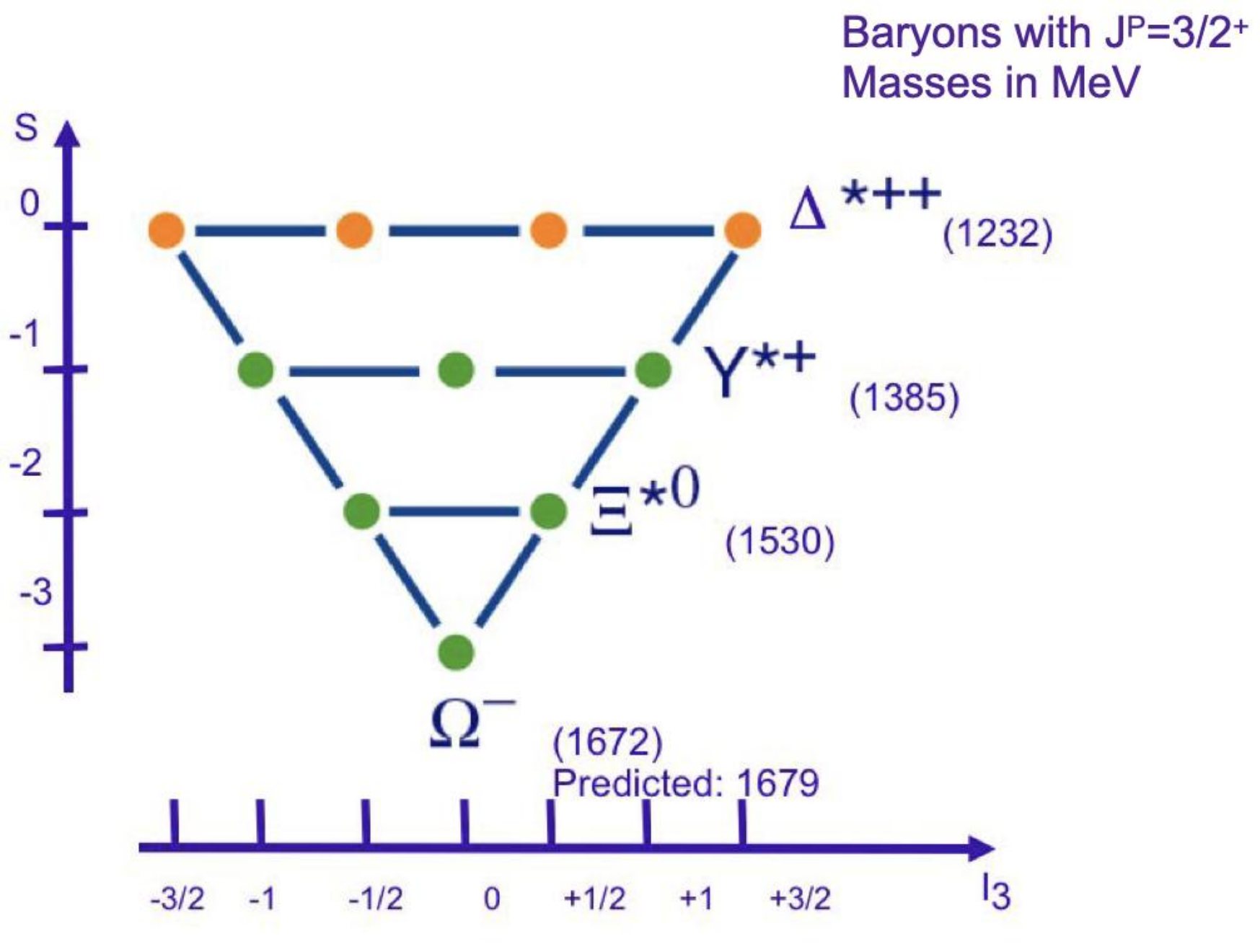}
 \caption{\footnotesize{The Eight Fold Way:(left) the octet of spin-$1/2$ baryons; (right) the decuplet of spin-$3/2$ baryons.}}
\label{8way2}
\end{center}
\end{figure}

Gell-Mann and  Ne'eman observed that the same octet representation could describe the lowest-lying spin-$1/2$ baryons (this is the {\it Eigthfold Way}), Fig.~\ref{8way2}(left). In addition, a ten-dimensional representation (decuplet) could accomodate the then observed spin-$3/2$ baryon resonances, the $ \Delta$ and the negative strangeness $Y^*$ hyperons, Fig.~\ref{8way2} (right), together with other, not yet observed, particles: a $S=-2$ isospin doublet  ($\Xi^*$) and and an $S=-3$ isosinglet ($\Omega^- $)

Assuming $SU(3)$ broken by octet interaction (see later and Lect.~\ref{sec:8fold}), Gell-Mann and Okubo derived  a very well verified mass-formula for the 
 octet baryons:
 
 \be
 \frac{N+\Xi}{2}~(1128\mev)=\frac{3\Lambda+\Sigma}{4}~(1136\mev)\label{gmok}
 \ee
 In the same approximation, decuplet masses are equally spaced: from $\Delta$ and $\Sigma^*$ masses one could predict $\Xi^*$ and $\Omega$ masses. 

The discovery of two $\Xi^*$ particles was presented at the Ginevra Conference, 1962, and Gell-Mann observed there that their mass verify the equal spacing rule. $\Omega^-$ was discovered in 1964 with the expected mass:

\begin{itemize}
 \item Predicted (Gell-Mann, 1962) $M=1679\mev$;
\item Observed (BNL, N. Samios and Coll., 1964) $M= 1672~\mev$.
\end{itemize}

It was the first prediction of mass and quantum number in particle physics, $SU(3)$ symmetry was established. 

 \subsection{Interactions of Fundamental Particles, a first look}

In general terms, three interactions are operative at particle level, distinguished by strength and selection rules.
\begin{enumerate}
\item Strong interactions, $\mathcal{O}(1)$: act on hadrons. Conserved quantities:
\begin{itemize}
\item $P$ (parity), $C$ (charge conjugation), $T$ (time reversal);
\item  $I$ (isospin), symmetry group $SU(2)$: $I^2=I(I+1)$, $-I \leq I_3 \leq +I ,~n(I)=2I+1$
\item $Q$ (electric charge), $S$ (strangeness), $B$ (baryon number);
\item  also introduced $Y$ (hypercharge) with:
\be
Q=I_3+\frac{S+B}{2}=I_3+\frac{Y}{2}~\text{(Gell-Mann~and~Nishijima~formula)}
\ee 
\item $L_{e,\mu,\tau}$ (lepton numbers) and $B$ (baryon number)
\item typical lifetime $\sim 10^{-23}~\text{sec}$ ($\Gamma \sim 100\mev$).
\end{itemize}
\item Electromagnetic interactions, $\mathcal{O}(1/137)$: act on hadrons and charged leptons ($e,~\mu$).  E.m. interactions violate $I$ and conserve:
\begin{itemize}
\item $P$, $C$, $T$; 
\item $Q$ and $S$;
\item $L_{e,\mu,\tau}$ and $B$;
\item typical lifetimes $\sim 10^{-18}~\text{sec}$.
\end{itemize}
\item Weak Interactions (Fermi, 1932) $\mathcal{O}(10^{-5})$. WI act on all particles, including neutrinos,  with: 
\begin{itemize}
\item $P$, $CP$ and $T$ violated; $CPT$ conserved 
\item  $B$ conserved,
\item WI violate: $S$, $L_{e,\mu\,\tau}$ (because of neutrino oscillations, but total $L$ is conserved).
\item typical lifetimes $\sim 10^{-12}~\text{sec}$ or longer.
\end{itemize}
\end{enumerate}

\subsection{Quarks!}

 $SU(3)$ representations and symmetry breaking can be studied by pure group theory, but quarks are much simpler to handle.

Quarks were introduced as hypothetical spin-$1/2$ particles in the basic $SU(3)$ triplet, the first fundamental representation,\footnote{ We denote $SU(3)$ representations by their dimensionality and indicate the complex conjugate representation with a bar.} {\bf  3}. Antiquarks are in an {\it anti-triplet} the second fundamental representation, ${\bf {\bar 3}}$:
\be
q= \left(\begin{array}{c} u \\ d\\s \end{array}\right), \quad\bar q=\left( \bar u,~\bar d,~\bar s\right)
\ee

\begin{table}[b]
\centering
    \begin{tabular}{||c|c|c|c|c||}
     \hline
quark & $I_3$& $Y$ & ($S$) & $Q$ \\ \hline
$u$ & $1/2$ & $1/3$ & ($0$) & $+2/3$ \\ \hline
 $d$ &$-1/2$ & $1/3$ & ($0)$ & $-1/3$ \\ \hline
  $s$ &$0$ & $-2/3$ & ($-1)$ & $-1/3$ \\ \hline
\end{tabular}
 \caption{\footnotesize {Quark quantum numbers.}}
\label{qqn}
\end{table}

With quarks and antiquarks of spin $1/2$, we should be able to construct all hadrons (forget Fermi statistics for a while, we'll come back). How do we make mesons and baryons? Following Gell-Mann, baryons are ($qqq$), see Eq.~\eqref{threeqs} below, while mesons are  ($q\bar q$)~\cite{GellMann:1964nj}. The same combinations have been independently suggested by George Zweig, who used the name {\it aces} for the particles of the fundamental triplet~\cite{Zweig:1981pd}.

\emph{\bf{Irreducible $SU(3)$ tensors and multi quark/antiquark constructions.}} $SU(3)$ is a group with two commuting generators: $I_3$ and $Y$. 
The irreducible multiplets are characterised by two integers $n_1$, $n_2$ 
and are represented by {\it standard tensors} with $n_1$ upper and $n_2$ lower indices, symmetric in the upper and in the lower indices and traceless, see~\cite{Coleman:1985rnk}:
\be
\hat T^{a_1 a_2\dots a_{n_1}}_{b_1 b_2\dots b_{n_2}},\qquad \hat T^{a a_2\dots a_{n_1}}_{a b_2\dots b_{n_2}}=0
\ee
 (sum over repeated indices understood). Quarks (antiquarks) are represented by three-dimensional vectors: $q_b$ ($\bar q^a$).
Products of quarks and antiquarks are tensors with both upper and lower indices, in general neither symmetric nor traceless. To express them as sums of standard tensors, we project with the $SU(3)$ invariant  operations: 
\begin{itemize}
\item symmetrisation/antisymmetrisation
\item contraction with $\delta^a_{b}$ (eliminates one upper and one lower indices)
\item contraction with $\epsilon^{abc}$ (transforms 1 (2)  lower indices in 2 (1) upper indices)
\item contraction with $\epsilon_{abc}$  (same for upper indices into lower indices).
\end{itemize}

To warm up, we work out explicitly a few tensor product decompositions. We put the hat over the standard tensors, symmetric and traceless.
\begin{itemize}
\item MESONS: ${\bf 3}\otimes {\bf{\bar 3}}$
\begin{align}
q_b\bar q^a & =\hat{T}^a_b+\frac{1}{3} \delta^a_b(\delta^c_d T^d_c)=\hat{T}^a_b+\frac{1}{3} \delta^a_bT,~ (\hat{T}^a_b~{\rm traceless}) \notag \\
{\rm dim}~\hat{T}&=3\cdot 3-1,~{\rm i.e.} ~{\bf 3}\otimes {\bf{\bar 3}}={\bf 8}\oplus {\bf 1}  \notag 
\end{align}
\item TWO QUARKS: ${\bf 3}\otimes {\bf 3}$
\begin{align}
q_a q_b & =S_{\{ab\}}+ A_{[ab]}, \quad S ={\rm symmetric},~A={\rm antisymmetric}
 \notag \\
A_{[ab]}&=\epsilon_{abc} \hat T^c  \notag
\end{align}
A symmetric tensor with two indices has 6 independent components, so that: 
\be
{\bf 3}\otimes {\bf 3}={\bf 6}\oplus{\bf{\bar 3}} \notag
\ee
\item PRELIMINARY TO THREE QUARKS: ${\bf 3}\otimes {\bf 6}$
\bea
&&q_a T_{\{bc\}}=\frac{1}{3} T_{abc}+\left(\epsilon_{abd}T^d_c + \epsilon_{acd}T^d_b\right)\notag \\
&&{\rm where}:~T^d_c=\frac{1}{3}\epsilon^{kld}q_kT_{\{lc\}}, ~{\rm note:}~T^d_d=0    \notag
\eea
so that:~${\bf 3}\otimes {\bf 6}= {\bf 8}\oplus {\bf 10}$ (why the fully symmetric tensor $T_{\{abc\}}$ has 10 dimension?\footnote{ Start from the tensor component $T_{333}$ (1 component), then consider $T_{33i}, i=1,2$ (two components), then $T_{3ij}, i,=1,2$: these are the same as the components of a spin 1 (3 components), finally  $T_{ijk},~i,j,k,=1,2$ these are the  components of a spin $3/2$ (4 components). We have enumerated all independent components of the {\it fully symmetric} tensor with three indices. In total: $1+2+3+4=10$.})
\end{itemize}

\emph{\bf{Are quarks real?}}
States with three quarks are composed according to
\be
{\bf 3}\otimes {\bf 3}\otimes {\bf 3}=\left({\bf {\bar 3}}\oplus {\bf 6}\right)\otimes {\bf 3}={\bf 1}\oplus {\bf 8}\oplus {\bf 8}\oplus {\bf {10}} \label{threeqs}
\ee
This is really an extraordinary result: 
\begin{itemize}
\item three quarks reproduce the baryon multiplets with negative or vanishing strangeness (octets, decuplets, singlets), as required by data; 
\item all observed baryons are treated equally (in agreement with {\it nuclear democracy}).
\end{itemize}
Two puzzling features remain: 
\begin{itemize}
\item quarks are fractionally charged: the lightest quark is absolutely stable; stable quarks liberated from cosmic rays high energy collisions should remain dispersed  in matter;
\item however, Millikan's like experiments have not been able to observe fractionally charged quarks in the environment;
\item no fractionally charged stable or metastable particles has ever been observed in high energy collisions.
\end{itemize}
The other problem arises from the observation that the lowest resonance $\Delta^{++}$, reasonably supposed to be an $S$-wave state (symmetric under coordinate exchange), should be composed as:
\be
\Delta^{++}=u^{\uparrow}u^{\uparrow}u^{\uparrow}
\ee
What about Fermi statistics? 

The opinion prevailing  in the Sixties was to consider quarks purely as a mathematical shorthand to summarise the solutions of the (unknown) basic equations of strong interactions, in presence of an $SU(3)$ symmetry. 

\subsection{Fermi Statistics leads to quark colour and QCD}

The Fermi statistics problem hints at the existence of further quantum numbers for $u$ (and $d$ and $s$) quarks. This was first suggested by Han and Nambu (HN) who made the hypothesis that the additional indices could be associated with a hidden gauge group, they called $SU(3)^\prime$, responsible for the strong interactions among quarks.  

In 1972, Gell-Mann, Fritzsch and Leutwyler reproposed the HN scheme with the additional assumption that the hidden gauge group commutes  with the gauge group  $SU(2)_L\otimes U(1)_Y$ introduced by Weinberg and Salam for electroweak unification.  
The quantum number of the hidden group was  called {\it colour} and the group $SU(3)_{colour}$  by G-MFL. The authors introduced the term  Quantum ChromoDynamics, QCD, to underline the analogy to Quantum ElectroDynamics of the colour forces generated by the exchange of the quanta of $SU(3)_c$.

The notion of hidden {\it three colours} was almost immediately supported by experiments at $e^+ e^-$ colliders, ADONE in Frascati and SPEAR in Stanford, that showed that the opening of the threshold for the production of pairs of quarks with a given flavour (u, d, s and, later, charm) was associated with an increase of the cross-section approximately equal to
\be
\Delta \sigma_Q= 3\times Q^2\times  \sigma(e^+ e^- \to \mu^+\mu^-)  \notag
\ee
where $Q$ is the electric charge reported in Tab.~\ref{qqn}.

\begin{figure}[t]
 \begin{center}
   \includegraphics[width=10.6truecm]{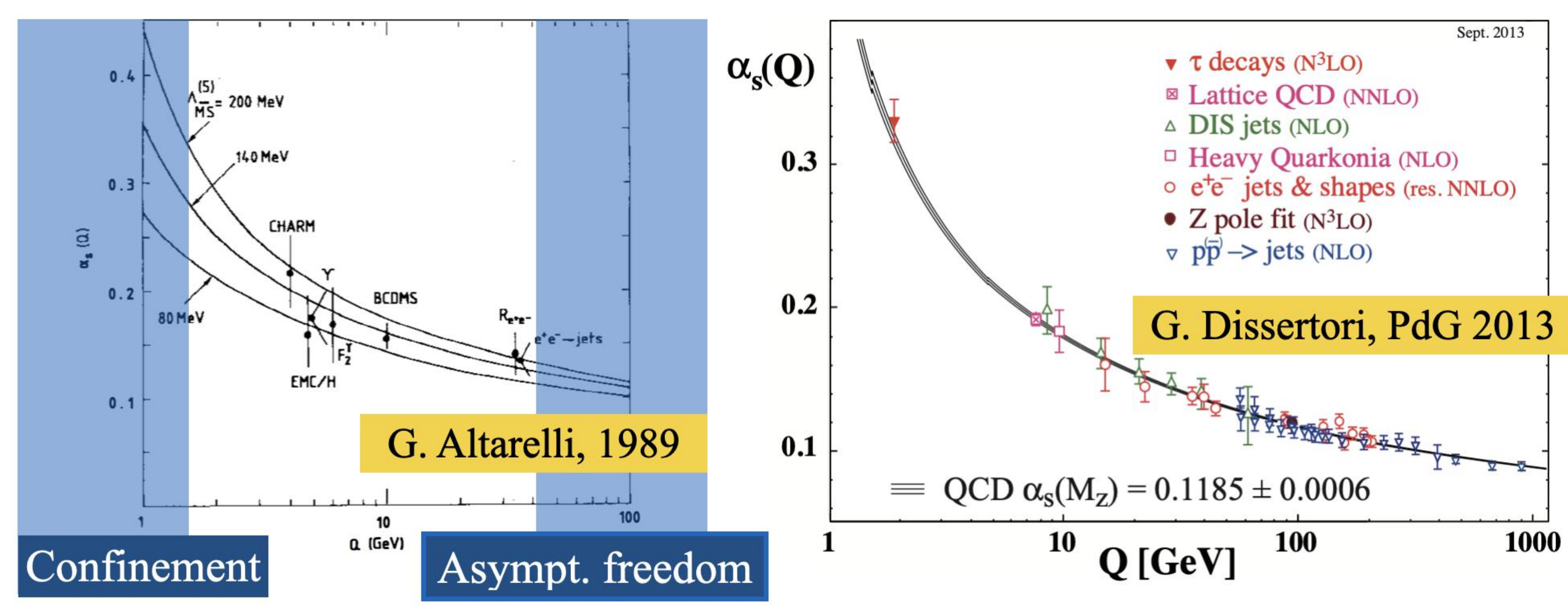}
  \caption{\footnotesize{Asymptotic Freedom is signalled by the decrease of the coupling constant at large $q^2$.}}
\label{asfreed}
\end{center}
\end{figure}

In 1973, Gross and Wilczek \cite{Gross:1973id} and, independently,  Politzer \cite{Politzer:1973fx}, discovered that non-abelian gauge groups, such as  $SU(3)_c$, are {\it asymptotically free} at large euclidean momenta, if the number of fermions is not too large. The result showed the potential for QCD  to explain the scaling rules observed in deep inelastic scattering processes of electrons on protons, which had indicated quasi-free behaviour of the constituents of the proton observed at large, negative values (euclidean!) of the momentum transfer squared, $q^2$.
A proton at rest is made by 3 quarks dressed by strong QCD interactions. Increasing $q^2$, quarks radiate gluons (the Altarelli-Parisi picture of scaling violations). At large $q^2$, we see quarks and neutral gluons as almost free partons. 
Already Han and Nambu had  observed that quark states in a color non-trivial representation should most likely have an infinite energy. This would imply the permanent confinement of color triplet quarks inside finite-energy, color singlet bound states.
Confinement in QCD has  not yet been proved but it is made  plausible by several  qualitative considerations. There are in fact  indications that a gauge Yang-Mills theory coupled with matter fields can live in two phases only: a spontanously broken Higgs phase if coupled to scalar particles (such as the electroweak sector), or a confined phase (such as QCD).

\begin{figure}[t]
 \begin{center}
\ 
   \includegraphics[width=16.0truecm]{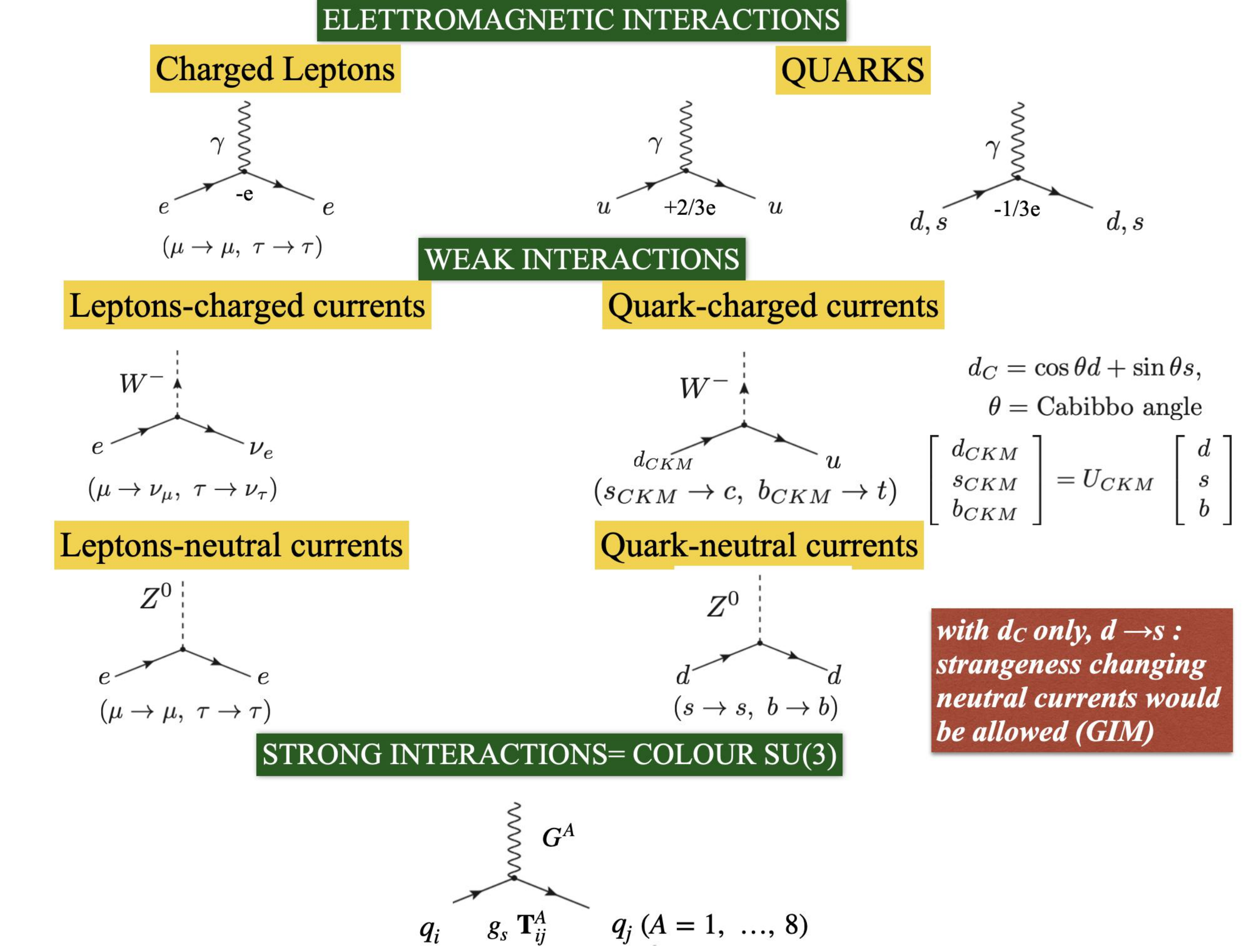}
   \end{center}
 \caption{\footnotesize{Elementary interactions of quarks with the gauge  fields of the Standard Theory, with gauge group $SU(2)_L\otimes U(1)_Y\otimes  SU(3)_c$.}}
\label{stint}
\end{figure}

The asymptotically free behaviour of  QCD has been confirmed  by the data of deep inelastic hadron production, at LEP in $e^+e^-$ collisions, and at the LHC (proton-proton collisions), as shown in Fig.~\ref{asfreed}.
Quarks carry  $SU(3)$ colour symmetry and are confined inside color singlet hadrons, e.g. $\Delta^{++}=\epsilon^{\alpha \beta\gamma}u^\uparrow_\alpha u^\uparrow_\beta u^\uparrow_\gamma$.  Fermi statistics is obeyed. There is only one way to make a color singlet with three quarks, so there is no particle proliferation due to the new quantum number.
The elementary interactions of quarks with the gauge  fields of $SU(2)_L\otimes U(1)_Y\otimes  SU(3)_c$, that we indicate as the Standard Theory of fundamental interactions, are illustrated in Fig.~\ref{stint}.


\subsection{The Constituent Quark Model }

\emph{\bf{Forces between coloured objects in the one gluon exchange approximation.}}

The one-gluon exchange process between two coloured objects gives an amplitude of the form
\be
{\cal M}_{1,2}=\frac{g_s^2}{k^2}\cdot \left\langle R\,\middle|\,T^A_1\,T^A_2\,\middle|\,R\right\rangle\label{oge}
\ee
where $T^A_{1,2}$ are the  matrices representing the generators of $SU(3)_c$ on objects 1 or 2, and the sum over the repeated indices A is understood
\footnote{ $T^A_{1,2}$ act on different spaces. One should write, more precisely $T^A_1=T^A_1\otimes {1}_2$,  $T^A_2={1}_1\otimes T^A_2 $. $T^A_1$ and $T^A_{2}$  obviously commute.}.  $R$ indicates the colour representation of initial and final states, which is the same since colour is conserved.
The generator of the whole $1+2$ system is, of course
\be
T^A_{1+2}=T^A_1+ T^A_2
\ee
and obviously (sum over $A$ understood):
\be
T^A_{1+2}\, T^A_{1+2}=T^A_1\, T^A_1+ T^A_2\,T^A_2 +2\, T^A_1\,T^A_2  \notag
\ee
so that
\be
\left\langle R\,\middle|\, T^A_1\,T^A_2 \,\middle|\,R\right\rangle=\frac{1}{2}\left[C_2(R)-C_2(1)-C_2(2)\right] \label{casimir}
 \ee
where for a generic color representation $R$, we have defined
\be
C_2(R)=T^A_R\, T^A_R
\ee
$C_2(R)$ is called the {\it quadratic Casimir operator} of the representation $R$. 

The amplitude \eqref{oge} is finally written as
\be
{\cal M}_{1,2}=\frac{g_s^2}{k^2}\cdot \frac{1}{2}\left[C_2({R})-C_2(1)-C_2(2)\right]=\frac{g_s^2}{k^2}\cdot \lambda({R})
\ee

\begin{table}[b]
\centering
    \begin{tabular}{||c|c|c||}
     \hline
$R$ & $C_2(R)$ & $\lambda(R)$  \\ \hline
${\bf 1}$ & $0$ &$-4/3$ \\ \hline
 ${\bf 8}$ &$3$ & $+1/6$\\ \hline
  ${\bf 3},~{\bf{\bar 3}}$ &$4/3$& $-2/3$ \\ \hline
   ${\bf 6}$ & $10/3$&$+1/3$ \\ \hline
\end{tabular}
 \caption{\footnotesize {Casimir eigenvalues and effective charge values, depending on the representation of the two (anti)quarks.}}
\label{casi2}
\end{table}

As indicated in Tab.~\ref{casi2}, the Casimir dependence of the amplitude shows  an interesting pattern of forces vs. $R$.
\begin{enumerate}
\item quark-antiquark:  
\begin{itemize}
\item $R=\text{singlet}$: attractive ($-4/3$) 
\item $R=\text{octet}$: repulsive ($+1/6$) 
\end{itemize}
\item quark-quark
\begin{itemize}
\item $R=\text{anti-triplet}$ attractive ($-2/3$) 
\item $R= \text{sextet}$: repulsive ($+1/3$)
 \end{itemize}
 \end{enumerate}
In conclusion: quark-antiquark pairs bind in color singlet mesons, while diquarks may bind:
\begin{itemize}
\item to another quark, to make a color-singlet baryon 
\item to an antidiquark, to make a color-singlet tetraquark.
\end{itemize}

Non perturbatively, color lines of force are supposed to condense in strings going from quarks to antiquarks.

Color string forces produce an overall spin-independent potential that confines quarks inside a definite volume (bag), with some wave functions.
Residual quark-quark or quark-antiquark interactions are local chromomagnetic, spin-spin, interactions of the form~\cite{Godfrey:1985xj}: 
\be
H_{ij}=-\frac{2}{3}\frac{g_s^2}{m_i m_j}\left(T^A_1\,T^A_2 \right)\left({\bf s}_1\cdot {\bf s}_2\right) ~\delta^{(3)}({\bf x}_1-{\bf x}_2)  \notag
\ee
                            
The $T^A$ and $\bf s$ are color and spin operators, and $g_s$ the strong coupling. This form is derived from the non relativistic limit of QCD. 
If $i,j$ are in a color representation $R$, the formula simplifies to;      
\be
{\bf H}_{ij}=2\kappa_{ij}~\left({\bf s}_i\cdot {\bf s}_j\right),\quad\kappa_{ij}=-\lambda(R) \times \frac{g_s^2}{3m_im_j}|\psi(0)|^2\label{hfcoup}
\ee

The Hamiltonian can be developed to first order in the small mass difference $m_s-m_{u,d}$,
there is also a first order contribution from the spin-spin interaction, which is very crucial for the mass difference of $\Sigma$-$\Lambda$ baryons.
The formula works well for mesons and baryons, as we shall see now.

Few parameters: $m_{u, d}$, $m_s$, $m_c$, $m_b$, $\kappa_{ij}$ reproduce masses of different hadrons with a $\sim 30\mev$ accuracy, $\kappa_ {ij}$  scale approximately like $1/m_im_j$ and are not far from scaling with the color factors.

One usually assumes that the wave function overlap is the same for all mesons and for all baryons, but this is dubious in the case of hadrons with a very heavy quark, $c$ or $b$. We consider explicitly the cases of light mesons and baryons. The basic ingredient is the formula
\be
2({\bf s}_1\cdot {\bf s}_2)={\bf J}^2-{\bf s}^2_1-{\bf s}^2_2=J(J+1)-s_1(s_1+1)-s_2(s_2+1)=
\left\{\begin{array}{c} +1/2 ~(J=1)\\ -3/2 ~(J=0)\end{array}\right.
\ee

\emph{\bf{Light meson masses}}

We assume ($q=u,~d$):  
\begin{itemize}
\item $\pi=(q\bar q)$, S-wave spin-singlet, $J=0$; 
\item $\rho,~\omega= (q\bar q)$, S-wave spin-triplet, $J=1$ 
\end{itemize}
 and find:
\be
m_\pi=2m_q-\frac{3}{2} \kappa_{q\bar q};\qquad m_\rho=m_\omega =2m_q+\frac{1}{2} \kappa_{q\bar q}
\ee

We leave out $\eta$, mixed in a complicated way with $\eta^\prime$. Similarly for  $(q\bar s)$ mesons, we find
\begin{align}
m_K&=m_q+m_s-\frac{3}{2}\kappa_{q\bar s};\qquad m_{K^*}=m_q+m_s+\frac{1}{2} \kappa_{q\bar s}\notag\\
m_\phi&=2m_s+\frac{1}{2}\kappa_{s\bar s}
\end{align}
With 5 parameters (2 masses and 3 kappas) and 6 masses, we get the equality of $\rho$ and $\omega$ masses.
However, to first order in the mass difference: $\kappa_{q\bar s}-\kappa_{q\bar q}=\delta_\kappa$ and $\kappa_{s\bar s }-\kappa_{q\bar q}=2\delta_\kappa$. This leaves us with one parameter less and another relation. In conclusion we find (masses in\mev):
\bea
&& m_\rho(775)=m_\omega(783)\notag \\
&& m_\phi-m_{K^*} (128)=m_{K^*}-m_\rho (117)
\eea

\emph{\bf{Octet, spin 1/2 baryons.}}
\begin{enumerate}
\item $p$, $uud: J=1/2$ and ${\bf{s}_{u_1}}+ {\bf{s}_{u_2}} =1$ (Fermi statistics, $u_1$ and $u_2$ are antisymmetric in color, so must be symmetric in spin!)
\bea
&&H_N=3m_q+2\kappa_{qq} \Big[{\bf{s}_1}\cdot{\bf s}_2+ {\bf s}_d({\bf{s}_{u_1}}+ {\bf{s}_{u_2}})\Big]\notag\\
&& m_N=3m_q+\kappa_{qq}\Big[(2-\frac{3}{2})+(\frac{3}{4}-2-\frac{3}{4})\Big]=3m_q-\frac{3}{2}\kappa_{qq}
\eea
\item $\Sigma^+, uus$: same trick as $p$, with $d \to s$ 
\be
m_\Sigma=2m_q+m_s +\frac{1}{2}\kappa_{qq}-2\kappa_{qs}
\ee

\item $\Lambda, uds$: $ud$ is in Isospin 0 $\to {\bf s}_u+{\bf s}_d =0$ (Fermi statistics, again) 
\be
m_\Lambda=2m_q+m_s-\frac{3}{2}\kappa_{qq}+(\frac{3}{4}-\frac{3}{4})\kappa_{qs}= 2m_q+m_s-\frac{3}{2}\kappa_{qq}
\ee
\item $\Xi, uss$, with $ss$ in spin 1: proceed like $\Sigma$, with $u\to s$
\be
m_\Xi=m_q+2m_s-2\kappa_{qs}+\frac{1}{2}\kappa_{ss}
\ee
\end{enumerate}
 
We have 4 masses ($N,\Lambda,\Sigma,\Xi$) and 4 parameters: $m_q-1/2\kappa_{qq}, m_s-m_q,\kappa_{qs}-\kappa_{qq}, \kappa_{ss}-\kappa_{qq}$. 
However, if we use the relation $\kappa_{qs}-\kappa_{qq}=\delta_\kappa$, and $\kappa_{ss}-\kappa_{qq}=2\delta_\kappa$, we obtain one relation: the Gell-Mann-Okubo relation, anticipated in Eq.~\eqref{gmok}
\be
 \frac{N+\Xi}{2}~(1128~{\rm MeV})=\frac{3\Lambda+\Sigma}{4}~(1136~{\rm MeV})\notag
 \ee

\emph{\bf{Decuplet, spin 3/2 baryons.}}

\begin{itemize}
\item $\Delta = uuu$:
\be
 m_\Delta= 3 m_q+\frac{3}{2}\kappa_{qq}
 \ee
This gives  the right combination to separate $m_q$ from $\kappa_{qq}$ and we can compute all decuplet masses in term of known parameters
\item we find equally spaced masses with spacing
\be
\Delta m =m_s -m_q +\kappa_{qs} -\kappa_{qq} =139
\ee
\item experimentally
\bea
&& m_{\Sigma^*}-m_\Delta=153\label{eqspac}\\
&&m_{\Xi}-m_{\Sigma^*}=145\notag\\
&&m_{\Omega}-m_{\Xi}=142\notag
\eea
\item The addition to Gell-Mann Okubo is the prediction of the mass difference in \eqref{eqspac} (i.e. 139=153), of similar quality as the other octet and decuplet mass relations: 8 masses, 4 parameters, 4 relations, satisfied within 20 MeV.  
\end{itemize}

\subsection{Spectroscopy of charm and beauty mesons and baryons }

\begin{figure}[htb!]
 \begin{center}
\ 
   \includegraphics[width=4.0truecm]{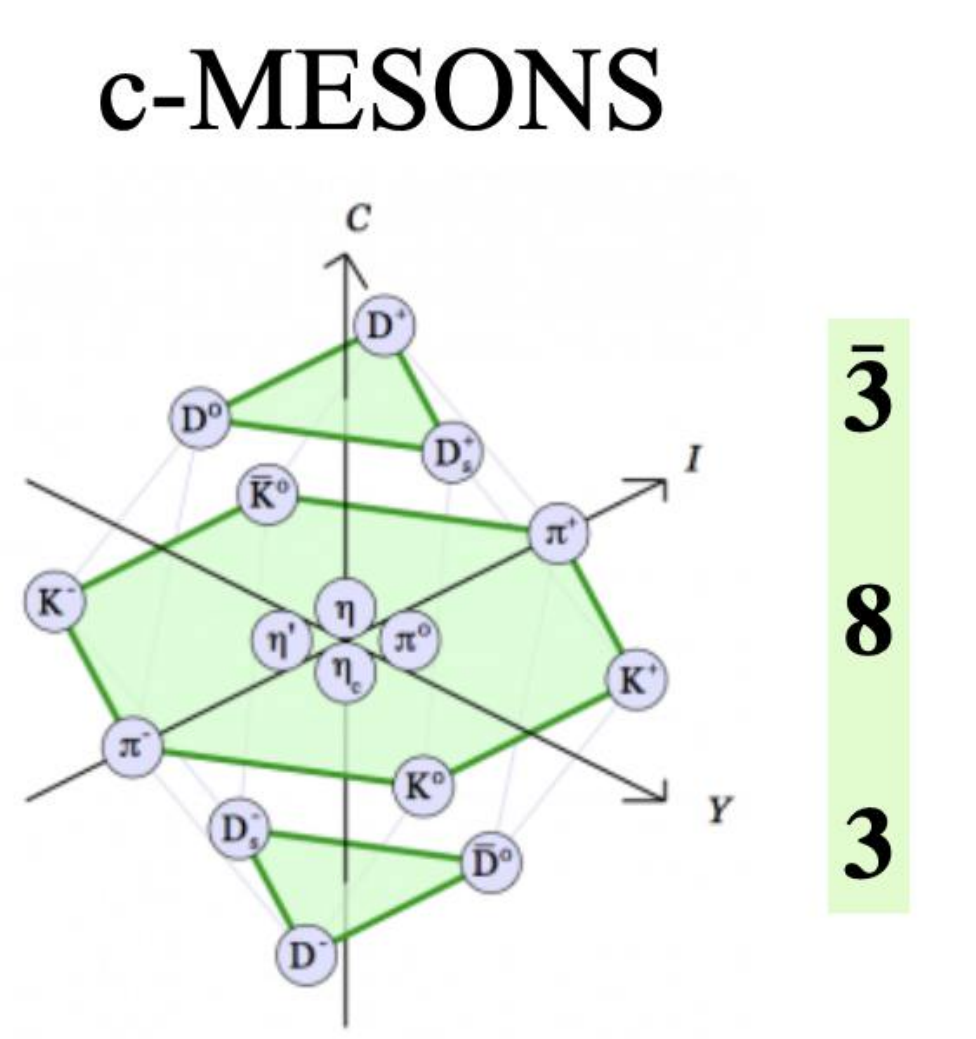}
   \end{center}
 \caption{\footnotesize{}}
\label{meso}
\end{figure}

\begin{figure}[htb!]
 \begin{center}
\ 
   \includegraphics[width=12.0truecm]{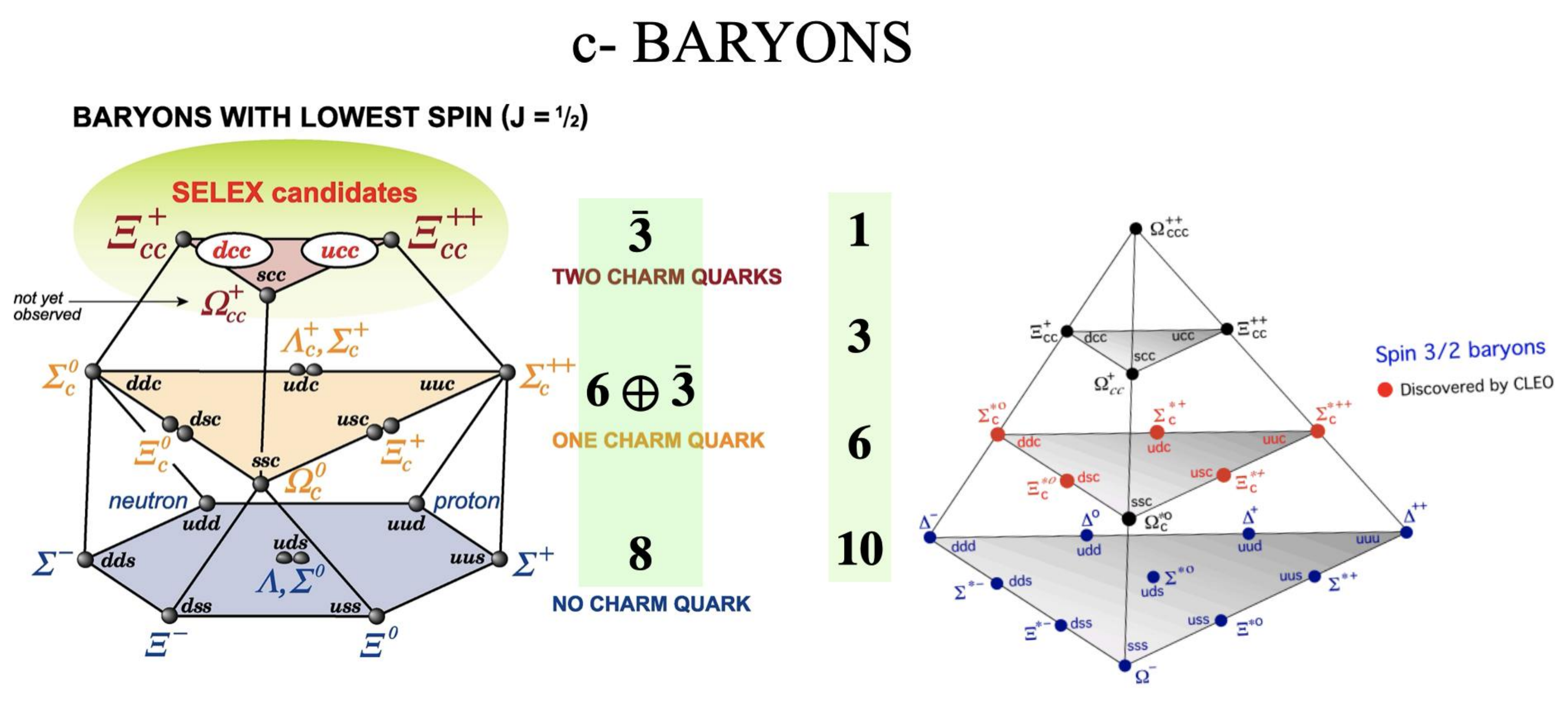}
   \end{center}
 \caption{\footnotesize{}}
\label{bary}
\end{figure}

Particle states are displayed in the 3 dimensional space of diagonal quantum numbers: $ I_3, Y$, Charm ($c$) or Beauty ($b$).
\begin{itemize}
\item {\bf Charmed mesons}: quark-antiquark states; lowest states ($J^P=0^-$) and first resonances ($J^P=1^-$) fall both in ${\bf 15}\oplus {\bf 1}$ dimensional multiplets, Fig.~\ref{meso}.
\item Particles made by a pair with the same quark flavor are neutral and fall in the center of the multiplets. 
\item {\bf Baryons}. Are 3 quark states, classified in two different {\bf 20}-dimensional multiplets, Fig.~\ref{bary};
\item c and b baryons are being observed in several different experiments, multiplets not yet completely filled.
\end{itemize}

\emph{\bf{B and D meson masses}}

Considering $D,~D^*$ mesons, we have three new couplings ($m_c$ and 2$ \kappa$s) and two new masses ( with respect to the $c=0$ case), and we find one relation
\bea
&&\frac{3m_{D_s^*}+m_{D_s}-(3m_{D^*}+m_D}{4}=101=\notag\\
&&= \frac{3m_{K^*}+m_{K}-(3m_\rho+m_\pi)}{4}=177
\eea
Similarly, for $B$ mesons
\bea
&&\frac{3m_{B_s^*}+m_{B_s}-(3m_{B^*}+m_B)}{4}=90=\notag\\
&&= \frac{3m_{K^*}+m_{K}-(3m_\rho+m_\pi)}{4}=177
\eea
The discrepancies seem to indicate that the mass of the strange quark, in presence of a heavy antiquark, {\it  is smaller }. 

Karliner and Rosner attribute the effect to the different QCD interactions of the strange quark with heavy quark inside the hadrons \cite{Karliner:2014gca}. The strange quark goes closer to a heavy than to a light quark and the stronger interaction detracts from its rest energy, reducing the observed  {\it constituent mass}.

A similar effect is also seen for the value of the strange quark mass, derived from $c=0$ and $c=1$ baryons and for the charm quark mass derived from $c=1,~b=0$ and $c=1,~b=1$ mesons.

\emph{\bf{Quark masses summary}} 

Quark masses and spin-spin couplings are summarised in the following tables.

 \begin{table}[htb!]
\centering
    \begin{tabular}{||c|c|c|c|c||} \hline
 & q &  s &  c & b \\ \hline
MESONS: quark mass (MeV) & $308$ & $484$ & $1667$ & $5005$ \\ \hline
BARYONS: quark mass (MeV) & $362$ & $540$ & $1710$ & $5044$ \\ \hline
difference (MeV)& 54 & 56 &  43  &  39\\ \hline
\end{tabular}
 \caption{\footnotesize {$S$-wave Mesons and Baryons: constituent quark masses.}}
\label{mas}
\end{table}

\begin{table}[htb!]
\centering
    \begin{tabular}{||c|c|c|c|c|c|c|c|c|c||} \hline
MESONS & $(q\bar q)_1$ &  $(q\bar s)_1$ & ($s \bar s)_1$&  $(q\bar c)_1$ & $(s \bar c)_1$ &$(c\bar c)_1$& $(q \bar b)_1$ & $(s\bar b)_1$ & $(b\bar b)_1$ \\ 
\hline
$\kappa_{ij}$ (MeV) & $318$ & $200$ & $103$ & $70$ & $72$& $56^a$ & $23$ & $24$  & $30^b$   \\ 
\hline
$\kappa_{ij} m_i m_j/\Lambda_{QCD}^3$ & $1.9$ & $1.9$ & $1.5$ & $2.3$  & $3.7$&$9.9 $& $2.2$ & $3.8$& $ 49$\\ \hline \hline
BARYONS & $(qq)_{\bar 3}$ &  $(q s)_{\bar 3}$ & ($s  s)_{\bar 3}$&  $(q c)_{\bar 3}$ & $(s  c)_{\bar 3}$ &$(cc)_{\bar 3}$& $(q b)_{\bar 3}$ & $(s b)_{\bar 3}$& $(b b))_{\bar 3}$ \\ 
\hline
$\kappa_{ij}$ (MeV) & $98$ & $59$ & $23$ & $15$ & $50$& $28^a$ & $2.5$ & $38$ & $15^b$   \\ 
\hline
$\kappa_{ij} m_i m_j/\Lambda_{QCD}^3$ & $0.82$ & $0.74$ & $0.43$ & $0.57$ & $2.7$& $5.2$ & $0.29$ & $6.6$& $24  $\\  \hline \hline
Ratio: $\frac{\kappa_{MES}}{\kappa_{BAR}}$ & 3.2 & 3.4 & 4.5 & 4.7 & 1.6 &$ 2.0$ & 9.2 & 0.6& $2.0$ \\ \hline
\end{tabular}
 \caption{\footnotesize {$S$-wave Mesons and Baryons: spin-spin couplings. $\Lambda_{QCD}=250$~MeV.}}
  \label{spin}
\end{table}

For future use, we have taken:
\bea 
&&\kappa_{c\bar c}=1/2  (J/\psi - \eta_c);~\kappa_{cc}=1/2 \kappa_{c\bar c};\notag \\
&&\kappa_{b\bar b}=1/2  (\Upsilon- \eta_b);~\kappa_{bb}=1/2 \kappa_{b\bar b}.\notag
\eea
 
 A few comments.
 \begin{itemize}
\item The third line of the Tab.~\ref{mas} supports the existence of a {\it constant} difference between the constituent quark mass of a given flavor in Mesons and Baryons \cite{Karliner:2016zzc}. 
\item Mass differences of constituent quarks with different flavor, e.g. $m_s - m_q$, taken in Mesons and Baryons are much the same, as if the effect of string self energy would cancel, giving results more similar to the quark bare masses, e.g. those derived from current algebra.
\item The scaling law implied by \eqref{hfcoup} applies well to the light quarks, persists to charm and gets lost for beauty. 
\item The ratio of spin-spin couplings in Mesons, $q \bar q$ in color ${\bf 1}$, and Baryons,  $q q$ in color ${\bar {\bf 3}}$, should reflect the ratio of the $C_2$ values and be equal to $2$, as indicated in Tab.~\ref{casi2}. Tab.~\ref{spin} gives a value $3$ to $4$ for light to charm quarks, supporting sizeable, but not dramatic, non perturbative corrections.
\end{itemize}
 All in all, the picture delineated in the Constituent Quark Model with one gluon exchange gives a reasonable description of the lowest lying hadrons with different flavours.

%% file: chapts/Tetraqs.tex

In the quark model, scalar mesons are to be expected as $q\bar q$, $P$-wave bound states,  predicted in the early days of the quark model~\cite{BORCHI1965352}. 
The first scalar mesons to be established are $f_0(980)~(I=0)$ and $a_0(980) ~(I=1)$. Lightest scalar states, with $S=0$ and $S=1$ and  have been going in and out until reasonably established as $f_0(600)~(S=I=0)$, often called $\sigma_0$, and $K^*_0(700)~(S=1,~I=1/2)$, often called $\kappa$. 

\subsection{Light  scalar mesons and tetraquarks}
\begin{figure}[htb!]
 \begin{center}
   \centering
   \includegraphics[width=11.0truecm]{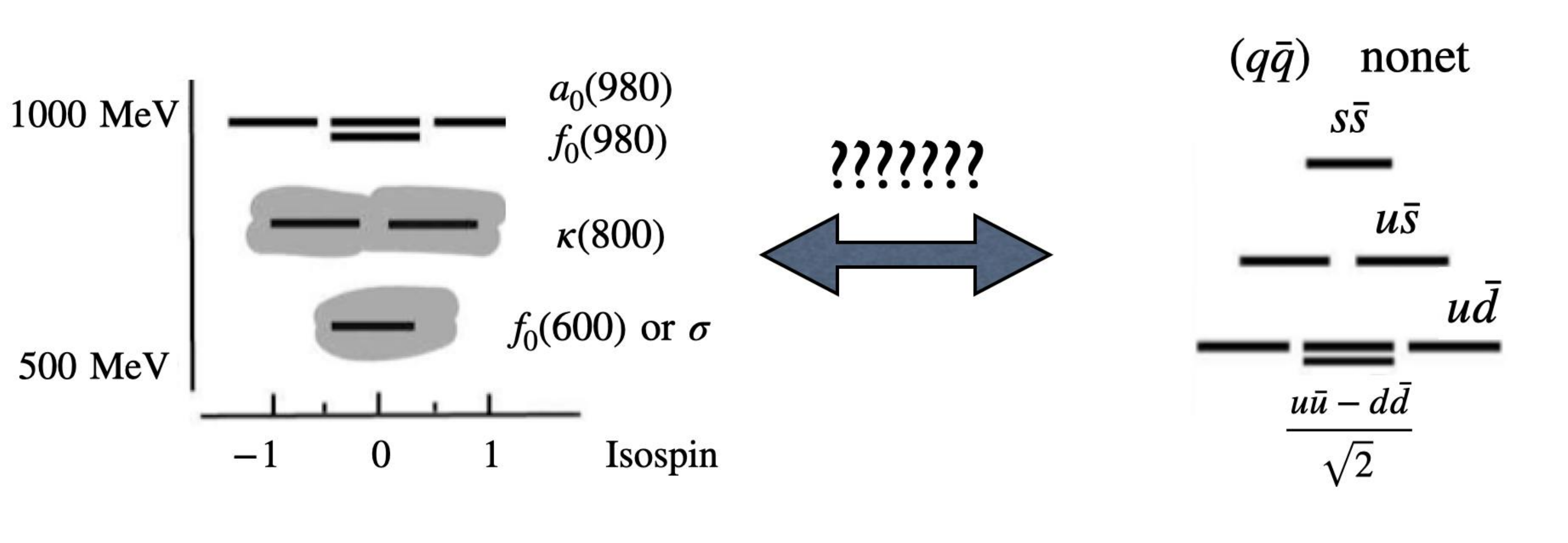}
 \caption{\footnotesize{Mass spectra of the scalar multiplets vs. a typical $(q\bar q)$ nonet. }}
\label{8way3}
\end{center}
\end{figure}
\begin{figure}[htb!]
 \begin{center}
   \includegraphics[width=12.0truecm]{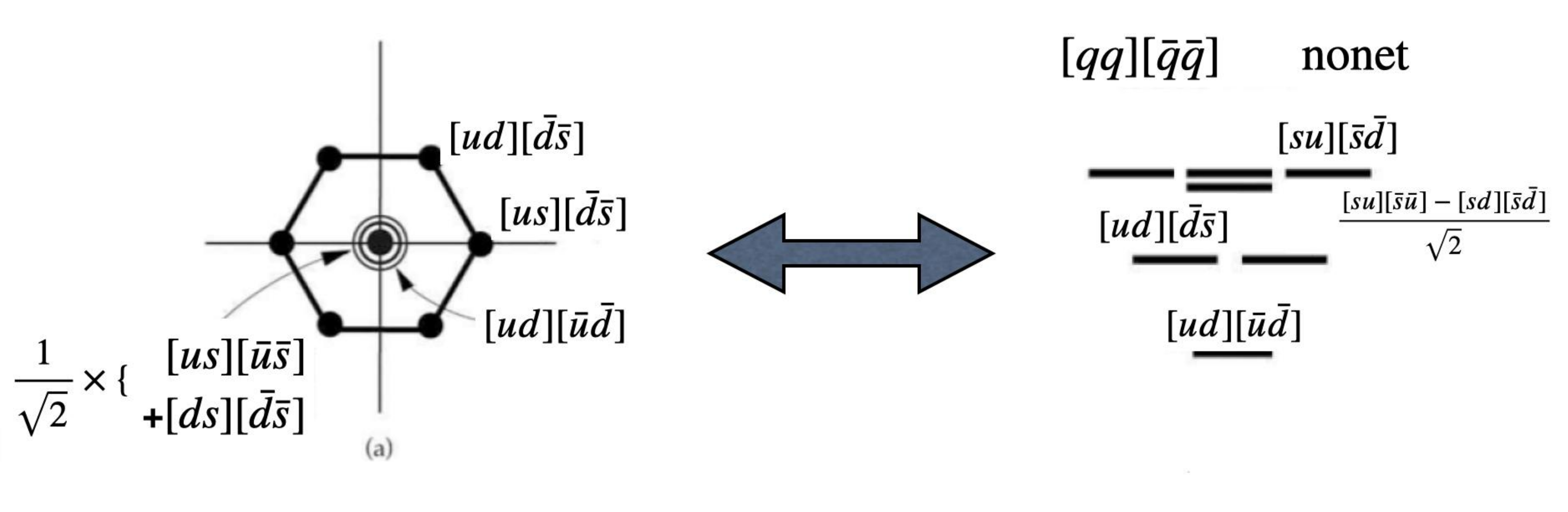}
 \caption{\footnotesize{Mass spectrum of Jaffe's tetraquark.}}
\label{scalar}.
\end{center}
\end{figure}
Compared to the pseudoscalar and vector mesons  mass spectra, Fig.~\ref{8way3}, the scalar meson mass distribution reserved the surprise that the $I=1$ isovector, supposed to be the light quark $q\bar q$ state, is heavier than the $s\bar q$ state that should describe the $\kappa$ meson, contrary to what happens in the established $q\bar q$ nonets.
 As noted by R. Jaffe~\cite{Jaffe:1976ig,Jaffe:2003sg} (see also~\cite{Ali:2019roi}), tetraquarks could reproduce neatly the anomalous behaviour, as shown in Fig.~\ref{scalar}. 
 
 To avoid states with I=2, Jaffe assumes that  {\it good diquarks} (${\rm colour}~{\bf{\bar 3}}, ~{\rm spin}_{qq}=0$)
 do bind and {\it bad diquarks} (${\rm colour}~{\bf{\bar 3}}, ~{\rm spin}_{qq}=1$) do not. Due to Fermi statistics, the good diquarks have the quark pair  antisymmetric in $SU(3)$ indices, i.e: {\it diquarks are in ${\bf {\bar 3}}$ and antidiquarks in ${\bf { 3}}$} $SU(3)$ representations (see Lect.2). 
 As a consequence, the diquark-antidiquark state is in a $SU(3)$ nonet.  
 
 The trick works very well. The $I=1$ state is made by
 \be
a_0^+=\frac{[u s][\bar d\bar s]}{\sqrt{2}}
\ee
and is heavier that the strange member
\be
\kappa^+=[ud][\bar d \bar s]
\ee
There are two $I=0$ states,  one light, $\sigma$, and one heavy, $f_0$, degenerate with $a_0$, see Fig.~\ref{scalar}:
\be
\sigma=[ud]\bar u\bar d],~~f_0=\frac{[us][\bar u\bar s]-[ds][\bar d\bar s]}{\sqrt{2}}
\ee

\emph{\bf{Tetraquarks vs Hadron Molecules: a first look.}}

To form hadrons, good or bad diquarks need to combine with other colored objects to form color singlets:
\begin{figure}[htb!]
 \begin{center}
   \includegraphics[width=10.0truecm]{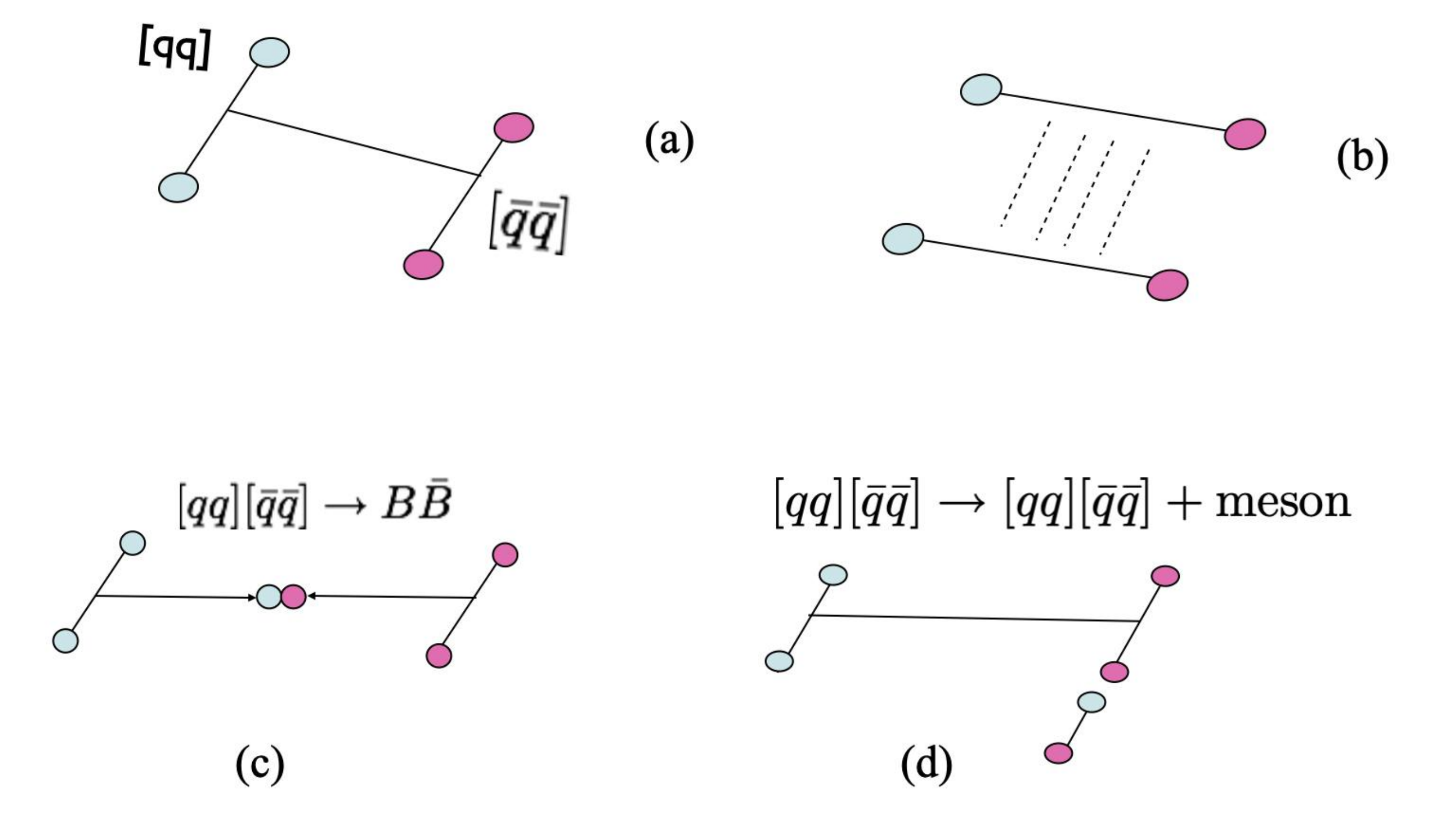}
 \caption{\footnotesize{.}}
\label{Tconf}
\end{center}
\end{figure}
\begin{itemize}
\item with another quark, to produce a baryon in a Y-shape configuration
\item with an antidiquark to form e.g. a scalar meson, an H-shape configuration as proposed by Rossi and Veneziano, Fig.~\ref{Tconf}(a);
\item We expect many diquark-antidiquark states: the string joining diquarks may have radial and orbital excitations;
\item the tetraquark string topology is more related to $B\bar B$  and would favour the decays $T\to B\bar B$, Fig.~\ref{Tconf} (c).
\item an allowed decay is also the de-excitation of a tetraquark with meson emission, Fig.~\ref{Tconf} (d)
\end{itemize}

The possibility of bound states of colourless hadrons was raised long ago by De Rujula, Georgi and Glashow. Meson-meson molecules have a different string topology, as shown in Fig.~\ref{Tconf} (b).

 Nuclei obviously {\it are} hadron molecules, being made by color singlet protons and neutrons. We expect very few molecular states: similarly to nuclei, no orbital or radial excitations are expected.

Do hadron molecules exist? We come back later in this lecture to a comparison  of nuclei vs. putative hadron molecule production in high energy collisions.

\subsection{Tetraquark constituent  picture of unexpected quarkonia}

We start with a number of assumptions.

We consider hidden charm tetraquarks of the form 
\be
[cq]^{{\bf{\bar 3}}}_{S=0,1} [\bar c\bar q^\prime]^{{\bf{3}}}_{S=0,1}
\ee 
in $S$-wave (i.e.  positive parity) and $q=u,d$ is a light quark.
 The total spin of each diquark may be $S=0,1$: heavy-light bad diquarks are admitted.

The neutral states are mixtures of isotriplet and isosinglet. Mass splittings are described by the Hamiltonian:
\be
{\bf H}=2m_{cq}+\sum_{ij} 2 \kappa_{ij}({\bf s}_i\cdot{\bf s}_j)
\ee
where $m_{cq}$ is the (constituent) diquark mass.

For fixed flavour, in the basis of the states $|S,~\bar S>_J$ with definite diquark and antidiquark spin ( $S,~\bar S$) and total angular momentum $J$, we have the following states
\bea
&&J^P=0^+,~C=+:~X_0=|0,0>_0~X^\prime_0=|1,1>_0\notag \\
&&J^P=1^+,~C=+:~X_1=\frac{|1,0>_1+|0,1>_1}{\sqrt{2}}\notag \\
&&J^P=1^+,~C=-:~Z=\frac{|1,0>_1-|0,1>_1}{\sqrt{2}};~Z^\prime=|1,1>_1;\notag\\
&&J^P=2^+,~C=+:~X_2^\prime=|1,1>_2. \label{spinbasis}
\eea

We propose to identify:
\bea
&&X(3872)=X_1;       Z(3900),  Z(4020)={\rm linear ~combinations ~of} ~Z,~Z^\prime~{\rm that~ diagonalize}~ {\bf H}\notag \\
&& X^\pm(4050)=X^\prime_0~??
\eea

\subsection{Can we guess tetraquark's spectrum from the spin-spin couplings of mesons and baryons ??}

Recall the formula given in Lect. 2:
\be\
\kappa_{ij}=CF(R) \times \frac{g^2}{m_im_j}|\psi(0)|^2 \notag
\ee
Spin-spin (or hyperfine, hf) interactions are proportional to the overlap probability $|\psi(0)|^2$  of  the two quark/antiquark constituents involved. 
No symmetry principle says that the overlap functions in tetraquarks have to be the same as in baryons or mesons.
Spin-spin couplings in tetraquarks are free parameters to be determined from the mass spectrum.

The effect of a dominant $q\bar q^\prime$ hyperfine interaction, as for mesons, is easily seen in a basis in which the $q\bar q^\prime$ and $c\bar c$ spin are diagonal. 

Such a basis is obtained  from the basis in \eqref{spinbasis} by a transformation analogous to  the Fierz Transformation introduced  in the  Fermi theory of weak interactions. A useful shortcut is to use Charge Conjugation. The basis in which  $q\bar q^\prime$ and C-conjugation are diagonal is easily found since, for fermion antifermion pairs in $S$-wave and total spin $s$, we have: ${ C}= (-1)^s$. One finds:
\bea
&& C=+1:~X_1 \leftrightarrow[\bar q^\prime q]_1 [\bar c c]_1 ~(+1/2) \notag \\
&& C=-1:~Z,~Z^\prime \leftrightarrow [\bar q^\prime q]_1[\bar c c]_0~ (+1/2),~ [\bar q^\prime q]_0[\bar c c]_1~(-3/2)
\eea
In parenthesis we have reported the corresponding eigenvalue of  the hyperfine operator $2<\frac{{\bf \sigma}_q}{2}\cdot\frac{{\bf \sigma}_{q^\prime}}{2}>$, see Lect. 2.

Neglecting charmed quarks hf interaction (inversely  proportional to mass squared),  one concludes that
a dominant $q\bar q^\prime$ hf interaction  leads to: {\it degenerate} $X_1$ and $Z$ ($s_{\bar q^\prime q}=1$) and {\it lighter } $Z^\prime$ ($s_{\bar q^\prime q}=0$).

The observed pattern instead is: $X(3872)\sim Z(3900) < Z(4020)$, i.e. {\it degenerate} $X,~ Z$ and {\it heavier}  $Z^\prime$! 

\emph{\bf{The right ansatz.}}
The correct mass ordering is obtained under the hypohesis that {\it dominant spin-spin interactions are those inside the diquark or the antidiquark}~\cite{Maiani:2014aja}. 

A simple explanation of the dominance of intra-diquark interaction is that diquarks and antidiquarks are at a relatively large distance in the hadron, so as to suppress the overlap probability of constituents in different diquarks, unlike what happens, e.g., in the usual baryons.

With this hypothesis, H is diagonal in the diquark spin basis used before and the Hamiltonian takes the form:
\be
{\bf H}=2m_{cq}+2\kappa_{cq}({\bf s}_c\cdot {\bf s}_q +{\bf s}_{\bar c}\cdot {\bf s}_{\bar q} )\label{ansatz}
\ee
In words: ${\bf H}$ counts the number of spin 1 diquarks/antidiquarks.

\emph{\bf{Parameters.}}

Assuming
\bea
&& X(3872)= X_1;~M(3872)=2m_{cq}-\kappa_{cq}\notag \\
&& Z(3900)=Z;~M(X_1)=M(Z)\notag\\
&&~Z(4020)=Z^\prime;~M(4020)=2m_{cq}+\kappa_{cq} \notag
\eea
From the experimental masses we find:
\be
m_{[cq]}=1980~{\rm MeV},~~\kappa_{cq}=67~{\rm MeV}\label{param}
\ee
The difference: $|M(X_1)-M(Z)|\sim 18~{\rm MeV}$ quantifies the error of our ansatz~\eqref{ansatz}.

\begin{figure}[htb!]
\begin{minipage}[c]{7.2cm}
   \centering
   \includegraphics[width=7.2truecm]{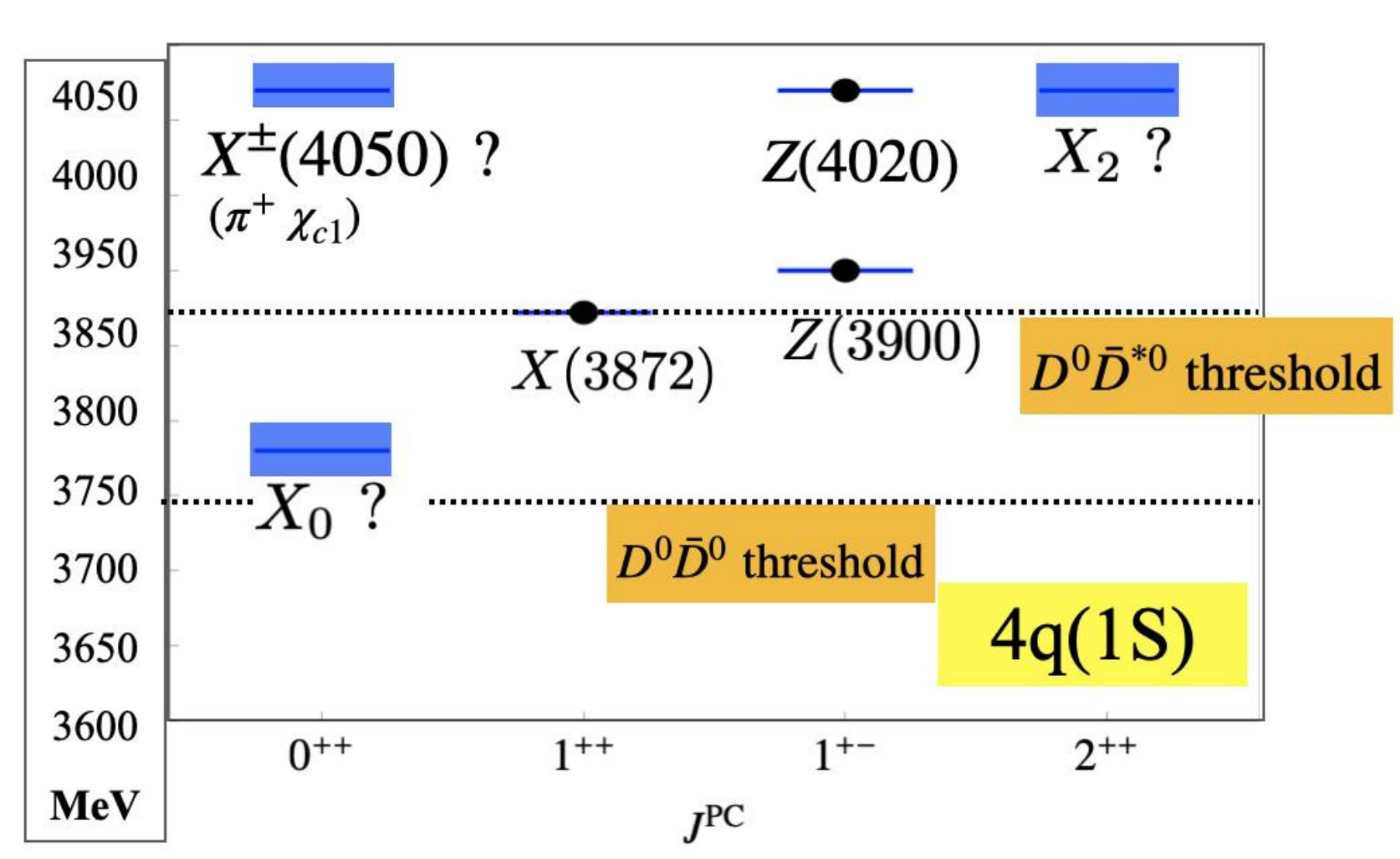}
    \end{minipage}%
 \begin{minipage}[c]{7.2cm}
\centering
   \includegraphics[width=6.26truecm]{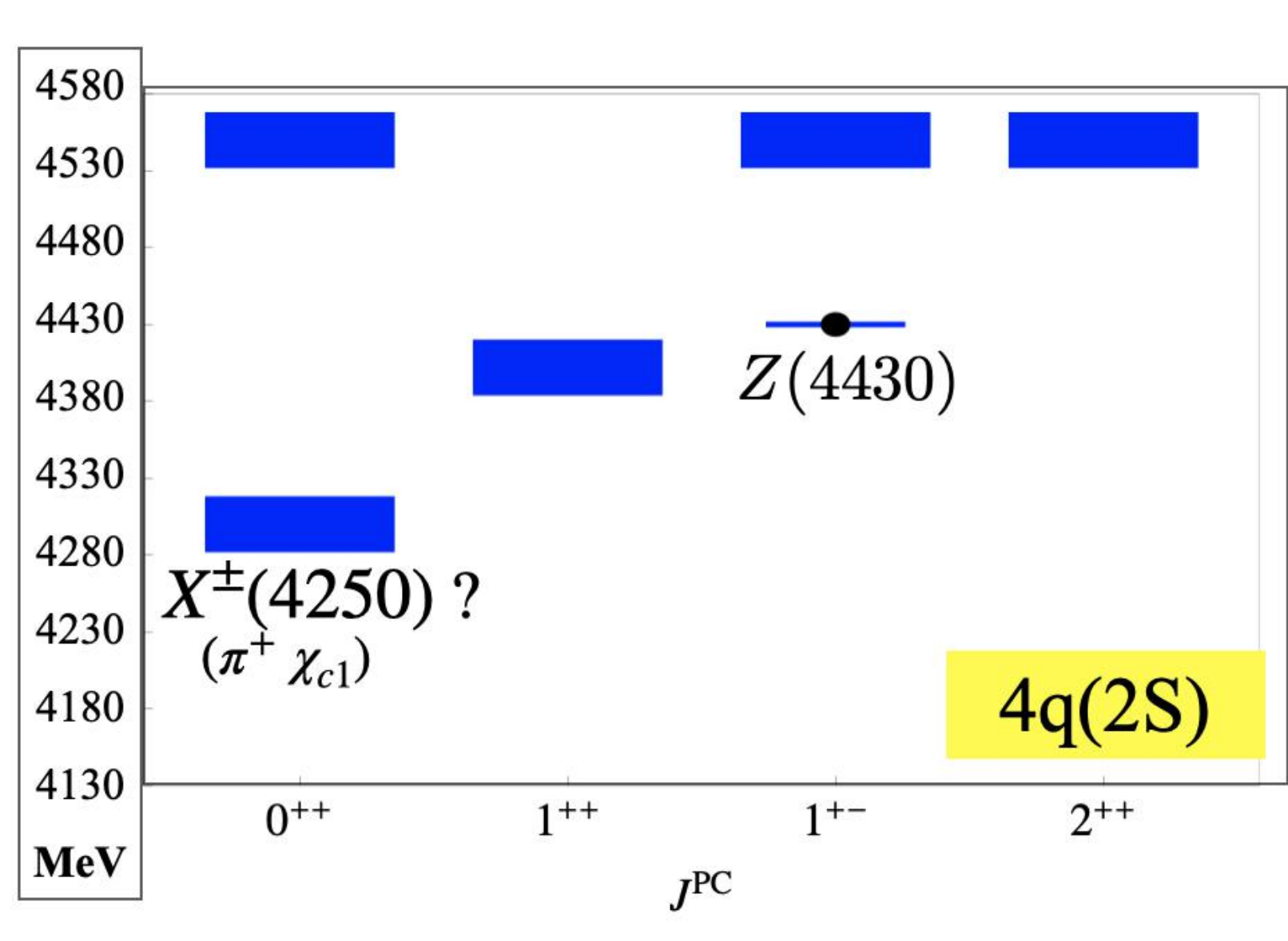}
 \end{minipage}%
\caption{\footnotesize{}}
 \label{Swave}
\end{figure}

For the other states, we predict
\bea
&& M(X_0)=M(X_1)-2 \kappa_{cq}=3738~{\rm MeV}\notag\\
&&M(X^\prime_0)=M(X_2)=M(Z^\prime)=4020~{\rm MeV}\notag
\eea
The pattern of the $1S$ states predicted in the simple ansatz~\eqref{ansatz} is illustrated in Fig.~\ref{Swave} (left panel).
\begin{itemize}
\item The lowest spin 0 tetraquark, $X_0$ could decay into a $D\bar D$ pair. No resonance has been seen thus far in this channel. The  $D^0\bar D^0$ threshold, however, is at $3730$~MeV,  and non leading corrections to Eq.~\eqref{ansatz} could easily bring the  $X_0$ mass below threshold.
\item The positive parity, charged exotic meson reported in Pdg as $X^\pm(4050)$:
\be
X^\pm(4050)\to \pi^\pm \chi_{c1}(1P),~M(4050)=4051^{+24}_{-40}
\ee
could very well be identified with $X^{ \prime \pm}_0$ or $X^\pm_2$.
\end{itemize}

The first radially excited states, $2S$, Fig.~\ref{Swave} (right panel) are shifted up by a common quantity, the radial excitation energy, $\Delta E_r$, expected to be mildly dependent on the diquark mass: $\Delta E_r (cq) \sim \Delta E_r (cs)$.
From the mass of $Z^\pm(4430)$:
\be
\Delta E_r(cq)=530~{\rm MeV}.
\ee
We predict further the masses (in MeV):
\be
X_0(2S)=4296;~X_1(2S)\sim Z(2S)=4430;~X_0^\prime(2S)=Z^\prime(2S)=X_2(2S)=4564
\ee
The positive parity, charged exotic meson reported in Pdg as $X^\pm(4250)$:
\be
X^\pm(4250)\to \pi^\pm \chi_{c1}(1P),~M(4250)=4248^{+190}_{-50}
\ee
could  be identified with $X^\pm_0(2S)$

To summarize:
\begin{itemize}
\item A simple ansatz reproduces the ordering of Z states: {\it spin-spin interaction is dominated by inter-diquark interaction};
\item constituents are not uniformly mixed in the hadron bag, rather clump into two separate entities: diquarkonium. 
\item The spectrum of 1S ground states is characterised by the diquark mass, $m_{cq}$  and the spin-spin interaction, $\kappa_{cq}$. 
\item The first radially excited, $2S$, states are shifted up by a radial excitation energy, $\Delta E_r$, very similar to the radial excitation energy of charmonium.
\end{itemize}

\subsection{New $J/\psi-\phi$ structures} 
In 2016, LHCb has reported the observation of four new resonances decaying in $J/\psi-\phi$
The structures have positive parity, $J=0$ and $1$, and obviously positive charge conjugation~\cite{LHCb:2016axx}. The lowest one, $X(4140)$ had been seen previously by CDF, D0, CMS and by Belle.

\begin{figure}[htb!]
\begin{minipage}[c]{7.2cm}
   \centering
   \includegraphics[width=7.2truecm]{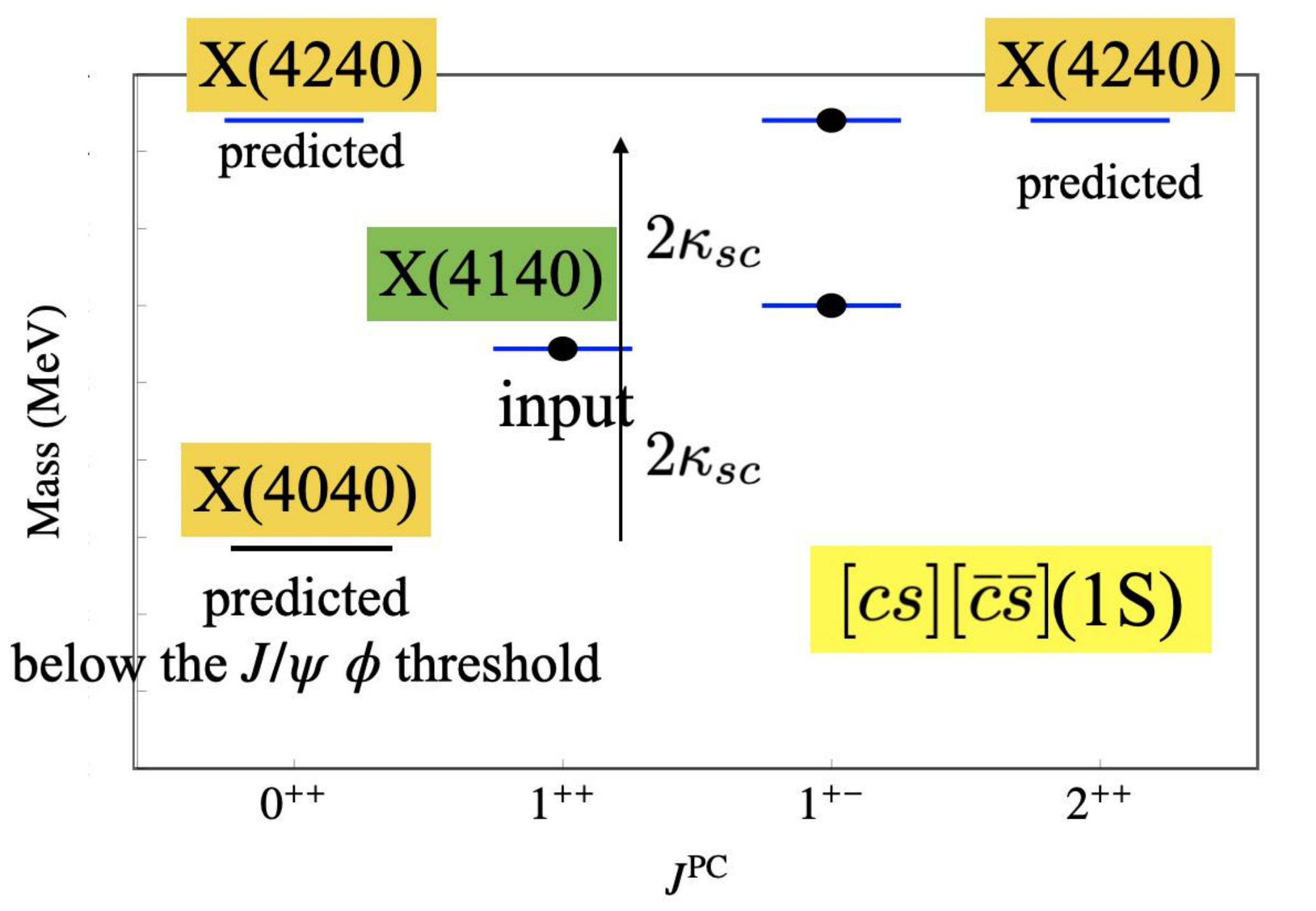}
    \end{minipage}%
 \begin{minipage}[c]{7.2cm}
\centering
   \includegraphics[width=7.26truecm]{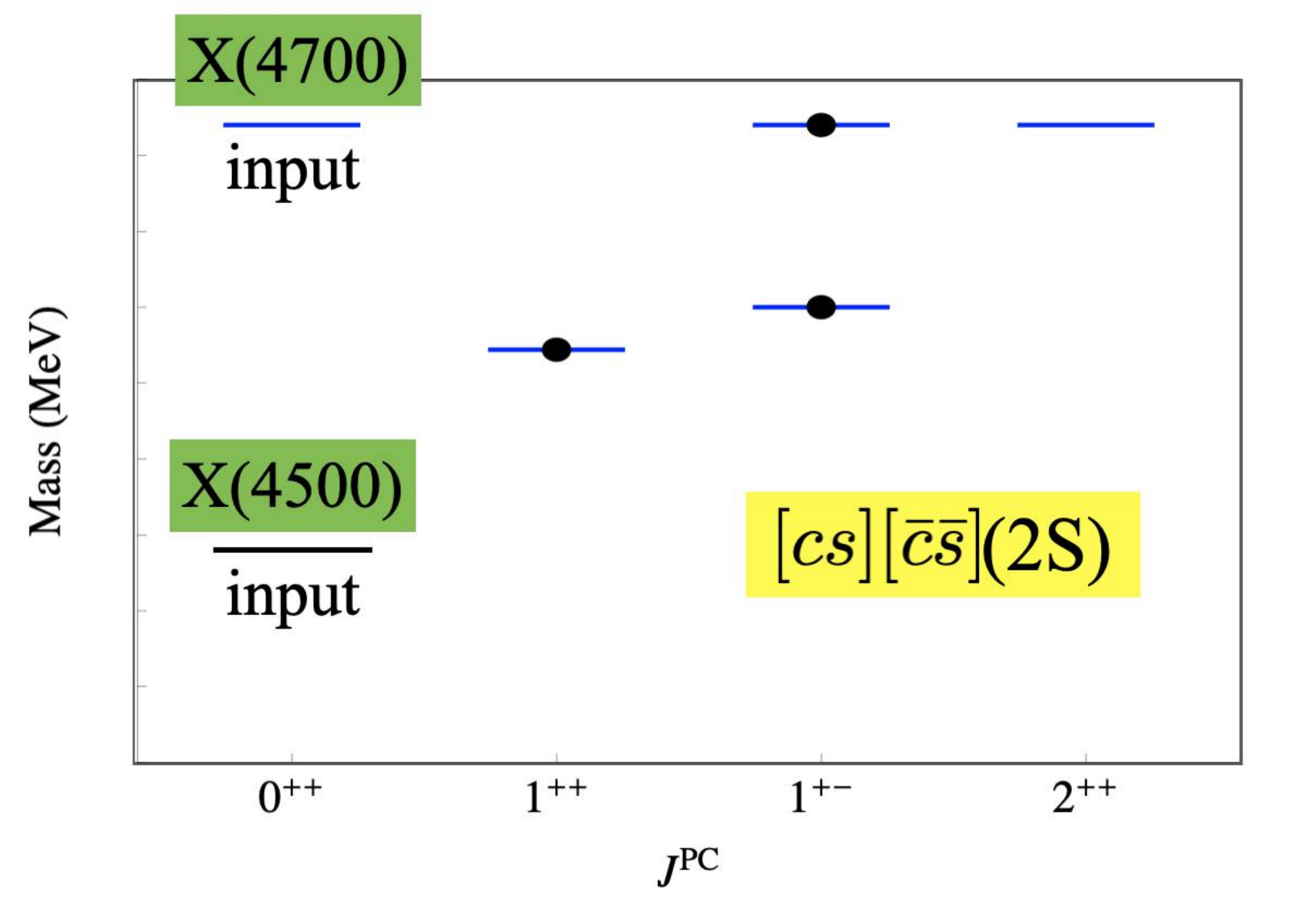}
 \end{minipage}%
\caption{\footnotesize{}}
 \label{Sphi}
\end{figure}

Judging from the masses, we suggest to fit the structures into {\it two new tetraquark multiplets}, S-wave ground state and first radial excitations and flavour composition $[cs][\bar c\bar s]$, illustrated in Fig~\ref{Sphi}. Together with the previously identified $[cq][\bar c\bar q]$ multiplet, the new resonances would make a step towards a full nonet of S-wave tetraquarks made by $c \bar c$ and a pair of light ($u, d, s$) quarks~\cite{Maiani:2016wlq}. We found a surprise and made a proposal.

Applying the ansatz~\eqref{ansatz} with $q\to s$, from the three input masses, marked in green in Fig.~\ref{Sphi}, we find:
\bea
&&\Delta m=m_{cs}-m_{cq}=129~{\rm MeV}.\label{dmqs}  \\
&&\kappa_{cs}=50~{\rm MeV}\label{hfs}\\
&& \Delta_r(cs)=460~{\rm MeV}\label{radcs}
\eea
With these value, we can predict all $C=+$ members of the two tetraquark multiplets. Using~\eqref{hfs}
For $1S$, we obtain
\be
 X_0(1S)=4040;~X^\prime(1S)=X_2(1S)=4240 \label{cplus}
\ee
$X_0(1S)$ is predicted below the $J/\psi-\phi$ threshold.

LHCb quotes four resonances. Beside the three we have used as input (in green in Fig~\ref{Sphi}), they quote  $X(4274), J^P=1^+$. The mass is in the $1S$ range but among the $1S$ tetraquark there is only one $1^+$ state, already identified with $X(4140)$.

A second $1^+$ would be present for tetraquarks made by diquarks in colour ${\bf 6}$. We consider this an unlikely possibility, given the repulsive force obtained in Lect. 2, Tab~\ref{casi2}. Rather, we suggest that the peak identified with $X(4274)$ is produced by the two degenerate states predicted  at $4240$ in Eq.~\eqref{cplus}, to be fitted as a superposition of $0^+$ and $2^+$ states. Non conclusive evidence, pro or con, has been produced until present. 

\emph{\bf{First hint of a nonet.}}

We note the result:
\be
X(4140)-X(3872)= 270 ~{\rm MeV}
\ee
to be compared to:
\be
\phi-\rho=244~{\rm MeV}
\ee
The near equality of the mass differences nicely confirms the hypothesis that $X(4140)$ and $X(3872)$ are the $s\bar s$ and $q\bar q$ components of the same $SU(3)$ nonet, similar to $\rho$ and $\phi$ of the  vector meson nonet. 

This leads to a firm prediction of the mass of the strange components of the tetraquark~\cite{lmaia16} nonet, with composition: $X_{s}=[cs][\bar c \bar q ]$, $\bar X_{s}=[cq][\bar c \bar s]$:
\be
M(X_s)=\frac{X(3872)+X(4140)}{2} = 4010~{\rm MeV}\label{zscmass}
\ee
(more about that in Lect.4)

\subsection{Few remarks about molecules}

The most studied Exotic Hadron is no doubt the $X(3872)$, yet we have not achieved a conclusive evidence in favour or against its molecular structure.

The present mass value: $M=3871.69 \pm 0.17$~MeV is very close to the $D\bar D^*$ threshold, so as to suggest its nature as a molecular state:
\be
M(D^0)+M(\bar D^{*0}) =3871.68;~M(D^+)+M( D^{*-})=3979.91
\ee
 if X is a $D^0\bar D^{*0}+\bar D^0 D^{*0}$ molecule, its binding energy is very small indeed
 \be
| B| <0.17~{\rm MeV}
\ee
and, correspondingly, its radius very large:
\be
R\sim \sqrt{2 \mu |B|} \sim 10 ~{\rm fm} ~\sim 15 \times~  ({\rm proton~ radius})\label{xradius}
\ee 
($\mu$ is the $D-D^*$  reduced mass  $\sim M(D)/2$).

How about $X(3872)$ production in collisions at large $p_T$ ?
The issue has been studied  in~\cite{Esposito:2015fsa} comparing  data on the production of light nuclei in Heavy Ion high energy collisions (from ALICE Collaboration), with data about $X(3872)$ production in $p-p$ collisions at large $p_T$ (from CMS Collaboration).

\begin{figure}[htb!]
 \begin{center}
   \includegraphics[width=8.0truecm]{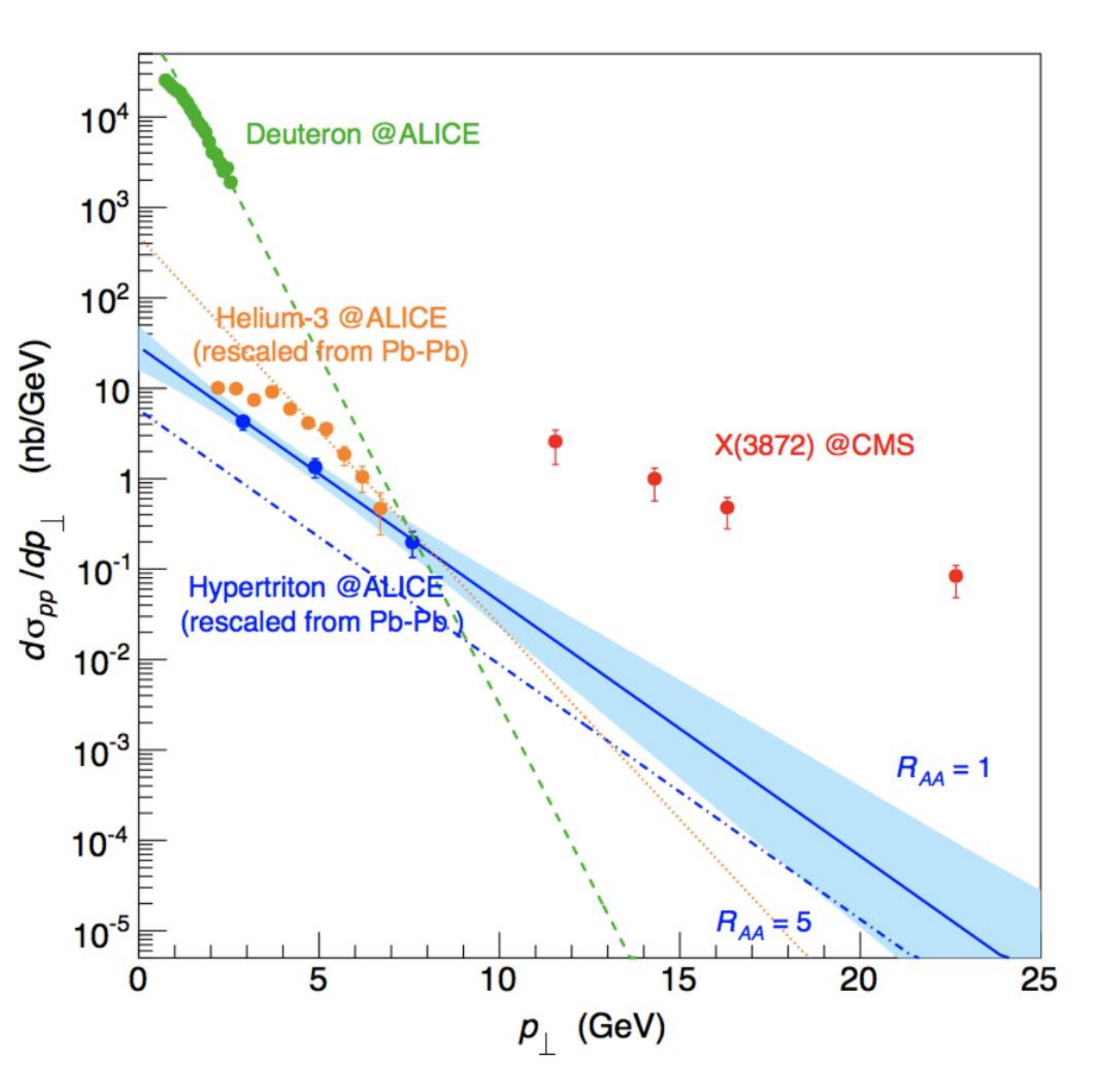}
 \caption{\footnotesize{}}
\label{pperp}
\end{center}
\end{figure}

The heavy ion cross sections have been translated in $p-p$ cross sections using Glauber's model and put together with the CMS cross section in Fig.~\ref{pperp}. (very similar results are obtained with other extrapolation methods, such as the blast-wave function extrapolation, see~\cite{Esposito:2015fsa}.)

There is a vast difference in the probability of producing X(3872) and that of producing light nuclei, true {\it hadronic molecules}, in high energy collisions at large $p_T$. 
The crosse section of light nuclei falls exponentially with $p_T$, as expected for loosely bound objects with a radius
\be
R \sim 1~ {\rm fm}>>\frac{1}{p_T} \sim  \frac{0.20}{p_T({\rm GeV})}~{\rm fm}
\ee
On the contrary, the $X(3872)$ cross section falls slowly and is of the same order as the $\psi(2S)$ cross section, i.e a compact system with elementary components held together by confining QCD forces.

To save the molecular interpretation the suggestion has been made of a strong ($\sim 30\%$) mixing of a molecular state with a charmonium state with the same quantum numbers, e.g. $\psi(2P)$. The charmonium component would be produced strongly, but the final asymptotic state would reconstruct the full molecule. The result in \eqref{xradius} implies however a very large $c\bar c$ distence on the molecule, making its overlap with any charmonium state quite unlikely.

The high energy production of suspected exotic hadrons from quark-gluon plasma in Heavy Ion collisions at colliders can be  a very effective tool to discriminate different models, particularly for charged states wihch cannot mix with charmonia. 

We have already a long list of suspects: $f_0(980), Z^\pm(3900), Z^\pm(4020), Z^\pm(4430), X(4140), \dots$.

%% file: chapts/8fold.tex

\subsection{QCD and light  quark masses.}

\bea
&& {\cal L}_{QCD}=\frac{1}{4}Tr[G_{\mu\nu}G^{\mu\nu}]+\sum_{u,d,s}(i\bar q_i Dq_i +m_i\bar q_i q_i)\notag \\
&& {\cal L}_{QCD}={\cal L}^0_{QCD}+{\cal L}_{m} \label{QCD}
 \eea
 The light quark mass Lagrangian can be rewritten as:
 \bea
 &&{\cal L}_{m}=\bar m \sum_{u,d,s} \bar q_i q_i+m_{3}(\bar u u-\bar d d)+ m_8(2 \bar ss-\bar u u-\bar d d)=\notag\\
 &&={\cal L}_{\bar m}+{\cal L}_3+{\cal L}_8\notag
  \eea
  with
  \be
  m_3=\frac{m_u-m_d}{2};~~m_8=\frac{2m_s-m_u-m_d}{6}
  \ee
  ${\cal L}^0_{QCD}$ is invariant under the light flavours symmetry (the EightFold Way)  and creates complete multiplets of $SU(3)_f$, with the heavy quarks $c, b$ behaving as $SU(3)_f$ singlets.

${\cal L}_3+{\cal L}_8$ are color singlets and can be treated as perturbations that split the degenerate $SU(3)_f$ multiplets created by ${\cal L}^0_{QCD}$ , like an external, weak magnetic field that splits the atomic levels.

Quark mas differences are universal, i.e equal for mesons and baryons. Neglecting Isospin breaking, we found in Lect.2:
\be
(m_s-m_q)_{baryon}\sim (m_s-m_q)_{meson}=180~{\rm MeV}
\ee
Constituent quark masses, on the other hand,  are sensitive to the energy stored in the QCD field, and have larger variations, as we have seen in Lect.~2:
\begin{itemize}
\item Mesons: $(m_q\sim 308,~ m_s\sim 484)$
\item Baryons: $(m_q\sim 362,~ m_s\sim 540)$
\end{itemize}

\emph{\bf {Nuclei are not like that.}} Forces exchanged between color singlets, e.g. nuclei, are strongly dependent from the mass of the exchanged particle, e.g. $\pi,~ \rho,~\eta \dots$ that is they are strongly flavour dependent.
Bound states, nuclei or hadron molecules, are thus not expected to form multiplets with a regular pattern of mass differences.

Assuming flavour singlet forces between color singlets, as done sometime, has no fundamental basis.

\emph{\bf{Exotic mesons: the New Wave.}} Starting from 2016, new kinds of exotic hadrons have been discovered: $J/\psi~\phi$ resonances, di-$J/\psi$ resonances, open strangeness Exotics: $Z_{cs}(3082)$ and $Z_{cs}(4003)$. 
There are no pion exchange forces that could bind them as hadron molecules of color singlet mesons: molecular models have to stand on the existence of  {\it phenomenological forces} with undetermined parameters.
 
 The New Exotics arise very naturally as $([cq]^{\bf \bar 3}[\bar c\bar q^\prime]^{\bf 3})_{\bf 1}$ bound in color singlet.
The compact tetraquark model makes a firm prediction: hidden charm tetraquarks must form  complete multiplets of flavour SU(3), with mass differences determined by the quark mass difference. With $Z_{cs}(3082)$ and $Z_{cs}(4003)$ we can almost fill two tetraquark nonets with the expected scale of mass differences. 

\emph{\bf{Tetraquarks with strangeness and hidden charm:}}  The first resonance with valence quarks $(cs\bar c\bar u)$ has been seen in 2021 by BES III in~$e^+ e^-\to K^+\dots$: an excess over the known contributions of conventional charmed mesons is observed near the $D_s^{*-}D^0$ and $D^-_s D^{*0}$ thresholds, in the $K^+$ recoil mass spectrum~\cite{BESIII:2020qkh}.
 A similar resonance, $Z_{cs}(4003)$, has been observed by LHCb in $B^+$ decay~\cite{LHCb:2021uow}: 
 \be
 B^+\to \phi +Z_{cs}(4003)\to \phi+K^++\psi
 \ee
 LHCb may have a third candidate in the $K^+~\psi$ channel, $Z_{cs}(4220)$. 
 
 \subsection{Hidden charm tetraquarks make $SU(3)_f$ nonets}

The mass spectrum of the vector meson nonets with mass diagonal $\omega-\phi$ mixing , can be simply plotted versus the number of strange quarks or antiquarks
\be
m=C(n_s+n_{\bar s})(m_s-m_u)\label{props}
\ee
 with $C$ a constant of order unity. 

Eq.~\eqref{props} gives the equal spacing rule, well obeyed in vector, $1^{--}$, and tensor, $2^{++}$, mesons, e.g.

\bea
&&\Delta(1^{-})=\phi(1020)-\rho(775)= 244~{\rm MeV}; ~\frac{\rho(775)+\phi(1020)}{2}-K^*(892)= 4~{\rm MeV}\label{vect}\\
&& \Delta(2^{+})=f_2^\prime(1525)-a_2(1320)=199~{\rm MeV};~\frac{f_2^\prime(1525)+ a_2(1320)}{2}-K^*_2(1430)=-10~{\rm MeV}\label{tens}
\eea

We expect a  similar pattern to be obeyed by the hidden charm tetraquarks listed in \eqref{spinbasis}, when we let $q,q^\prime=u,d,s$. 
 Concentrating on the better known states, we shall consider the three nonets associated, respctively, to $X(3872)$, Z(3900) and $Z(4020)$, focussing initially to the first two, to which we propose to associate $Z_{cs}(4003)$ and $Z_{cs}(3985)$.
  We find two solutions~\cite{Maiani:2021tri}.
  
  {\emph{\bf{Solution 1.}} Has the LHCb resonance associated with the $X(3872)$ nonet,  Fig.~\ref{sol1}. The mass of $Z_{cs}(4003)$ fits very well with the equal spacing prediction  given in Eq.~\eqref{zscmass} and completes the steps  in the mass ladder. The BESIII resonance also fits reasonably in the nonet of $Z(3900)$. The spacing in this nonet, $Z_{cs}(3982)-Z_c(3900)\sim 82$~MeV, is comparable to $\Delta(2^+)/2$, the spacing of the tensor mesons, and it is still in the range of  the mass difference of strange to light quark. The equal spacing rule for the $Z(3900)$ nonet predicts its  $s\bar s$ component to the mass
\be
X_{s\bar s}(J^P=1^{+-})=4076~{\rm MeV}  \label{pred1ssbar}
\ee

\begin{figure}[htb!]
 \begin{center}
   \centering
   \includegraphics[width=13.0truecm]{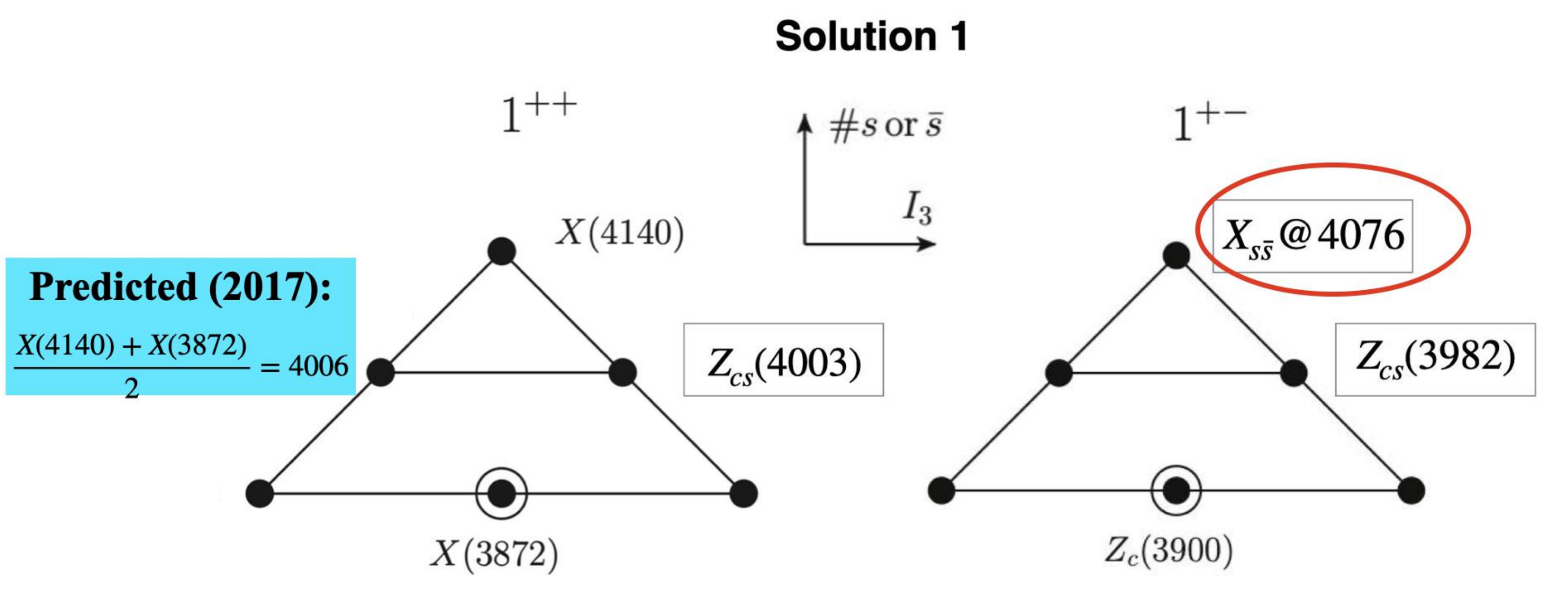}
 \caption{\footnotesize{}}
\label{sol1}
\end{center}
\end{figure}

  {\emph{\bf{Solution 2.}}  The alternative association of $Z_{cs}(3982)$ to the $X(3872)$  nonet, disagrees by  $\sim 27$~MeV from the equal spacing rule~\eqref{zscmass}. This is larger than the violation of the rule in the vector meson none}, but  we think it could still still be acceptable. Associating  $Z_{cs}(4003)$ to the  $Z(3900)$ nonet, we predict:
\be
X_{s\bar s}(J^{PC}=1^{+-})=4121~{\rm MeV}  \label{pred2ssbar}.
\ee
\begin{figure}[htb!]
 \begin{center}
   \centering
   \includegraphics[width=11.0truecm]{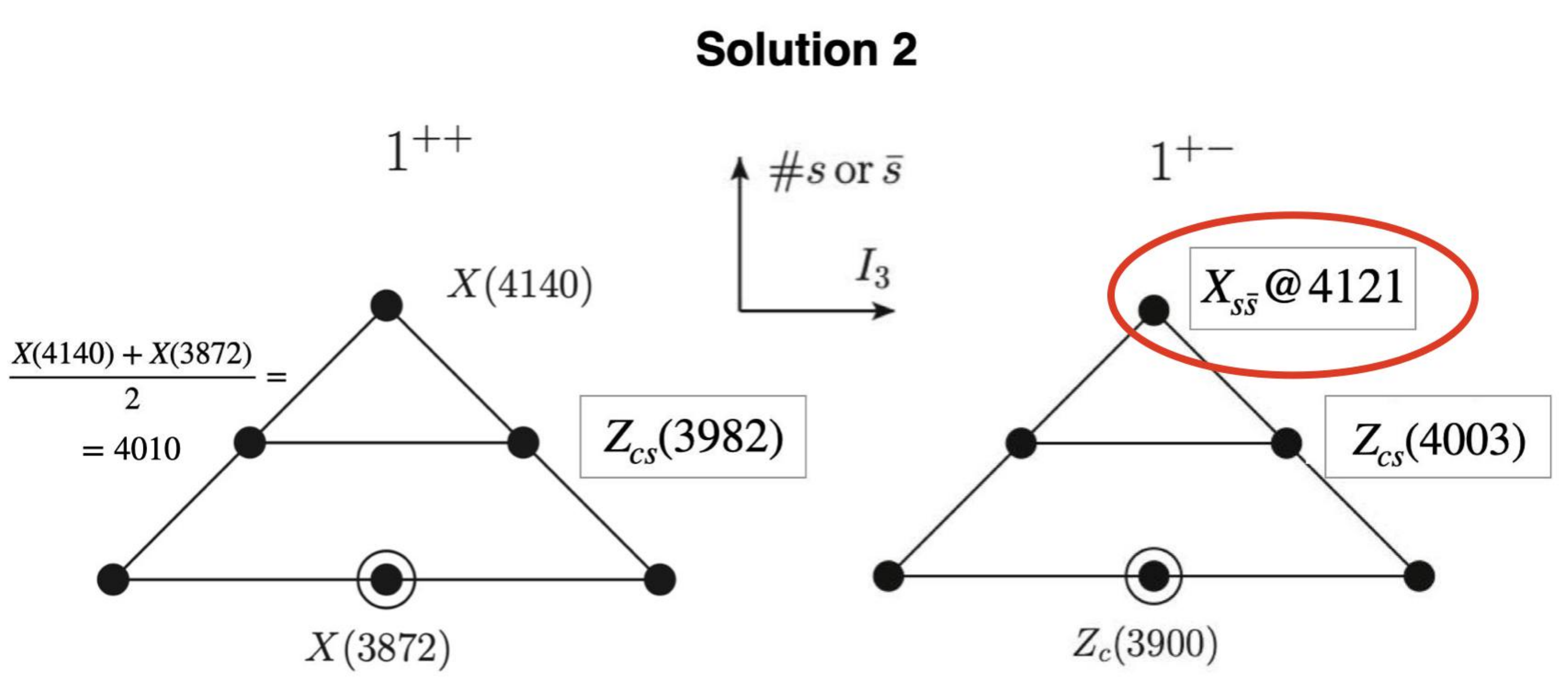}
 \caption{\footnotesize{}}
\label{sol2}
\end{center}
\end{figure}

For both solutions, $X_{s\bar s}(J^{PC}=1^{+-})$ should be seen in the decay channels $\eta_c~\phi,~\eta ~ J/\psi$.  Concerning $\bar D_s^*D_s$ decay, the  mass value predicted by Solution 1 is below   threshold ($M_{D^*_s}+M_{D_s}=4080$~MeV). The decay is possible for the mass value predicted by Solution 2.

\emph{\bf{The third nonet.}}  $Z_c(4020)$ must be part of a third nonet,  therefore there must be a third $Z_{cs}$ with $J^{P}=1^{+}$ at a somewhat higher energy then the two just discussed. Taking Solution 1 as a basis, we would expect a mass of $4150 - 4170$ MeV.  Indeed LHCb sees a $Z_{cs}(4220)$ with $J^P=1^+ {\rm or}~ 1^-$. Is it too heavy ? 

This problem could go together with the fact that, in our preferred Solution 1, $Z_{cs}(3982)$ is a perhaps too light to be associated  with $Z(3900)$. A bold possibility is that there is a mixing between the two $Z_{cs}$ belonging to negative charge conjugation nonets: after mixing, levels repel each other, and the effect is that one $Z_{cs}$  goes down and the other goes up, Fig.~\ref{repel}, in better agreement with the mass of their partners. A small mixing angle $\theta\sim 30^0$ is needed to do the job as shown by a simple calculation and illustrated by Fig.~\ref{repel}.
Take:
\be
M=M_{0}+\Big(\begin{array}{cc}a&b\\b&-a\end{array}\Big)
\ee
\be
M_{diag}=M_0+U^\dagger(\theta)M U(\theta); ~~U(\theta)=\Big(\begin{array}{cc}\cos(\theta/2)&-\sin(\theta/2)\\\sin(\theta/2)&\cos(\theta/2)\end{array}\Big) \notag
\ee
diagonalising $M$ requires
\bea
&&\tan\theta=\frac{b}{a};\notag \\
&& \Delta  M_d=(M_{diag})_{11}-(M_{diag})_{22}= 2a\sqrt{1+(\frac{b}{a})^2} \notag
\eea
Requiring: $2a=(4160-4038);~\Delta  M_d=(4220-3982)$ one finds
\be
\tan\theta\sim 1.7 ~~\to~~\theta\sim 60^0,~i.e. ~{\rm mixing~angle}=\theta/2 \sim 30^0 \notag
\ee

\begin{figure}[htb!]
 \begin{center}
   \centering
   \includegraphics[width=11.0truecm]{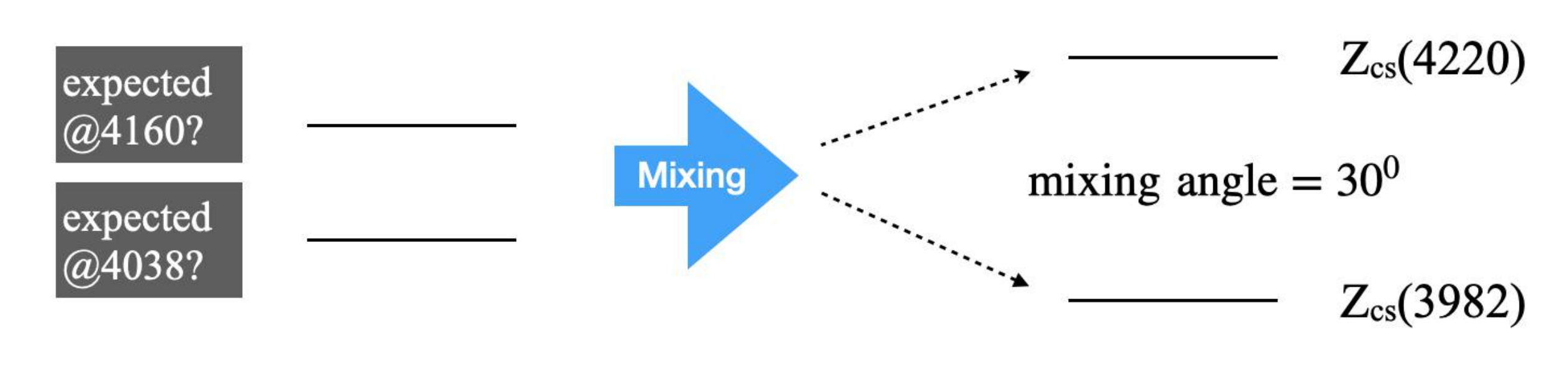}
 \caption{\footnotesize{}}
\label{repel}
\end{center}
\end{figure}

 \subsection{Charge conjugation for $SU(3)_f$ nonets (mesons or tetraquarks)}
 
 The $Z_{cs}$ we have considered are associated with two nonets with the neutral components, $X(3872)$ and $Z_c^0(3900)$, of opposite charge conjugation. One may wonder if this fact has consequences on production or decay modes that can distinguish between Solution 1 and Solution 2. We consider this matter in exact and first order broken $SU(3)_f$ symmetry.
 
 A charge conjugation quantum number can be given to each self conjugate SU(3) multiplet under the transformation:
 \be
 {\cal C}T{\cal C}=\eta_T {\tilde T},~{\tilde T}={\rm transpose~matrix},~\eta_T=\pm 1
  \ee
$\eta$ is the sign taken by neutral members, but it can be attributed to all members of the multiplet. 
$\eta = -1$ is given to the electromagnetic current . 

For the pseudoscalar mesons: ${\cal C} P {\cal C}= +{\tilde P}$;
\be
 \Bigg(\begin{array}{ccc} \frac{\pi^0}{\sqrt{2}} +\frac{\eta}{\sqrt{3}}&\pi^+ &K^+\\
  \pi^- & -\frac{\pi^0}{\sqrt{2}}+\frac{\eta}{\sqrt{3}}&K^0\\
 K^- &\bar K^0& -\frac{2\eta}{\sqrt3}  \end{array}\Bigg);~~~ {\tilde P}={\cal C}P{\cal C}=+\Bigg(\begin{array}{ccc}+\frac{\pi^0}{\sqrt{2}}+\frac{\eta}{\sqrt{3}}&\pi^-&K^-\\\pi^+&-\frac{\pi^0}{\sqrt{2}}+\frac{\eta}{\sqrt{3}} & \bar K^0\\
 K^+& \bar K^0 & -\frac{2\eta}{\sqrt3}
  \end{array}\Bigg)
 \ee
 
\emph{\bf{ Trilinear couplings for nonets}} Given three nonets $A, ~B, ~C$  with C-signs $\eta_A, ~\eta_B,~ \eta_C$,
there are two $SU(3)_f $ invariant couplings: ${\rm Tr}(A B C), {\rm Tr}(A C B)$, and one may form two $C$-invariant combinations (i.e. C-conserving effective couplings):
\bea
&& D={\rm Tr}\Big(A\{B,C\}\Big),~{\rm for}~\eta_A \eta_B \eta_C=+1\notag \\
&&F={\rm Tr}\Big(A[B,C]\Big),~{\rm for}~\eta_A \eta_B \eta_C=-1
\eea
these couplings are used in the classical applications of $SU(3)_f$ symmetry to Vector ($C=-1$)  and Tensor ($C=+1$) meson decays and mixing (see e.g.~\cite{Glashow:1965pf}; mixing of two Kaons belonging to different nonets is considered  in~\cite{Kane:1967trm}), where the allowed couplings are:
\begin{itemize}
\item ${\rm Vector}\to 2~ {\rm Pseudoscalars}$: allowed coupling ${\rm Tr}(V^\mu[P,\partial_\mu P])$
 \item ${\rm Tensor}\to 2~ {\rm Pseudoscalars}$: allowed coupling ${\rm Tr}(T^{\mu \nu}\{P,\partial_\mu \partial_\nu P\})$
\end{itemize}
 
\emph{\bf{ Production of $Z_{cs}(3982)$.}} BES III observes the reaction:  $e^+ e^-\to K^+ + Z^-_{cs}(3985)$. Consider the two cases:
\begin{enumerate} 
\item Production from the  virtual photon continuum
\item Production by a $Y,~J^{PC}=1^{--}$ resonance, assumed to have Isospin=0.
\end{enumerate}
\begin{enumerate}
\item Production from continuum: ${\rm photon}\to K^+ + Z^-_{cs}(3985)$:
\begin{itemize}
\item $C_Z=+1: H_\gamma=i{\rm Tr}(Q[K,Z_{cs}])=i(\frac{2}{3}+\frac{1}{3})(K^+Z_{cs}^- -h.c.)$;
\item $C_Z=-1: H_\gamma={\rm Tr}(Q\{K,Z_{cs}\})=\frac{1}{3}(K^+Z_{cs}^- +h.c.)$;
\end{itemize}
\item  Production from  a $Y(cs\bar c\bar s)$ resonance: $ {\rm photon}\to Y\to K^+ + Z^-_{cs}(3985)$;
\begin{itemize}
\item $C_Z=\pm 1: H_Y={\rm Tr}(Y[K,Z_{cs}]_\mp)= K^-Z^+ \pm  h.c.$
\end{itemize}
\end{enumerate}
Measuring  the energy dependence of the cross section, BES III can distinguish production from continuum (which favours Solution 2) from resonant production (no preference).

\emph{\bf{ ${\bf {Z_{cs}(4003) \to \psi K}}$ decay observed by LHCb.}} For the two possible assignments of  $Z_{cs}(4003)$ and taking into account that $\psi$ has $C=-1$, we have
\begin{itemize}
\item $Z_{cs}, C=-1:~{\cal H}_I=\lambda \mu ~\psi ~({\rm Tr}\{Z,K\}) =\lambda~ \mu~ [Z_{cs}^-~(\psi K^+)  + c.c.],~([\mu]={\rm mass});$

\item for the $J^{PC}=1^{++}$ nonet, the decay may occur to first order in $SU(3)_f$ breaking. Denoting the symmetry quark mass matrix by $\epsilon_8 ={\rm Diag}(m_u,~m_d,~m_s)$ , one finds

 $Z_{cs},C=+1:~{\cal H}_I=\lambda~i\psi ~{\rm Tr}(\epsilon_8[Z,K])\sim \lambda~ (m_s-m_u)~i[Z_{cs}^-~(\psi K^+) -c.c.]$.
\end{itemize}
for $\mu =0.1-1$~GeV there should be no much difference between the two cases and LHCb should be able to see also $Z_{cs}(3982)$.

\subsection{Di-$J/\psi$ resonances: a game changer}

Baryon-antibaryon molecules? Pion exchange can occur between a pair of $\Xi_{cc}=[ccu]$ . However, the bound state would be out of scale:
$2 M_{\Xi_{cc}}\sim 7242~{\rm MeV}$.
Meson-Meson molecule: option tried by some authors assuming $SU(3)_f$ singlet exchange, the only possibility being $J/\psi$ exchange. This leads, however to force range of fractions of fm,  in the full domain where color forces are supposed to dominate.

\emph{\bf{ Tetraquark constituent  picture of di-$J/\psi$ resonances.}} Based on valence, four-charm tetraquarks, $[cc][\bar c\bar c]$. Fermi statistics with color antisymmetric pairs implies diquarks and  antidiquarks in spin 1 state. $S$-wave states are
\begin{itemize}
\item $C=+1$ states: $J^{PC}=0^{++},~2^{++}$, decay in $2 J/\psi$, $2 \eta_c$ (not the $2^{++}$), charmed quark pairs+hadrons
\item $C= -1$ state: $J^{PC}=1^{+-}$, decays in $\eta_c+J/\psi$ and charmed quark pairs+hadrons.
\end{itemize}

Mass spectrum computed using QCD inspired potential (Coulomb+linear potential), gaussian wave functions in the three Jacobi coordinates, ${\bf \xi}_1,~{\bf \xi}_2,~{\bf \xi}_3$~\cite{Bedolla:2019zwg}. Parameters of the gaussians are determined by minimizing the energy.

Authors report the computation of the energy levels of radial and orbital excitations. 
The values of predicted mass include an {\it a priori} unknown additive constant (necessary to fix the zero of the energy for confined states), which is to be determined from one mass of the spectrum. 

In the paper by Bedolla {\it et al}., the constant was taken (provisionally) from calculations of meson masses.
What we can do now is to readjust the additive constant using a known mass as input. Bedolla {\it et al}.  then predict the mass differences.
For illustration, In Fig.~\ref{shift4c}, we show the predictions of few levels obtained from the original article with the above procedure.We have defined the additive constant by identifying the $6900$ MeV peak with the $N=1,~ L=0,~2^{++}$  state, and shifted the other levels accordingly.
The result is tantalizing.

\begin{figure}[htb!]
 \begin{center}
   \centering
   \includegraphics[width=14.0truecm]{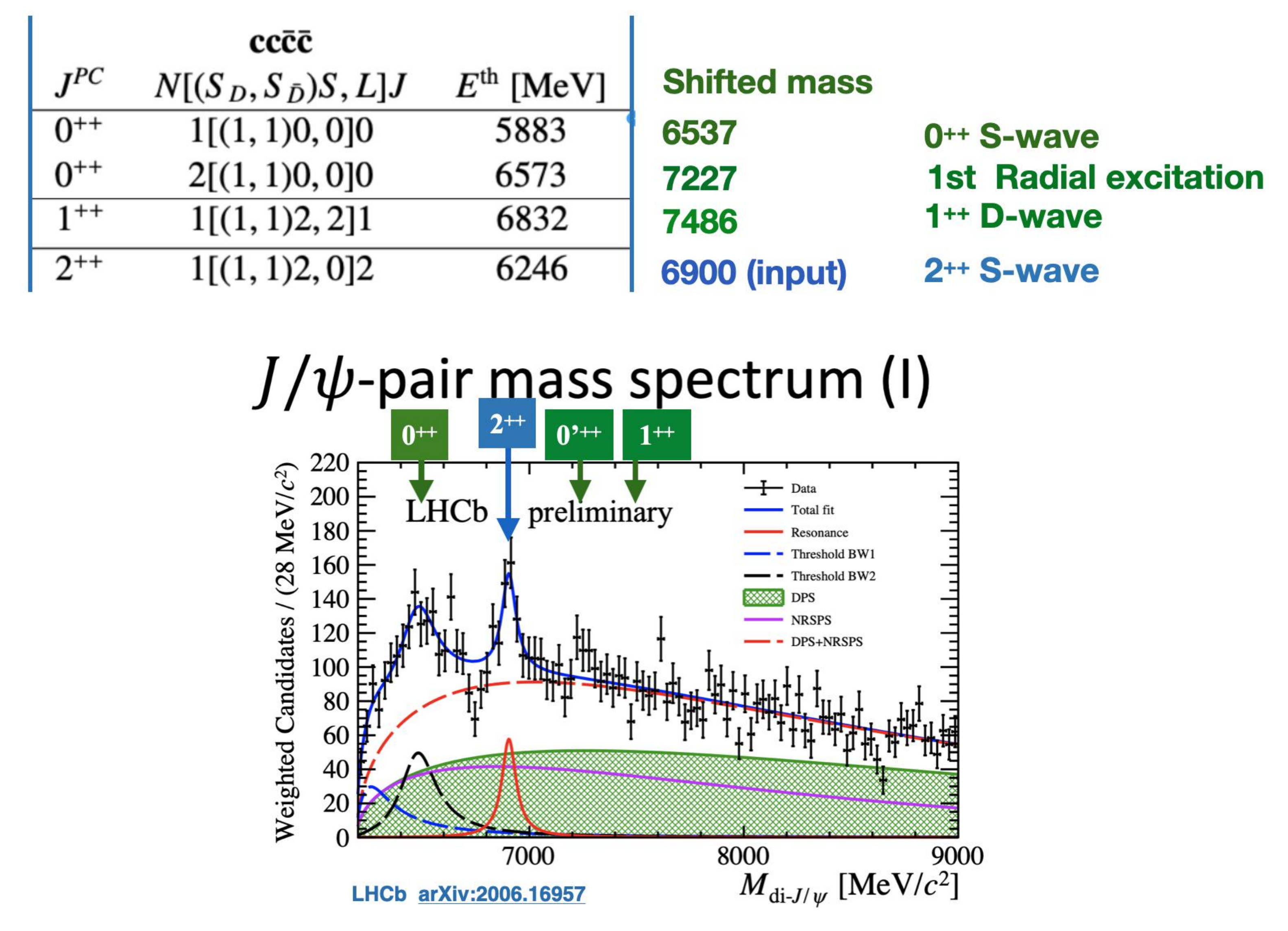}
 \caption{\footnotesize{}}
\label{shift4c}
\end{center}
\end{figure}

\subsection{Decays and branching fractions}

Decays take place via $c\bar c$ annihilation. The starting point is to bring $c\bar c$ pairs together by a Fierz rearrangement, including spin and colour\footnote{from here we follow~\cite{Becchi:2020uvq}.}. One obtains

\bea
&&{\cal T}(J^{PC}=0^{++})=|(c c)^1_{\bf {\bar 3}}(\bar c\bar c)^1_{\bf 3}>^0_1=\notag \\
&&=-\frac{1}{2}\left(    \sqrt{\frac{1}{3}} \left| \left( c^{} \bar{ c}^{} \right)^{\;1}_{\bf{1} }  \left(  c^{}  \bar{ c}^{} \right)^{\;1}_{\bf{1} } \right\rangle^{\;0}_{\bf{1} }   -\sqrt{\frac{2}{3}} \left| \left(  c^{} \bar{ c}^{} \right)^{\;1}_{\bf{8} }  \left(  c^{}  \bar{c}^{} \right)^{\;1}_{\bf{8} }   \right\rangle^{\;0}_{\bf{1} }  \right)+ \notag\\
&&+\frac{\sqrt{3}}{2} \left(    \sqrt{\frac{1}{3}}  \left| \left(  c^{} \bar{ c}^{} \right)^{\;0}_{\bf{1} }   \left(  c^{}  \bar{ c}^{} \right)^{\;0}_{\bf{1} } \right\rangle^{\;0}_{\bf{1} } -\sqrt{\frac{2}{3}} \left| \left(  c^{}  \bar{ c}^{} \right)^{\;0}_{\bf{8} }  \left(  c^{}  \bar{ c}^{} \right)^{\;0}_{\bf{8} }   \right\rangle^{\;0}_{\bf{1} }  \right).\label{mastercol}
\eea
Superscript indicate spin, subscripts the dimensionality of color representation.

there are four possible annihilations
\begin{enumerate}
\item First terms in the second and third  lines of Eq.~\eqref{mastercol}: a color singlet pair of spin 1 or 0 annihilates into a $J/\psi$ or $\eta_c$, the other pair rearranges into the available states (near threshold, $J/\psi$ or $\eta_c$ again);
\item Second term  a color octet, spin 1 pair annihilates into a pair of light quark flavours, q=u,d,s and the latter recombine with the spectator pair to produce a pair of  lower-lying, open-charm mesons. 
\item First term in the third  line of Eq.~\eqref{mastercol}: the similar process from color octet spin 0 pair is higher order in $\alpha_s$ and neglected.  
\end{enumerate}
 Rates are computed with the formula (well known in atomic physics) 
 \be
 \Gamma=|\Psi_{T}(0)|^2\cdot |{\bf v}|\cdot \sigma (c\bar c\to f)
 \ee
Branching fractions are independent from $|\Psi_{T}(0)|^2$. Total rates will be discussed later.

\emph{\bf{Results.}}

Branching fractions are summarised in Tabs.~\ref{gigi}. 

Branching ratios in 4 muons are more favourable in $4~c$ than in $4~b$ tetraquarks, by a factor 4-10. 

\begin{figure}[htb!]
 \begin{center}
   \centering
   \includegraphics[width=12.0truecm]{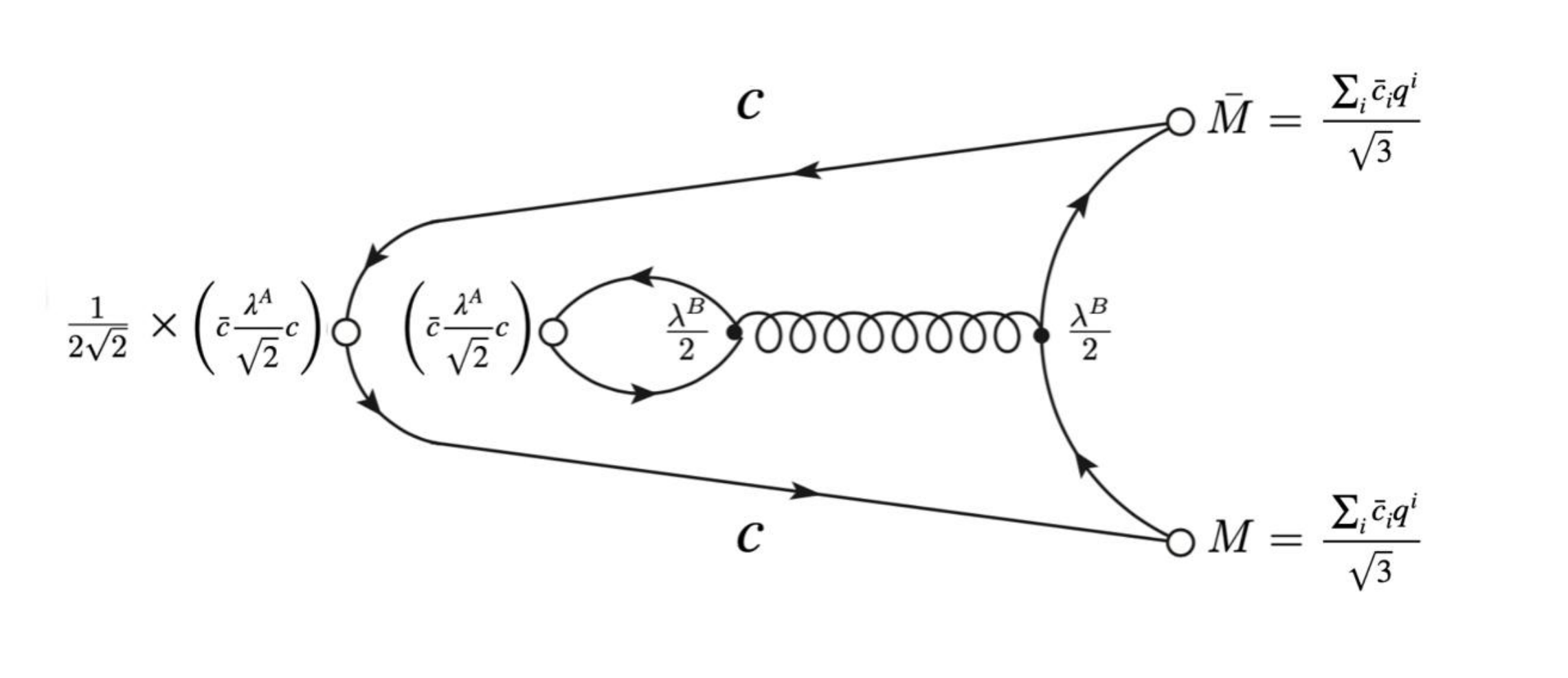}
 \caption{\footnotesize{Annihilation of color octet, spin 1 quark pairs.}}
\label{coloroc}
\end{center}
\end{figure}

Among $4 c$ states, the ratio is more favourable for the $2^{++}$. In addition a spin $2$ particle is produced in $pp$ collision with a statistical factor $2J+1=5$. In summary, the comparison of spin $0$ and $2$ is

  \begin{table}[htb!]
\centering
    \begin{tabular}{|c|c|c|c|c|c|c|}
     \hline
{\footnotesize $[cc\bar c\bar c]$}  &{\footnotesize $\eta_c$+ any} & {\footnotesize $D_q \bar D_q$ ($m_q<m_c$)} & {\footnotesize $ D^*_q \bar D^*_q$}& {\footnotesize $J/\psi$+ any} & {\footnotesize $J/\psi+\mu^+\mu^-$} &  {\footnotesize $4\mu$} \\ \hline 
 $J^{PC}=0^{++}$& {\color{blue}\footnotesize $0.75$}&{\color{blue} \footnotesize $0.021$}&{\color{blue} \footnotesize $0.061$}&{\color{blue}\footnotesize $7.3\cdot 10^{-4}$} & {\color{blue} \footnotesize $4.3\cdot 10^{-5}$} & {\color{blue} \footnotesize $2.58\cdot 10^{-6}$} \\ \hline
  $ J^{PC}=2^{++}$& 0&0&{\color{blue}0.247}& $ {\color{blue} 29.1\cdot 10^{-4}}$ & $ {\color{blue} 17.4\cdot 10^{-5}}$ &  ${\color{blue} 10.3 \cdot 10^{-6}}$ \\ \hline
\end{tabular}
 \caption{ \footnotesize {Branching fractions of fully-charmed tetraquarks, assuming $S$-wave decay.}}
 \label{gigi}
\end{table}  
  \be 
B_{4\mu}(2^{++}):B_{4\mu}(0^{++})\sim 6:1;~ \sigma(2^{++}):\sigma(0^{++})=5:1 \notag
\ee
which gives a visibility ratio 30:1 in favour of $2^{++}$!!

The largest decay fraction is in  charm-anticharm mesons, Fig.~\ref{coloroc}, perhaps accompanied by a tail of light mesons from gluons irradiated in the decay process\footnote{we thank M. Mangano for interesting remarks.}. 
Weak decays of charm-anticharm mesons give rise to a characteristic $e-\mu$ signal, a good signature for events containing the fully-charm tetraquark.

\emph{\bf{Widths.}} Total widths are proportional to the ratio: $\xi=|\Psi_{\it T}(0)|^2/|\Psi_{J/\Psi}(0)|^2$.
We determine $\xi$  from models, and use the spread of values as error estimate. We find $\xi=4.6 \pm 1.4$.

\emph{\bf{In conclusion...}}
The existence  of exotic $SU(3)_f$ multiplets with a characteristic scale of symmetry breaking is a distinctive prediction of compact tetraquarks.
The newly found strange exotics $Z_{cs}$ are close in mass, like X(3872) and Zc(3900), and fit into their nonets: a clear score in favour.
Much remains to be done, to produce more precise data and to search for still missing particles, some with well predicted mass and decay modes.
 It is a tough order: more luminosity, better energy definition, detectors with exceptional qualities... a lot of work...close exchange between theory and experiment is needed.

%% file: chapts/missingpart.tex

We have stressed several times that the compact tetraquark model requires complete $SU(3)_f$ multiplets. Therefore, if $X(3872)$ is a combination of $[cu][\bar c\bar u]$ and $[cd][\bar c\bar d]$, why don't we see its neutral partner (the orthogonal combination), or its charged partner $X^+=[cu][\bar c\bar d]$ ?

We present in this Lecture arguments that indicate that the reasons may be: a) resolution: there may be two still unresolved  lines under the X(3872) lineshape, b) statistics: to see the $X^\pm$ we need to push further down the upper  limits of branching fractions. 

\subsection{Isospin breaking}

The electromagnetic (e.m.) interaction violates Isospin symmetry, since proton and neutron or, equivalently, $u$ and $d$ quarks, have different charges.
A current-algebra calculation of the purely e.m. $\pi^+-\pi^0$ mass difference~\cite{Das:1967it} gives indeed
\be
m_{\pi+}-m_{\pi^0}\simeq 5.0~{\rm MeV},~({\rm expt.}~4.5936 \pm 0.0005)
\ee
This is not the whole story, however. Calculations of  the purely  e.m.  neutron-proton mass difference gave systematically the wrong result: $m_p -m_n > 0$ (has had to be expected from a  purely electrostatic origin).
On top of the e.m. corrections, S. Coleman and S. Glashow ~\cite{Coleman:1963pj}
made the hypothesis that there is a part of the strong-interaction lagrangian (the tadpole) responsible for an additional violation of Isospin. For baryons, the tadpole dominates the p-n mass difference, to give the observed  negative value $m_p-m_n\simeq -1.4~{\rm MeV}$, see e.g.~\cite{Gasser:1974wd}. 

In QCD, the tadpole hamiltonian of Coleman and Glashow is what we called ${\cal L}_3$ (Lect. 4)
\be 
{\cal L}_3= m_3(\bar u u-\bar d d),~m_3=\frac{m_u-m_d}{2};~~(\Delta I=1)
\ee

\emph{\bf{$u,~d$ and $s$ current quark masses}}
Current quark masses measure the degree of non-conservation of the Axial currents, e.g. $A_\mu^1(x)$, in  terms of the Pseudoscalar  current $P^1(x)$
\bea
&&A^1_\mu=\frac{1}{2}\left({\bar u}\gamma_\mu\gamma_5 d+{\bar d}\gamma_\mu\gamma_5 u\right);~P^1=(\bar u\gamma_5d+\bar d\gamma_5 u)
\notag \\
 &&\partial^\mu A^1_\mu=\frac{m_u+m_d}{2} P_1\notag
 \eea
 We introduce the full Scalar and Pseudoscalar octets: 
 \be
 S_i={\bar q}\lambda_i q;~~P_i=i {\bar q}\lambda_i \gamma_5q~(i,j= 0, 1,\cdots, 8)
\ee
with commutation relations:
\be
\left[V^i, S_j\right]=i f_{ijk}~S_k;~~\left[V^i, P_j\right]=i f_{ijk}~P_k;~~\left[A^i, S_j\right]=i d_{ijk}~P_k;~~~\left[A^i, P_j\right]=i d_{ijk}~S_k
\ee 
the fully antisymmetric coefficients  $f_{ijk}$ are the $U(3)_f$ structure functions; the fully symmetric coefficients $d_{ijk}$ are defined in terms of  Gell-Mann's matrices: ${\rm Tr}[\lambda_i\{\lambda_j,\lambda_k\}]=2d_{ijk}$.

The basic Ward identities are ($J^\mu= V^\mu,~A^\mu$):
\bea
&&\quad q_\mu \int d^4x~e^{iqx}<0|T\left[J^\mu(x) P_i(0)\right]|0>=  \notag \\
&&= i\int d^4x~e^{iqx}<0|T\left[\partial_\mu J^\mu(x) P_i(0)\right]|0>+  i\int d^3x~e^{-i{\bf qx}}<0|\left[ J^0({\bf x}, 0),P_i(0)\right]|0>\label{wardid}
\eea

In the limit of quark  masses =0, Chiral symmetry $SU(3)\otimes SU(3)$ is exact. Long ago, G.~Jona-Lasinio and Y.~Nambu~\cite{Nambu:1961tp,Nambu:1961fr} proposed chiral symmetry to be spontaneously broken, with pions the massless Goldstone Bosons. Quark mass would then give a non vanishing pion mass. 

If spontaneous breaking respects (vector) flavour symmetry, $SU(3)_f$, pion and kaons would be in an octet, splitted by quark mass differences due to ${\cal L}_8$, which is what we see.
This situation obtains if scalar densities take a vacuum-expectation-value, with
\be
<0|S_0|0> \neq 0,~<0|S_{3,8}|0>= 0
\ee

Relations between quark current masses and $\pi,~K$ masses has been worked out in 1968 by Gell-Mann, Oakes and Renner~\cite{Gell-Mann:1968hlm} and by Glashow and Weinberg~\cite{Glashow:1967rx}, using the basic Ward identities~\eqref{wardid}. For $m_q\neq 0$ here are no massless Goldstone bosons, the correlation functions have no pole for $q\to 0$ and in this limit the first line tends to zero. We find:
\be
\frac{m_u+m_d}{2} \int d^4x~e^{iqx}<0|T\left(P^1(x) P^1(0)\right)|0> = -i d_{110}<0|S_0|0>=-i A \notag
\ee
We approximate:
\be
 \int d^4x~e^{iqx}<0|T\left(P^1(x) P^1(0)\right)|0> = Z_\pi^2 \frac{i}{q^2-m_{\pi^+}^2} \notag
 \ee
and find:
\be
m_u+m_d= C m_{\pi^+}^2
\ee
with $C$ a constant. Using the other currents, we find, similarly:
\bea
&& m_u+m_d= C m_{\pi^+}^2\notag \\
&& m_u+m_s= C m_{K^+}^2\notag \\
&& m_d+m_s= C m_{K^0}^2 \notag
\eea
and
\be
\frac{m_u+m_d}{m_s+\frac{m_u+m_d}{2}}=\frac{m_{\pi^+}^2+ m_{\pi^0}^2}{m_{K^+}^2+m_{K^0}^2}=R\simeq 0.077~\to~\frac{m_u+m_d}{2}\simeq 6~{\rm MeV}, ~{\rm for}~m_s=150~{\rm MeV}.\label{quarkmass}
\ee

R. Dashen~\cite{Dashen:1969eg} showed that the combination:
\be
[ m_{K^+}^2-m_{K^0}^2] - [m_{\pi^+}^2- m_{\pi^0}^2] \notag
\ee
is not affected by e.m. corrections. Using the previous results, we find:
\be
\frac{m_u-m_d}{m_s+\frac{m_u+m_d}{2}}=2\frac{[m_{K^+}^2-m_{K^0}^2]-[m_{\pi^+}^2-m_{\pi^0}^2]}{m_{K^+}^2+m_{K^0}^2}\simeq -0.021 \notag
\ee
Finally
\bea
&&m_s=150~{\rm  MeV}:~m_u\simeq 4.3~{\rm MeV};~m_d\simeq 7.6~{\rm MeV}~~m_d-m_u=3.3~{\rm  MeV} \label{emdiffs1}\\
&&m_s=180~{\rm  MeV}:m_u\simeq 5.0~{\rm MeV};~m_d\simeq 8.8~{\rm MeV}~,~m_d-m_u=3.9~{\rm MeV}\label{emdiffs2}
\eea
The mass difference of up and down quarks is not small compared to their masses. The real world is close to isotopic spin symmetry not because  quark masses are very similar, like the masses of neutron and proton, but because both are roughly equal to zero (this result was first reported in~\cite{cm70}).

\emph{\bf{Current vs Constituent quark masses.}} In Lect.2 we found much larger constituent quark masses then the current masses just found.
The difference is believed to be due to the  mass associated to the QCD field that surrounds quarks in the bound states. Indeed constituent quark masses from baryon and meson spectrum are different and we may attribute the difference to the different QCD field configurations in mesons and baryons.
However, as remarked already in Lect. 2, the $q-s$ mass difference, which is determined by the Ward identity valid to all orders in QCD, turns out to be the same for baryon or meson constituent masses. 

The value $m_s-m_q=150$ MeV used in Eq.\eqref{emdiffs1} is an estimate from the equal spacing rule of the  baryon spin 3/2 resonances. 
In the value used in  \eqref{emdiffs2},  the breaking due to spin-spin interaction has been removed, and it gives a slightly larger value of Isospin breaking: $ m_d-m_u = 4$ MeV (we neglect the small difference and adopt \eqref{emdiffs1} in the following).
 
\emph{\bf{Isospin breaking and baryon masses.}} Isospin breaking effects have three components: the quark mass difference, the e.m. hyperfine interaction and the electrostatic repulsion between quarks.
Karliner and Rosner~\cite{Karliner:2017gml} fit isospin breaking differences in baryons to determine the relevant parameters,
in particular the electrostatic corrections, which scale with the radius of the diquark in the baryon:
\be
a=\frac{1}{137}<\frac{1}{R}>_{qq}^{{\rm Baryon}}\sim 2.83~ {\rm MeV}\rightarrow R_{(qq^\prime)_{{\bf{\bar 3}}}}^B\sim 0.5~\rm fm \label{radbar}
\ee

\subsection{Isospin breaking in tetraquarks}

\begin{figure}[htb!]
 \begin{center}
   \centering
   \includegraphics[width=12.0truecm]{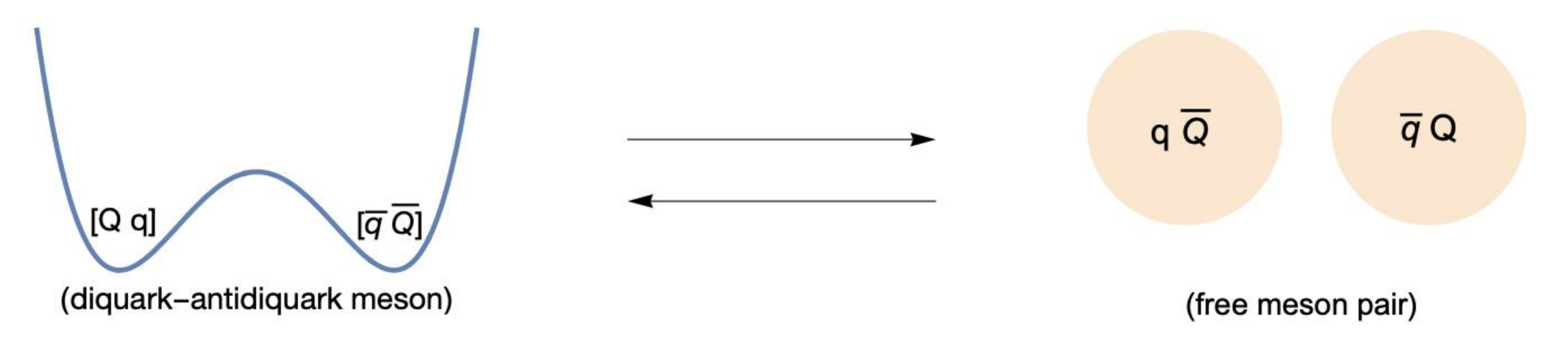}
 \caption{\footnotesize{}}
\label{wells}
\end{center}
\end{figure}

The attraction that produces the diquark implies that diquarks, or antidiquarks, are segregated in two different potential wells, separated in space~\cite{Maiani:2017kyi}.
 QCD confining forces prevail at large distances, where the diquarks see each other as QCD point charges. 
At shorter distances the internal structure is felt and the competing interactions that tend to dissociate the diquark, e.g. the attraction between quarks and antiquarks, produce repulsive forces between diquark and antidiquark and a rise in the potential~\cite{Selem:2006nd}.
A phenomenological basis is provided by the mass ordering in Z(3900) vs Z(4020):
\begin{itemize}
\item spin-spin interactions between light quark and antiquark located in different diquarks are definitely smaller than one would guess from the same interactions within mesons;
\item spin-spin interaction inside the diquark is about four times larger that the same interaction in the diquarks inside charmed baryon states.
\end{itemize}
Separated diquark-antidiquark potential wells are illustrated in Fig.~\ref{wells}. This configuration could explain the more important decay into open charm pairs compared to charmonium decays, due the difficulty of the heavy charm quark to tunnel below the potential barrier (see Lect. 6). 

If quark mass difference dominates: $M_{X_u}-M_{X_d}= 2(m_u-m_d)\sim -6~{\rm MeV}=\Delta m$, but we do not see two distinct lines around $3872$ MeV. However, one must account (mainly) for electrostatic interactions inside diquarks and between diquark and antidiquark.
 
Let us indicate with $R_{cq}$, and $R_X=\lambda R_{cq}$ the diquark radius and the $X$ radius, respectively. 
From the parameter determined by Karliner and Rosner, \eqref{radbar} and the ratio of hyperfine couplings, $\kappa_{cq}$ in baryons and tetraquarks, we determine $R_{cq}$ as:
\be
R_{cq}\sim R_{(cq^\prime)_{{\bf{\bar 3}}}}^B ~(\frac{\kappa_{cq}^B}{\kappa_{cq}^X})^{1/3}\sim 0.3~\rm fm \notag
\ee
$\lambda$ appears in the electrostatic interaction between quarks. Following K$\&$R~\cite{Karliner:2017gml}}\footnote{ The role of the electrostatic corrections goes as follows: the second term in the r.h.s. is the effect of the electrostatic $u-c$ repulsion vs $d-c$ attraction, which increases $M( X_u)$ with respect to 
$M(X_d)$, the third term is the effect of the $[uc]-[\bar u\bar c]$ vs. $[dc]-[\bar d\bar c]$ electrostatic attraction, which reduces $M(X_u)$ w.r.t. $M(X_d)$ and is suppressed for large $\lambda$.}
\be
M(X_u)-M(X_d)=\Delta m+\frac{4}{3}a^\prime-\frac{5}{3}\frac{a^\prime}{\lambda}+\dots \notag
\ee
where $a^\prime=a\frac{R_{(cq)_{{\bf{\bar 3}}}}^B}{R_{cq}}$ is the parameter $a$ in \eqref{radbar}, rescaled with the diquark radius.

With respect to the naive estimate, the size of the mass difference is suppressed for $ \lambda>1$. Numerically: $M(X_u)-M(X_d)\sim -1.2$ MeV for $\lambda=3$, i.e. $R_X\sim 1$ fm.

$X_u~{\rm and}~X_d$ may be degenerate within the X line,  observed at present with resolution larger than $1$ MeV. Is experimental resolution the key to the puzzle ?

\subsection{Searching for $X^+=[uc][\bar d\bar c]$}

The charged $X$ has been searched in $B^0$ and $B^+$ decays: $B\to K X^+$. 
Present experimental upper limits are inconsistent wth the hypothesis that X(3872)  has isospin$=1$, which however is a very restrictive hypothesis
To test consistency of data with the tetraquark picture, we have analysed~\cite{Maiani:2020zhr} the observed four ratios of decay rates
\bea
&& R^{(++)}=~\frac{\Gamma(B^+\to K^+ +X(3872) \to K^+ + 3\pi ~\psi )}{\Gamma(B^+\to K^+ +X(3872) \to K^+ +2\pi ~\psi )}\label{rat1};  \\
&& R^{(00)}=\frac{\Gamma(B^0\to K^0 +X(3872) \to K^+ + 3\pi ~\psi)}{ \Gamma(B^0\to K^0 +X(3872) \to K^+ + 2\pi ~\psi)}; \label{rat2} \\
&& R^{(++,00)}_{2\pi}=\frac{\Gamma(B^+\to K^+ +X(3872) \to K^+ +2\pi ~\psi )}{ \Gamma(B^0\to K^0 +X(3872) \to K^+ + 2\pi ~\psi)};\label{rat3}  \\
&&R^{(s\phi,00)}_{2\pi}=\frac{\Gamma ( B_s\to \phi + X(3872)  \to \phi +2\pi~ \psi)}{\Gamma( B^0\to K^0\, X(3872)  \to K^+ 2\pi ~\psi)}.\label{rat4}
\eea
We have updated the analysis using the recent value~\cite{LHCb:2022bly}:
\be
R^{(++)} =0.70\pm 0.4 \to 2.0 \pm 0.5
\ee
with  the other  values  unchanged:
\be
R^{(00)}=1.4\pm 0.6,~R^{(++,00)}_{2\pi}=2.0\pm 0.6,~R^{(s\phi,00)}\sim 1
\ee

\begin{figure}[htb!]
 \begin{center}
\begin{minipage}[c]{5.6cm}
   \includegraphics[width=5.6truecm]{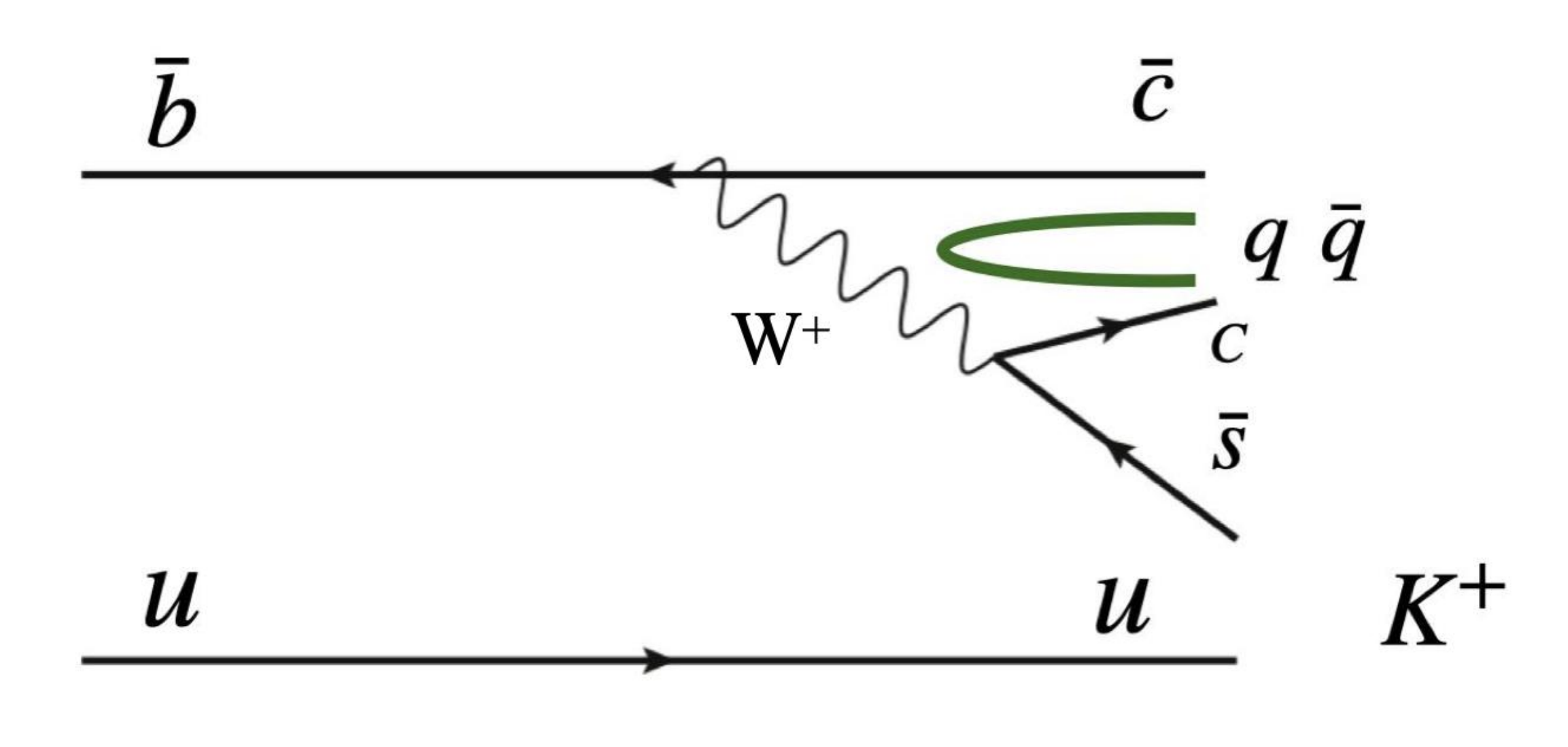}
  \end{minipage}%
\begin{minipage}[c]{10.6cm}
    \includegraphics[width=10.6truecm]{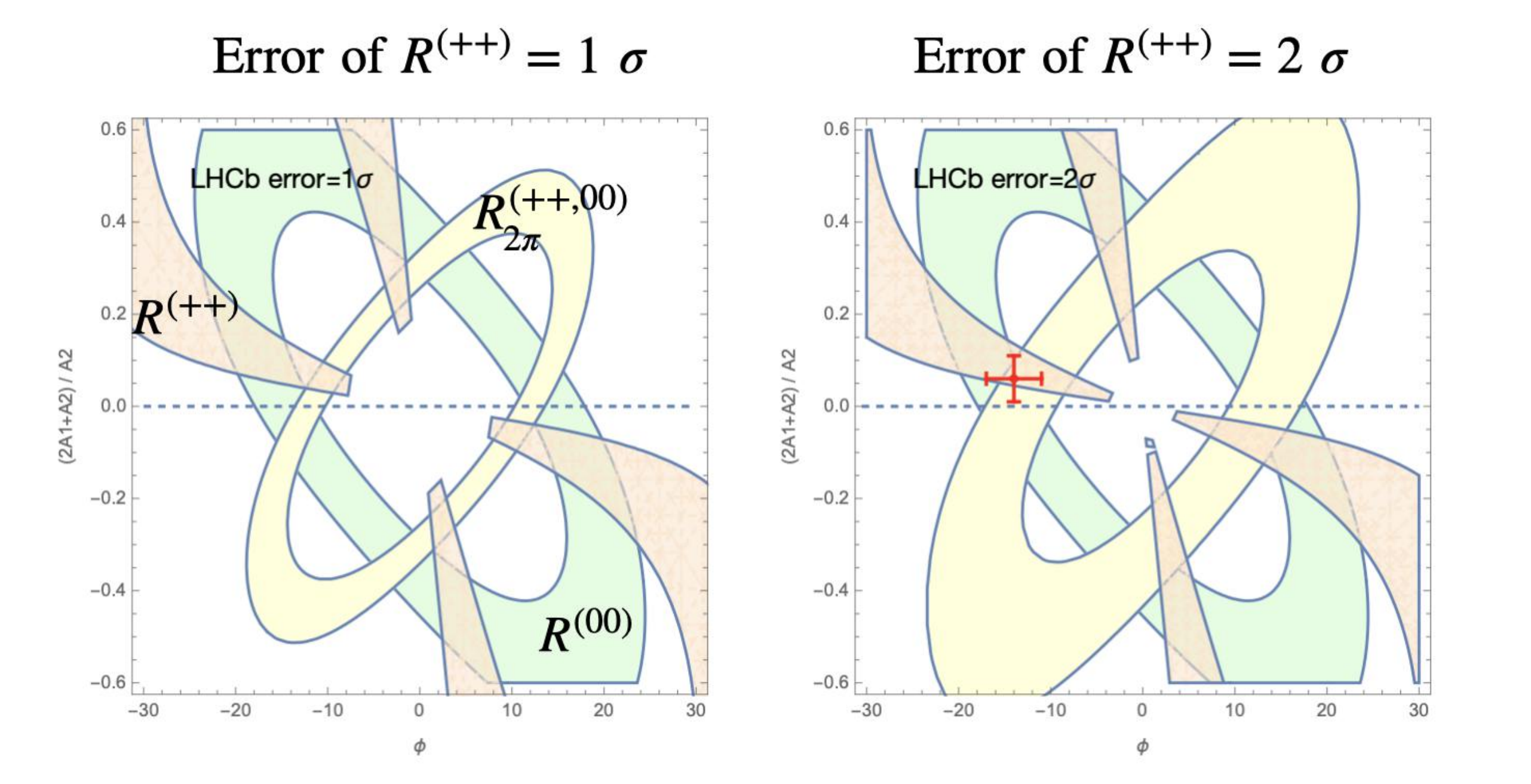}
   \end{minipage}%
 \caption{\footnotesize{Left: diagram for$B$-decay to a six-quark state. Right: contour plots of the ratios in \eqref{rat1}-\eqref{rat3}. }}
\label{bdec}
\end{center}
\end{figure}

Fig.~\ref{bdec}  (left panel) shows the diagram for the decay: $B^+\to K^++{\rm tetraquark}$. Restricting to non-strange tetraquarks $X_u ~{\rm and}~X_d$, the $K$ meson is formed by the $\bar s$ from weak decay and: either the spectator quark (amplitude $A_1$) or the quark from the sea ($A_2$). One has:
\bea
&&{\cal A}(B^+\to X_d\, K^+) = A_1,~{\cal A}(B^+\to X_u\, K^+)= A_1+A_2;~ {\cal A}(B^+\to X^+\,K^0 )= A_2\notag \\
&&{\cal A}(B^0\to X_d\, K^0) = A_1+A_2,~{\cal A}(B^0\to X_u\, K^0)= A_1;~ {\cal A}(B^0\to X^-K^+ )= A_2\notag
\eea

Physical tetraquarks $X_{1,2}$ are combinations of $X_{u,d}$ with a mixing angle:
\be
X_1=\cos\phi \frac{X_u+X_d}{\sqrt{2}}+\sin\phi \frac{X_u-X_d}{\sqrt{2}};~X_2=-\sin\phi \frac{X_u+X_d}{\sqrt{2}}+\cos\phi \frac{X_u-X_d}{\sqrt{2}}\notag
\ee
so that, e.g. 
\bea
&& A(B^+\to K^+X_1\to K^+ \rho^0+\psi)=A(B^+\to K^+X_1)\cdot A(X_1\to K^+ \rho^0+\psi) \propto \notag \\
&& \propto~\Big[(\cos\phi+\sin\phi)(A_1+A_2)+(\cos\phi-\sin\phi)A_1\Big]\sin\phi=\notag \\
&& =\Big((2A_1+A_2)~\cos\phi+A_2~\sin\phi \Big) \sin\phi \notag
\eea
 the observed rates are obtained by summing incoherently the amplitudes squared of the two unresolved lines inside X(3827), i.e. summing  the rates into $X_1~{\rm and}~X_2$. Apart from  phase space, ratios of rates are functions of $\phi$ and $z=A_2/(2A_1+A_2)$.

 \emph{\bf{Results}}
 
  The red cross in the figure on the right indicates a solution with errors estimated from the extension of the overlap:
\be
\phi=-14^0  \pm 3^0;~z=+0.06\pm 0.005
\ee
Parameters of this solution are rather similar to those given previously in ~\cite{Maiani:2020zhr}.
 
  Limits to $X^\pm$ production
\bea
&&PdG:~R_{2\pi}^{(0+,00)}< 1 \\
&& R^{(0+,00)}_{2\pi}<0.57~(\rm previous ~analysis) \\
&& R^{(0+,00)}_{2\pi}<0.26~({\rm present~analysis}).
\eea

%% file: chapts/bornoppe.tex

Doubly heavy hadrons, baryons or tetraquarks, are a relatively new sensation. 

A doubly charmed baryon, $\Xi^+_{cc}=[ccd]$ has been first observed by LHCb in 2018~\cite{LHCb:2018pcs}. Doubly heavy tetraquarks, e.r. $[cc][\bar q \bar q^\prime]$ have been proposed by Esposito {\it et al} in 2013~\cite{Esposito:2013fma} and later considered as possibly weakly decaying hadrons by Karliner and Rosner~\cite{Karliner:2017qjm} and by Eichten and Quigg~\cite{Eichten:2017ffp} in 2017.

As illustrated in Lect.~1, starting from a hadron containing one heavy antiquark, e.g. a $q\bar c$ meson, one can generate a QCD allowd configuration by replacing the anticharm with a $[cc]$ pair in a color  antisymmetric ${\bf {\bar 3}}_c$ configuration. In the limit of infinite  charm mass, the two charm quarks will be closer and closer and the spectator quark sees a QCD field configuration identical to the field in the original meson. In this limit, one has a symmetry linking the QCD field energies of double charm baryon and charmed meson. 
This is the  content of the
{\it single heavy-doubly heavy quark symmetry} introduced by Savage and Wise and by Brambilla, Vairo and Rosch~\cite{Savage:1990di,Brambilla:2005yk,Mehen:2017nrh}, which says that the masses of the two particles are the same, except for a calculable difference due to the different constituent quark mass and the hyperfine, spin-spin interaction. An example of this symmetry is the relation between hyperfine mass differences:
\be
(\Xi^*_{cc}-\Xi_{cc})=\frac{3}{4} (D^*-D) \notag
\ee
Constituent quark masses cancel separately in the l.h.s and r.h.s.; the ratio 3/4 arises from explicit calculations of hyperfine interactions (see Lect. 2)
\be
\frac{ (\Xi^*_{cc}-\Xi_{cc})}{(D^*-D)}=\frac{3\kappa_{(cq)_{\bf{\bar 3}}}}{2\kappa_{(q\bar c)_{\bf 1}}}\notag
\ee
after using the QCD relation $\kappa_{(q\bar c)_{\bf 1}}=2\kappa_{(cq)_{\bf{\bar 3}}}$.

\subsection{Born-Oppenheimer approximation in brief}

Consider a system with two heavy and (say, two) light particles, with Hamiltonian~\cite{Weinbergqm}:
\be
H= H_{heavy}+H_{light}=\frac{1}{2M}\sum_{heavy} P_i^2 + V({\bf x}_A,~{\bf x}_B) +\frac{1}{2m} \sum_{light}p_i^2+V_l({\bf x}_A,~{\bf x}_B,~{\bf x}_1,~{\bf x}_2)
\ee

First, consider the heavy particles as classical sources with fixed coordinates and quantum numbers, and find the ground state of the light particles, solving the eigenvalue equation: 
\be
H_{light}|f_0>={\cal E}|f_0>;~f_0=f_0({\bf x}_A,~{\bf x}_B,~{\bf x}_1,~{\bf x}_2), ~{\cal E}={\cal E}({\bf x}_A,~{\bf x}_B)
\ee
Then look for solutions of  the complete Schr\"odinger equation for wave functions of the form: 
 \be
  \Psi=\psi({\bf x}_A,{\bf x}_B)~ f_0({\bf x}_A,~{\bf x}_B,~{\bf x}_1,~{\bf x}_2) \notag
\ee
Applying $H$ to $\Psi$ one encounters terms of the kind:
\be
-i{\bf P}\Psi=\frac{\partial}{\partial x_A}\Psi=\frac{\partial \psi}{\partial x_A}f_0+\psi\frac{\partial f_0}{\partial x_A}\label{twoterms}
\ee
The Born-Oppenheimer approximation consists in neglecting systematically the second with respect to the first term. 
The error vanishes for $m/M \to 0$ (for Electrodynamics see Weinberg's book, we shall consider later the error in QCD, Sect.~\ref{error}).

The  upshot is the Born-Oppenheimer (BO) equation: 
\be
\left(\sum_{\rm heavy} \frac{P_i^2}{2M}+V_{BO}({\bf x}_A, {\bf x}_B)\right)\psi=E\psi  \notag
\ee
with the BO potential given by
\be
V_{\rm BO}({\bf x}_A, {\bf x}_B)= V({\bf x}_A, {\bf x}_B) + {\cal E}({\bf x}_A, {\bf x}_B) \notag 
\ee

\emph{\bf{Orbitals (borrowed from molecular physics).}}

We associate each light particle to one heavy particle with attractive mutual interaction, solving the corresponding Schr\"odinger equation and neglecting the interactions with the other particles. The wave function thus obtained is an {\it orbital} and we choose $f_0$ as the product of all orbitals. 
If there are identical light particles, we have to symmetrise (we consider here $S$-wave states, statistics will be taken care by symmetrising or anti-symmetrising in the internal degrees of freedom).
The interactions left-over from the orbitals, e.g. interactions between light particles, are taken to first order in perturbation theory and
\be
{\cal E}=<f_0| H_{light}|f_0> \notag
\ee
\emph{\bf {Warming-Up with  the ion $H_2^+(PPe^-)$}} 

The light particle interaction potential is
\be
V_I=-\alpha~\left(\frac{1}{|{\bf x}-{\bf x}_A|}+\frac{1}{|{\bf x}-{\bf x}_B|}\right)\label{ionh2}
\ee
It produces two orbitals: $e-P(x_A)$ and $e-P(x_B)$ with wave functions $f(x-x_A)$ and $f(x-x_B)$ ($f=$Hydrogen wave function) and eigenvalue 
\be
{\cal E}=-\frac{1}{2} m_e \alpha^2 \notag
\ee
 We take~(a standard reference on $\text{H}$, $\text{H}_2$ and $\text{He}$ molecules and ions is~\cite{pauling}): 
 \be
 f_0=\frac{f(x-x_A)+f(x-x_B)}{\sqrt{2(1+S)}};~~S=\int f^*(x-x_A)f(x-x_B)\notag
 \ee
 and
 \be
 V_{BO}=+\alpha\frac{1}{|{\bf x}_A-{\bf x}_B|}+<f_0|H_{light}|f_0>
 \ee
 
 \subsection{Colour gymnastic:  couplings \dots and strings}
 \label{colorgim}
 
 We start with the simplest case~(explicit derivations are found in~\cite{Maiani:2019lpu}).
 
 {\bf \emph{Doubly charmed baryon: $\bf c\bf c$ in ${\bar {\bf 3}}$}}.
In a colour singlet baryon, all pairs are in colour ${\bar{\bf 3}}$, and the colour couplings (see lect. 2) are distributed according to
\be
\lambda_{c c}=\lambda_{c q}= -2/3 \label{lambar}
\ee
{\bf \emph{Hidden charm tetraquark.}}
  We take the charm pair in {\it{\bf color~octet}}, so that
  \be
  {\rm T}= {\rm Tetraquark}=|(\bar c c)_{\bf 8}(\bar q q^\prime)_{\bf 8}>_{\bf 1}
  \ee
correspondingly, we compute the colour couplings as
\be
\lambda_{c\bar c}=[\frac{1}{2}(C_2({\bf 8})-2C_2({\bf 3}))]=+\frac{1}{6}=\lambda_{q\bar q^\prime}~({\rm repulsive})
\ee
$c\bar c$ repel each other as $q\bar q$ do, like protons and electrons in $H_2$  molecule.
Using a Fierz rearrangement formula (see the Appendix of~\cite{Maiani:2019lpu}),  we also obtain:
\be
 {\rm T}=|(\bar c c)_{\bf 8}(\bar q q^\prime)_{\bf 8}>_{\bf 1}=\sqrt{\frac{2}{3}} |(Qq)_{\bar{\bf 3}} (\bar Q \bar q)_{\bf 3}\rangle_1-\frac{1}{\sqrt{3}}|(Qq)_{\bf 6} (\bar Q \bar q)_{{\bar {\bf 6}}}\rangle_1\notag
  \ee
 so that
 \be
 \lambda_{\bar q c}=\lambda_{\bar c q}=\frac{2}{3}\lambda_{\bf 3}+\frac{1}{3}\lambda_{\bf 6}=\frac{2}{3}(-\frac{2}{3})+\frac{1}{3}(\frac{1}{3})=-\frac{1}{3}~({\rm attractive})\notag
 \ee
Similarly one finds
\be
\lambda_{cq}=\lambda_{\bar c\bar q}=-\frac{7}{6}~({\rm attractive})\notag
\ee

{\bf \emph{Double charm tetraquark ${\bf  {[cc]{\bar u\bar d]}}}$.}} 
We assume  $c c$ in ${\bar{\bf 3}}$.
The lowest  energy state corresponds to $ cc$ in spin one and light antiquarks in spin and isospin zero. 
The tetraquark state is:
\be T_{cc}=|(cc)_{\bar {\bf 3}}, (\bar u\bar d)_{ {\bf 3}} \rangle_{\bf 1}\label{tetra3}\notag
\ee
It can be Fierz transformed into
\be
T_{cc}=\sqrt{\frac{1}{3}}|(\bar u c)_{\bf 1},(\bar d c)_{\bf 1}\rangle _{\bf 1}-\sqrt{\frac{2}{3}}|(\bar u c)_{\bf 8},(\bar d c)_{\bf 8}\rangle _{\bf 1}\notag
\ee
with all attractive couplings
\be
\lambda_{cc}=\lambda_{\bar u \bar d}=-\frac{2}{3};\quad 
\lambda_{c\bar u}=\lambda_{c\bar d}=-\frac{1}{3}\notag
\ee

{\bf \emph{String tension.}} The full potential is obtained as a generalization of the Cornell potential, introduced in connection with charmonium spectrum.
For a heavy colour triplet pair, $Q\bar Q$, in an overall colour singlet state, the QCD potential is taken as:
\be
V(r)=-\frac{4}{3}~ \frac{\alpha_S}{r} + k r+V_0=V_{C}(r)+V_{\rm conf}(r)+V_0
\ee 
\begin{itemize}
\item $V_0$ is an unknown constant, to be determined from the mass spectrum, e.g. from the mass of the ground state;
\item the first term corresponds to the one-gluon exchange  approximation. It is generalised to any pair of coloured particles in a colour representation ${\bf R}$ by the combination of the Casimir coefficients introduced in Lect.2
\be
 -4/3 \to \lambda_{q_1q_2}\notag
\ee
\item the second term arises from quark confinement and dominates at large separations; 
in the simplest picture, it is due to the condensation of Coulomb lines of force into a string that joins quark to the antiquark;
in  this picture, it is natural to assume that the string tension, $k$, scales with the coefficient of the Coulombic interaction: $k_{q_1 q_2}\propto |\lambda_{q_1 q_2}|$.
\item for colour charges combined in an overall colour singlet, the assumption leads to $k\propto |C_2({\bf q})|$ (called Casimir scaling). 
\end{itemize}
In conclusion, we take $k$ from chamonium spectrum and:
\be
V(r)=\lambda_{q_1 q_2}~ \frac{\alpha_S}{r} +\frac{3 |\lambda_{q_1 q_2}|}{4}~k r+V_0=V_{C}(r)+V_{\rm conf}(r)+V_0 \label{potbo}
\ee
 Orbitals with non-vanishing triality are confined and we add to the BO potential the appropriate linearly rising potential, Fig.~\ref{ccbar}(left).
 
Triality zero orbitals, e.g. $(c\bar q)_{\bf 8}$, are NOT confined because color can be neutralised by extra gluons, and the BO potential vanishes for large separation of the heavy constituent~\cite{Maiani:2019lpu,Bali:2000gf}. 

The Cornell potential contains the additive constant $V_0$. In charmonium physics $V_0$ is determined from one physical mass of the spectrum, e.g. the ground state.  In some cases, we are able to determine $V_0$:
\begin{itemize}
\item for $Q \bar Q q \bar q$, $V_0$ is fixed by the ground state mass, which then cannot be predicted. The wave function gives  valuable information about the internal structure.
\item for Double charm baryon: $V_0$ is fixed by the boundary condition from heavy quark-heavy diquark symmetry; 
 \item for $QQ \bar q\bar q^\prime$ tetraquarks: $V_0$ is fixed by the condition that the potential gives rise to a two mesons state at infinite separation.
 \end{itemize}

\subsection{The mass of the doubly charmed baryon.} \label{bcc}

\begin{figure}[htb!]
 \begin{center}
\begin{minipage}[c]{7.6cm}
   \includegraphics[width=6.0truecm]{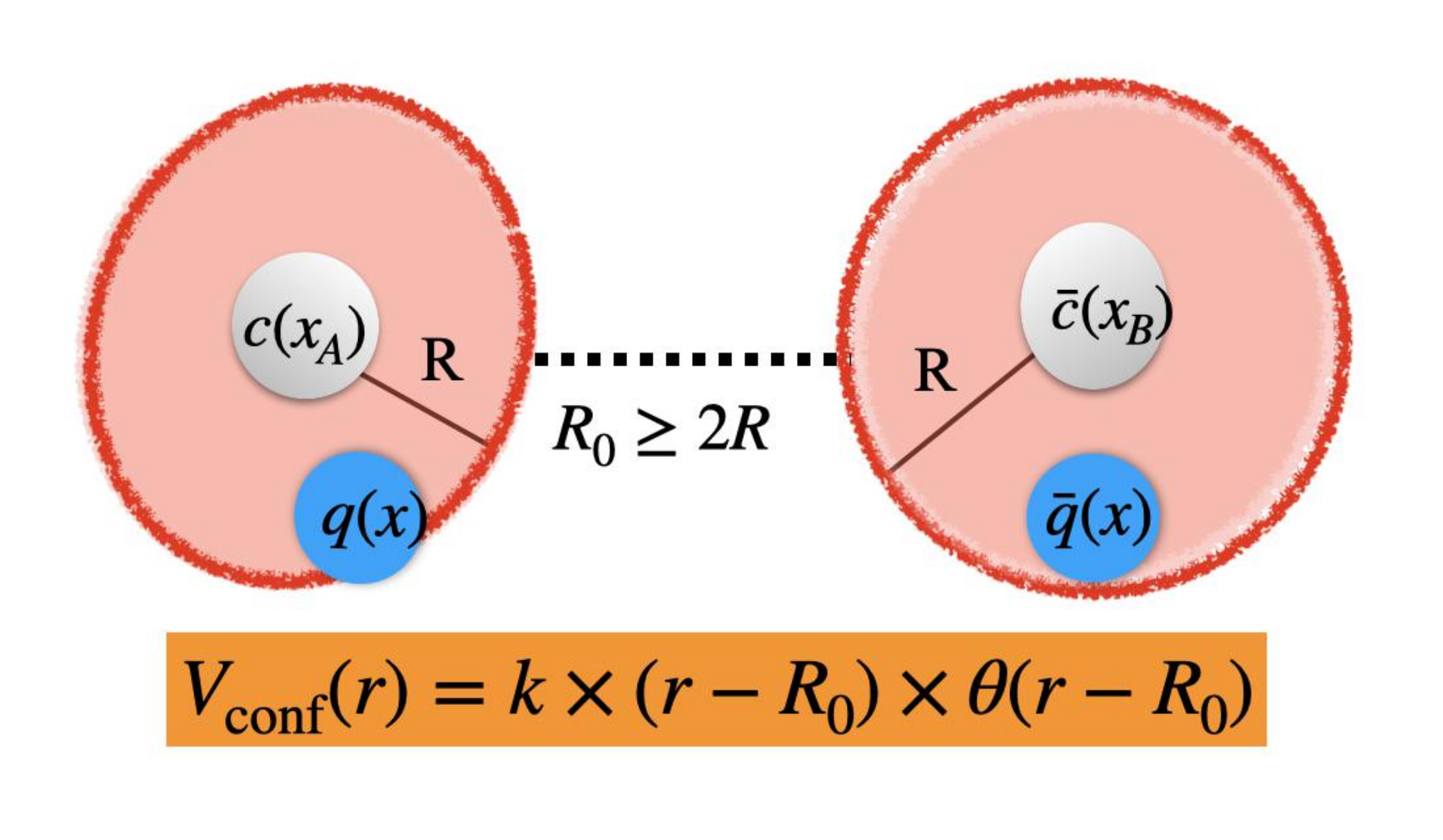}
  \end{minipage}%
\begin{minipage}[c]{7.6cm}
    \includegraphics[width=8.6truecm]{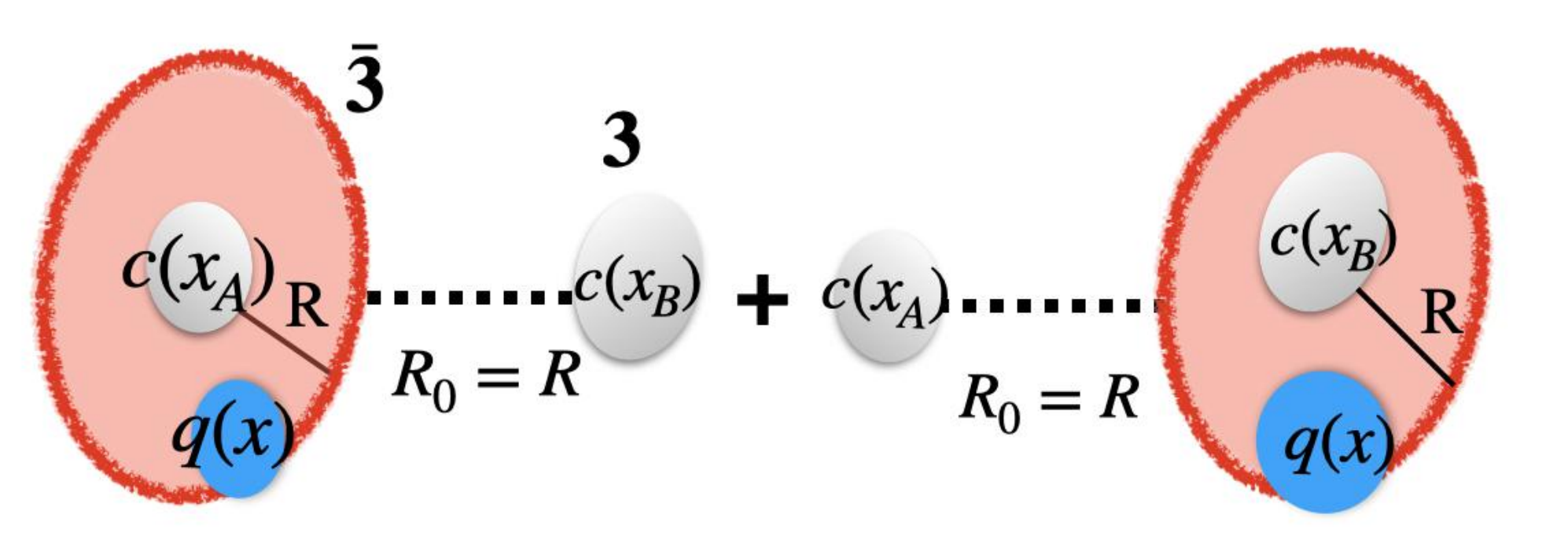}
   \end{minipage}%
 \caption{\footnotesize{(left) Confined orbitals.} (right) Ground state of the doubly charmed baryon.}
\label{ccbar}
\end{center}
\end{figure}

We treat it like the $H_2^+=(PPe^-)$  ion.
\be
H_{light}=\frac{p^2}{2m}
+\Big[-\frac{2}{3}\alpha_s\frac{1}{|{\bf x}-{\bf x}_A|}+V_{conf}(x-x_A)\Big]+
\Big[-\frac{2}{3}\alpha_s\frac{1}{|{\bf x}-{\bf x}_B|}+V_{conf}(x-x_B)\Big]
\ee
There are two orbitals, obtained by combining kinetic energy with either the first or the second term of the interaction: 
\be
{\rm orbital}(A,B)=f(x-x_{A,B})=f_{A,B}(x) \notag
\ee
Ground state, Fig.~\ref{ccbar}(right):
\be
f_0=\frac{f(x-x_A)+f(x-x_B)}{\sqrt{2(1+S)}};~~S=\int d^3x ~f_A(x)^*f_B(x)  \notag
\ee
Energy of the orbital: 
\be
E_0 =V_0+E_{0,orb}+2M_c+M_q \label{zeroen}
\ee

Calling $r_{AB}=|x_A-x_B|$, the first order energy is:
\bea
&&{\cal E}(r_{AB})=<f_0|H_{light}|f_0>=E_0+\Delta E(r_{AB}) =\notag\\
&&=E_0-\frac{2\alpha_S}{3}\frac{1}{1+S}\left[I_1(r_{AB})+I_2(r_{AB})\right] \notag
\eea
$I_{1,2}$ are functions of $r_{AB}$ defined in terms of the orbitals
\bea
&&I_1(r_{AB})= \int d^3x \,|f_A(x)|^2 \, \frac{1}{|{\bf x}-{\bf x}_B|}\notag\\
&&I_2(r_{AB})= \int d^3x~f_A(x)^*f_B(x)~ \frac{1}{|{\bf x}-{\bf x}_B|}\notag
\eea
$E_0$ is the sum of constituent quark masses plus all additive constants, that is the energy of the orbitals and the confinement constant $V_0$, given in Eq.~\eqref{zeroen}. 
When we let $r_{AB}\to 0$  and subtract one unit of $M_c$, the total energy should tend to the mass of the $c\bar q$ meson, which, in the constituent quark model equals $M_c+M_q$. In formulae:
\bea
&&E_0+\Delta E(0)-M_c=M_c+M_q, ~{\rm that~is}\notag\\
&&E_{0,orb}+V_0=-\Delta E(0)
\eea
this is the {\it {\bf single heavy quark-double heavy quark}} symmetry.
Adding the confinement potential, the full BO potential is determined as
\bea
&&V_{\rm BO}(r_{AB})=\notag \\
&&=2M_c+M_q -\frac{2}{3}\alpha_S\frac{1}{r_{AB}}+ [\Delta E(r_{AB})-\Delta E(0)]+k \times (r-R_0)\times \theta(r-R_0)\notag\\
&&=2M_c+M_q -\Delta E(0) +\Big\{-\frac{2}{3}\alpha_S\frac{1}{r_{AB}}+ \Delta E(r_{AB})+k \times (r-R_0)\times \theta(r-R_0)\Big\}\label{finalbo}
\eea
We solve numerically the Schr\"odinger equation for heavy particles with the potential given be the expression in curly brackets and denote by $E_{BO}$ the eigenvalue of the ground state. We find
\be
E_{BO}= -45^{+17}_{-7}~{\rm MeV}~{\rm for}~R_0=8\pm 2 ~{\rm GeV}^{-1}
\ee
\begin{table}[htb!]
\centering
    \begin{tabular}{|c|c|c|c|c|c|}
     \hline
     {\footnotesize{Terms}} &  {\footnotesize{Value (MeV) MPR}}&  {\footnotesize{Notes}} & {\footnotesize{Terms}} & {\footnotesize{Value (MeV) K$\&$R}}   &  {\footnotesize{Notes}} \\
     \hline
       {\footnotesize{$2M_c^m+M_q^m$}} & {\footnotesize{$3642$ }}& {\footnotesize{from mesons}}&{\footnotesize{$2M_c^b+M_q^b$}} &  {\footnotesize{$3784$}} & {\footnotesize{from baryons}}\\
    \hline
    {\footnotesize{$-\Delta E(0)+E_{BO}$}} & {\footnotesize{$+25^{+17}_{-7}$}} & {\footnotesize{BO eq. and eigenv.}} &{\footnotesize{cc binding}} &  {\footnotesize{$-129$}} &  {\footnotesize{from $J/\psi,~\eta_c$}}  \\
    \hline 
  {\footnotesize{partial sum}}& {\footnotesize{${\bf 3667}$}} & --&  {\footnotesize{partial sum}}  & {\footnotesize{${\bf 3655}$}} & --\\
     \hline
{\footnotesize{$1/2 \kappa_{cc}$}} &  {\footnotesize{$-14$}} & {\footnotesize{from $J/\psi,~\eta_c$}} & {\footnotesize{$1/2 \kappa_{cc}$}} & {\footnotesize{$-14.2$}}&   {\footnotesize{from $J/\psi,~\eta_c$}}   \\
 \hline
  {\footnotesize{$-2 \kappa_{qc}$}} &{\footnotesize{$-30$}} &  {\footnotesize{from $\Lambda_c$ mass}}&{\footnotesize{$-2 \kappa_{qc}$}} &  {\footnotesize{$-42.4$}}&  {\footnotesize{fit  to charm. bar. masses}}   \\
  \hline
  {\footnotesize{Total}} & {\footnotesize{${\bf 3651}$}}& --&{\footnotesize{Total}} &  {\footnotesize{${\bf 3627\pm 12}$}} &{\footnotesize{{\bf expt: }${\bf 3621.2\pm 0.7}$}}\\
  \hline
\end{tabular}
 \caption{\footnotesize BO calculation of the doubly-heavy baryon mass, first and second columns,  compared to the quark model prediction, fourth and fifth columns, and to the experimental value reported in the sixth column.}
\label{dhbar}
\end{table}
Given the result
\be
\Delta E(0)=-65~{\rm MeV}\notag
\ee
we find the value of the constant to add to the sum of the constituent masses in \eqref{finalbo}
\be
-\Delta E(0)+E_{BO}=+25^{+17}_{-7}~{\rm MeV}\notag
\ee

The result of our Born-Oppenheimer calculation are reported in Table.~\ref{dhbar} and compared to the quark model prediction~\cite{Karliner:2014gca}. It is remarkable that the estimate of the baryon mass in the two methods, prior to the addition of hyperfine corrections, differ  by $12$~MeV over corrections of the order of $100$~MeV, a $10\%$ accuracy! The further difference between quark model and us depends on the way one estimates $\kappa_{qc}$ and is in line with the constituent quark model precision.

A recent lattice QCD calculation~\cite{Mathur:2018rwu} gives
\be
M(\Xi_{cc})_{{\rm QCD Latt.}}=3634\pm 20~{\rm MeV}
\ee
The experimental value by LHCb~\cite{LHCb:2018pcs} is: 
\be
M(\Xi_{cc})_{{\rm Expt}}=3621.2\pm 0.7~{\rm MeV}
\ee
The consistency of results derived by alternative routes among themselves and with the experimental value is remerkable.

\subsection{Hidden charm tetraquark.}\label{hidch}

We apply the Born-Oppenheimer approximation to the configuration: $(c\bar c)_{\bf 8}~ (q\bar q^\prime)_{\bf 8}$.

\begin{figure}[htb!]
 \begin{center}
   \includegraphics[width=14.0truecm]{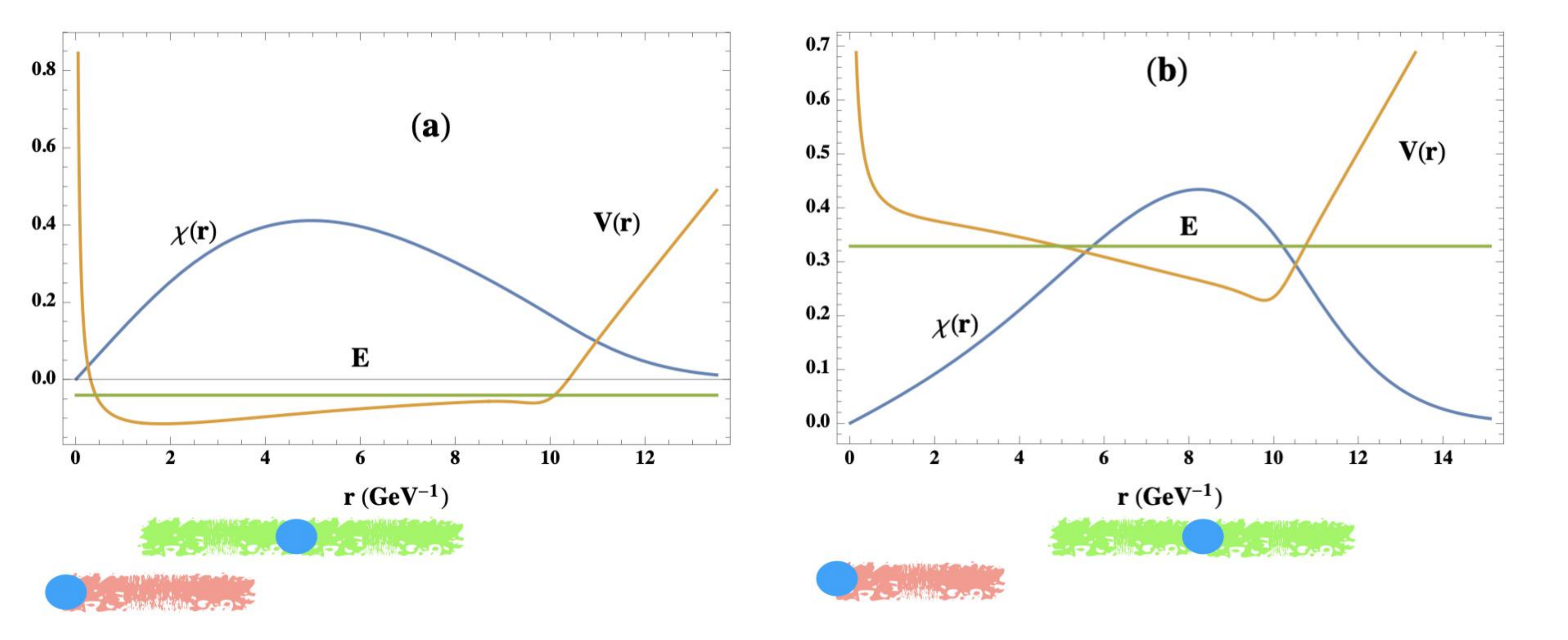}
 \caption{\footnotesize{.}}
\label{hctetra}
\end{center}
\end{figure}

We report in Fig.~\ref{hctetra} the $BO$ potential (yellow line), radial wave function (blue) and  Eigenvalue (green) for two cases.
\begin{itemize}
\item Left panel: $\Delta V(q\bar q^\prime)$ from one-gluon exchange. Diquark and antidiquark orbitals overlap in a configuration similar to the adjoint hadrocharmonium;
\item Right panel: we let $\Delta V(q\bar q^\prime) \to 2~\times$ one gluon exchange. Diquark and antidiquark are well separated (Lect. 5), there is 
a large potential barrier to $c\bar c$ annihilation
\end{itemize} 
The internal configuration depends crucially upon $q\bar q$ repulsion. For large repulsion, we reach configurations similar to those envisaged in Lect. 5. The $cq$ orbital is found to have a radius $R\sim (0.27~ {\rm GeV})^{-1}=0.74~ {\rm fm}$.

\subsection{Double charm or beauty  tetraquark}\label{doubch}

We assume the composition $[(QQ)_{\bf {\bar 3}}(\bar u\bar d)_{\bf 3}],~Q=b,~c$. 
We fix the heavy quark coordinates in $x_1$ and $x_2$ and indicate by $\psi_q$ the orbital $\bar q Q(x_1)$ and $\phi_q$ the orbital  $\bar q Q(x_2)$.
There are two possibilities, namely $\psi_u \phi_d$ and $\psi_d\phi_u$ and the unperturbed state is the superposition of two states with the roles of $\bar u$ and  $\bar d$ interchanged
\be
f_0=\frac{\psi_u\phi_d+\psi_d\phi_u}{\sqrt{2(1+S^2)}}
\ee
with $S$ the superposition integral encountered before, Sect.~\ref{bcc}.

Orbitals are superpositions of color octet and singlet  and their color can be screened by gluons, there is no string joining them and the BO potentail vanishes at infinity.

\begin{figure}[htb!]
 \begin{center}
   \includegraphics[width=14.0truecm]{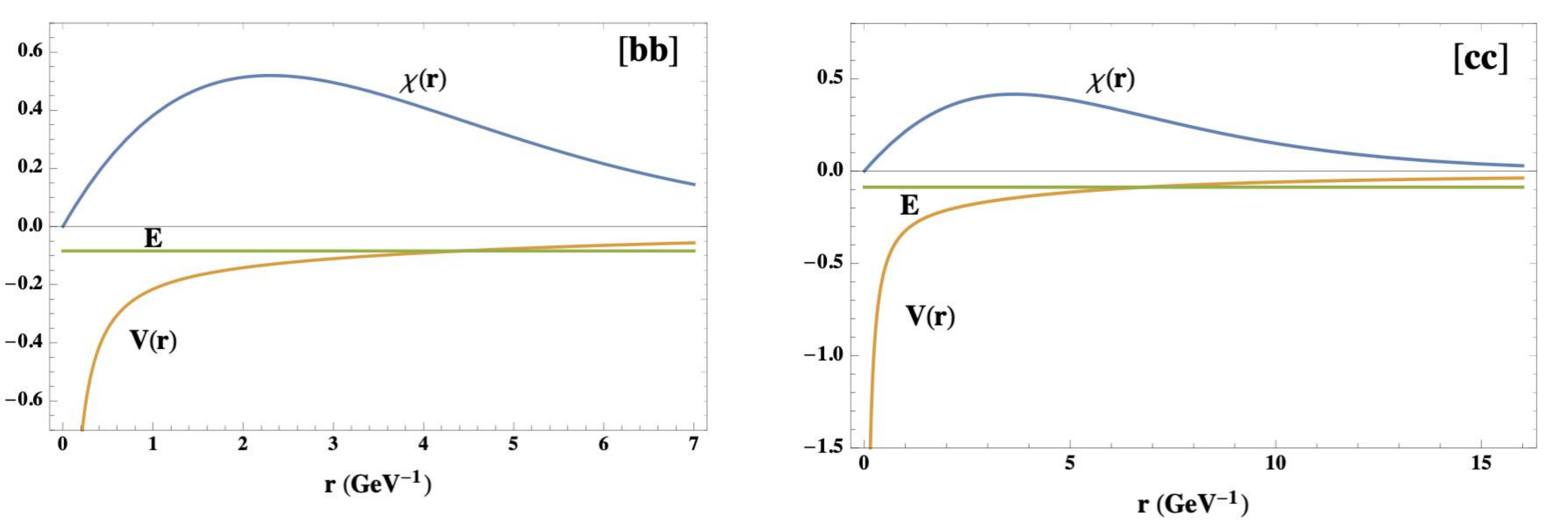}
 \caption{\footnotesize{.}}
\label{dof}
\end{center}
\end{figure}

We report in Fig.~\ref{dof} the $BO$ potential (yellow line), radial wave function (blue) and  Eigenvalue (green) for $T_{bb}$ (left) and $T_{cc}$ (right) tetraquarks. Indicating with $E_{BO}$ the eigenvalue of the $BO$ Schr\"odinger equation and specializing to the double charm case, the predictted $T_{cc}$ mass is given by
\bea
&& M(T_{bb})=2(M_c + M_q) + E_{BO}+\frac{1}{2}\kappa_{cc}-\frac{3}{2}\kappa_{qq} \notag\\
&& M_{PS}(c\bar q)=M_c + M_q -\frac{3}{2}\kappa_{b\bar q}\notag
\eea 
Correspondingly, the  $Q$-value with respect to the threshold of two-pseudoscalar mesons is:
\be
Q_{cc}=M(T_{cc})-2M_{PS}(c\bar q)=E_{BO}+\frac{1}{2}\kappa_{cc}-\frac{3}{2}\kappa_{qq}+3~\kappa_{c\bar q}\label{qval}
\ee
Since $E_{BO}<0$ and $\kappa_{Qq},~\kappa_{QQ}\to 0$ for $M_Q\to \infty$, Eq.~\ref{qval} indicates that the $Q$-value may become negative, for sufficiently large heavy flavour mass $M_Q$, a possibility raised by Karliner and Rosner~\cite{Karliner:2017qjm} and by Eichten and Quigg~\cite{Eichten:2017ffp} and Luo {\it et al.}~\cite{Luo:2017eub}. Preliminary Lattice QCD calculations seem to support this possibility,  within still considerable errors~\cite{Junnarkar:2018twb,Francis:2018jyb,Francis:2016hui,Leskovec:2019ioa}.

The evaluation of \eqref{qval} given in the MPR paper takes $\kappa_{QQ}$, $\kappa_{qq}~{\rm and}~\kappa_{Q\bar q}$ from baryon and meson spectrum (see Lect.~2) and gives (in MeV)
\be
Q_{cc}=+7 (-10);~~Q_{bb}=-138 (-156)
\ee
indicating a mass of $T_{cc}$ close to the $D\bar D$ threshold and $T_{bb}$ well below the $B\bar B$ threshold.
\begin{table}[ht!]
\begin{center}
\begin{tabular}{|c|c|c|c|c|c|}
\hline
{\footnotesize $QQ^\prime\bar u\bar d$}  & {\footnotesize MPR } & {\footnotesize K. and R.} & {\footnotesize E. and Q.}& {\footnotesize L.} & {\footnotesize Lattice QCD}  \\
\hline
{\footnotesize $cc\bar u\bar d $} &  $+7 (-10)$ &  $+140$  & $+102$    & $+39$  & $-23\pm 11$~{\footnotesize{Junn. et al.}}\\
{\footnotesize $cb\bar u\bar d$} &  $-6 (-74) $ &  $\sim 0$  &  $+83$  & $-108$ &  $+8\pm 23$~{\footnotesize {\rm Francis et al.} }\\ \hline 
{\footnotesize $bb\bar u\bar d$}  &  $-138 (-156)$ & $-170$  & $-121$  & $-75$ & 
$\begin{array}{cr}-143\pm 34&{\footnotesize{{\rm Junn.~ et ~al}.}}\\ -143(1)(3) & \footnotesize{{\rm Francis ~et~ al.}}\\
-82\pm 24 \pm 10&\footnotesize{ {\rm Leskovec ~et~ al.}}\\
 \end{array}$\\
\hline
\end{tabular}
\end{center}
\caption{\footnotesize $Q$ values in MeV for decays $T\to  {\rm PS~meson+PS~meson}+\gamma$ obtained with string tension $1/4 \,k$, see MPR and, in parentheses, with string tension  $k$. Models in~Karliner and Rosner, Eichten and Quigg, Luo are different elaborations of the constituent quark model we use throughout this paper, more details are found in the original references. In the last column the lattice QCD results
}
 \label{tabPRD}
\end{table}

However, preparing these lectures, I realised that taking $\kappa_{qq}~{\rm and}~\kappa_{QQ}$  from baryon and Quarkonia masses, as done to obtain the BO result in Tab~\ref{dhbar}, is not correct. As we have seen in Lects. 2 and 3, the hyperfine couplings depend crucially from the  overlap probability of the quark  pair, $|\psi(0)|^2$, which cannot be assumed to be equal, in tetraquarks, to the overlap probability of the same pair in baryons. 

In our Born-Oppenheimer scheme, however, we may estimate deviations from this hypothesis and obtain an improved estimate of the $Q$-values.

Light antiquarks are each bound to a heavy quark in orbitals with wave functions $\psi({\bf \xi })~{\rm and}~\phi({\bf \eta });~{\bf \xi}={\bf x}_{q_1}-{\bf x}_A,~{\bf \eta}={\bf x}_{q_2}-{\bf x}_B$
and the average distance of the light quarks is a function of the heavy quarks distance, $r_{AB}$, given by:
\be
D(r_{AB})= \int d^3\xi d^3\eta~
\frac{|\psi(\xi)|^2\,|\phi(\eta)|^2+\psi(\xi)\phi(\xi)\psi(\eta)\phi(\eta)}{\sqrt{1+S^2(r_{AB)}} }|({\bf \xi}+{\bf x}_A)-({\bf \eta}+{\bf x}_B)| \label{dist}
\ee
The average distance  in the tetraquark  is then:
\be 
D^{(QQ)}_{qq}=\int dr_{AB} ~\chi^2(r_{AB})D(r_{AB}) \notag
\ee
denoting by $\kappa^{(QQ)}_{qq}$, the h.f. coupling of the light quarks $qq$ in $T_{QQ}$, we have
\be
\kappa^{(QQ)}_{qq}=\kappa_{qq}~\Big(\frac{R_{qq}^{B}}{D^{(QQ)}_{qq}} \Big)^3 \notag
\ee
which leads to: 
\be
\kappa_{qq}^{cc}=+2.1~(+4.7);~~\kappa_{qq}^{bb}=+1.2~(+1.9) \notag
\ee
Where the estimates refer to orbitals computed with the string tension scaled from charmonium string tension by the ratio of the Casimir couplings, see Sect. \ref{colorgim} (in parenthesis evaluations with charmonium string tension).

We proceed analogously for the h.f. $QQ$ coupling in the tetraquark, specialising to the charm case and defining 
\be
D^{(cc)}_{cc}=\int dr_{AB} ~\chi^2(r_{AB})~r_{AB}
\ee
We scale with the quarkonium average radius, $R_{c\bar c}$, obtained from the wave function of the Cornell  potential
\be
V_{Cornell}=-\frac{4}{3} \frac{\alpha_s}{r}+ kr;~~(\alpha_s|_{M_c}=0.3,~k=0.15~{\rm GeV}^2) \notag
\ee 
to obtain
\be
\kappa^{(cc)}_{cc}=\kappa_{cc}~\Big(\frac{R_{c\bar c}}{D^{(cc)}_{cc}} \Big)^3  \notag
\ee
with $\kappa_{cc}$ from Tab.~\ref{spin}.

Numerical results are summarised in Tab.~\ref{masnew}. The string tension is taken  from \eqref{potbo}  with $|\lambda_{qQ}|=\frac{1}{3}$ (in parenthesis, values with  $k=1$). Eigenvalues $E_{BO}$ are from~the MPR paper and, following the  notation of Tab.~\ref{tabPRD}, the $Q$-value is taken from the  threshold of 2 -Pseudoscalar mesons: $Q= M_{BO}(QQ\bar q\bar q)- 2 M_{PS}(Q\bar q)$.

\begin{table}[htb!]
\centering
    \begin{tabular}{|c|c|c|c|c|c|c|}
     \hline
Heavy Quark&  $\kappa^{QQ}_{qq}$  MeV&  $\kappa^{QQ}_{QQ}$  MeV& $E $(MeV) & Q-value & Mass BO & Observed \\ \hline
cc &   $+2.1~(+4.7)$ &$+1.2~(1.9)$ & $-70~(-85)$ & $+137~(+119)$& $3872~(3854)$ & $3875$\\ \hline
bb &  $+2.4~(+7.7)$ & $+0.34~(+0.51)$ & $-67~(-85)$ & $-1.8~(-27.5)$ & 10556(10531) & ?? \\  \hline
\end{tabular}
 \caption{\footnotesize {}}
\label{masnew}
\end{table}

In conclusion, the new estimate of $T_{cc}$ mass is close to the observed mass of the double charm meson
\be
M(T_{cc}^{ BO})=3872 (3854) \leftrightarrow {\rm LHCb}: T_{cc}^+(3875)
\ee

A stable Double Beauty tetraquark is still, but only marginally, possible.

\subsection{About the error of BO approximation in QCD}\label{error}

When deriving the Born-Oppeheimer equation for the heavy particles, we considered Eq.~\eqref{twoterms}
\be
\-i{\bf P}\Psi=\frac{\partial}{\partial x_A}\Psi=\frac{\partial \psi}{\partial x_A}f_0+\psi\frac{\partial f_0}{\partial x_A}  \notag
\ee
and we choose  to neglect the second term with respect to the first one.

The ratio of the second (neglected) to the first (retained) is given approximately by
\be
\Lambda=\frac{1/a}{1/b} \notag
\ee 
where $a$ and $b$ are the lengths over which $f$ or $\psi$ show an appreciable variation. 

The length $a$ is simply the radius of the orbitals, which we determine by minimizing the Schr\"odinger functional of the light quark. We find typically $1/a=A\sim0.3$~GeV, {\it i.e.} $a\sim0.7$~fm.

The length $b$ has to be formed from the  dimensional quantities from which the Born-Oppenheimer equation depends. In the case of double heavy baryons and hidden heavy flavor tetraquarks, Sects.~\ref{bcc} and~\ref{hidch}, the BO equation depends upon $M$, $A$ and the string tension $k$, which has dimensions of GeV$^2$.
A quantity $b$ with dimensions of length can be formed as
\be
b=(M k A)^{-1/4} \notag
\ee

Therefore 
\be
\Lambda=A^{3/4}(k M)^{-1/4}\label{err1}
\ee
which is 0.57 for charm and 0.43 for beauty,
using $k=0.15$~GeV$^2$ and the constituent masses of charm and beauty from Tab.~\ref{mas}.
 
 We note in Sect~\ref{doubch} that the Born-Oppenheimer potential for double heavy tetraquarks does not depend on the string tension, which is screened by gluons for colour octet orbitals. In this case
 \be
b=\left(M  A\right)^{-1/2}\notag
\ee
 and 
  \bea
  \Lambda=\left(\frac{A}{M}\right)^{1/2}\label{err2}
\eea
giving 0.42 for charm and 0.24 for beauty.
It is worth noticing that the error we are estimating is the error on binding energies, which turn out to be around $100$~MeV or smaller, in absolute value. So, the errors corresponding to~\eqref{err1} and \eqref{err2} may be in the order of $20-50$~MeV.

%% file: chapts/N_infty.tex

There is no difficulty to write QCD in an arbitrary number, $N$, of colours. In the Seventies, `t-Hooft~\cite{tHooft:1973alw,tHooft:1982ltl} investigated the limit $N\to \infty$ and  found that, to leading order in $N$, correlation functions are dominated by the sum of {\it planar} diagrams and that this limit allows to describe in very simple terms the properties of mesons, the $q\bar q$ bound states. The description of baryons in large $N$ QCD requires a different approach, in that the three-quark description of a baryon requires in an essential way that there are only three colors. The description of baryons for arbitrary $N$ has been proposed and investigated by Witten in an equally seminal paper~\cite{Witten:1979kh}. In this lecture we describe two alternative descriptions of muktiquark fields in the $N\to \infty$ limit.

\subsection{QCD at large N in a nutshell}
We start from the $QCD$ coupling
\be
{\cal L}_{QCD}=g_{QCD}~\bar q\frac{\lambda^a}{2}A^a_\mu\gamma^\mu q
\ee
where $q ~{\rm and}~A$ are quark and gluon fields in color $SU(N)$. 

\begin{figure}[htb!]
 \begin{center}
\begin{minipage}[c]{6.6cm}
   \includegraphics[width=5.6truecm]{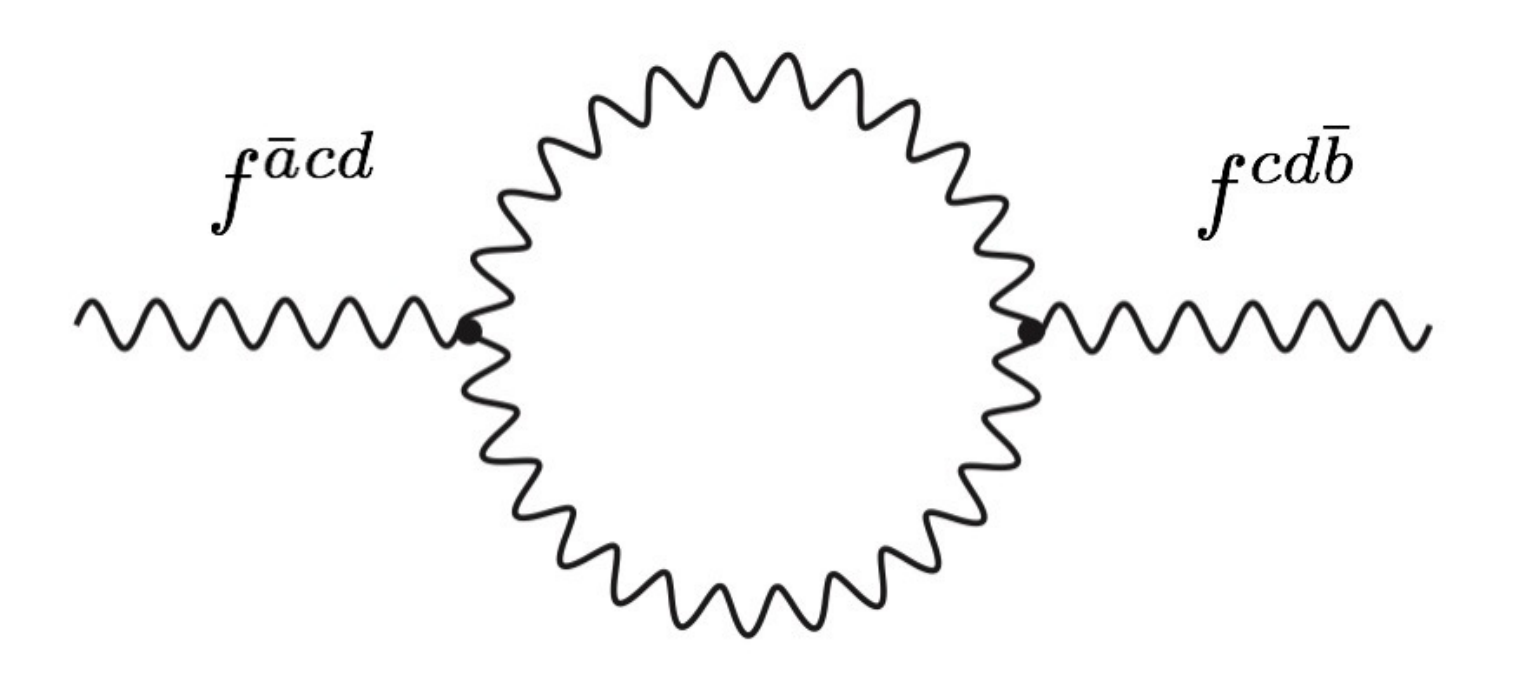}
  \end{minipage}%
\begin{minipage}[c]{6.6cm}
    \includegraphics[width=4.0truecm]{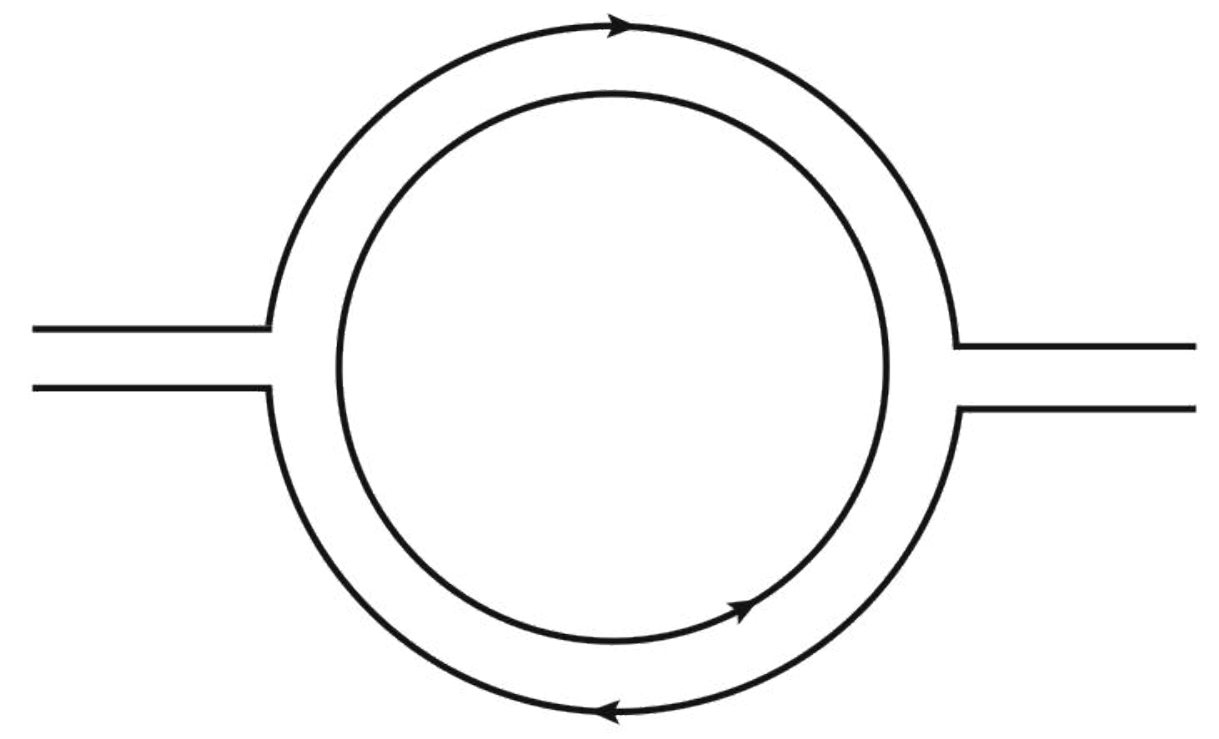}
   \end{minipage}%
 \caption{\footnotesize{ (left); (right) .}}
\label{loop}
\end{center}
\end{figure}

Consider the gluon loop, Fig.~\ref{loop} (left). The $N$ dependence arises from the structure functions, which appear in  the three-gluon coupling, after we sum over color indices ($cd$) that run in the loop, and is made explicit by the formula
\be
A\propto g^2_{QCD}~{\rm Tr}(T^a T^b)=g^2_{QCD}~2N\delta^{ab}; ~T_{ab}^c= i f^{acb}\label{loopamp}
\ee
The amplitude remains finite when we send $N\to \infty$ with $ g^2_{QCD}~N={\rm fixed}=\lambda$. $\lambda$ is often indicated as the 't-Hooft coupling.

Another way to get to the same result is to rewrite the gluon field in terms of the $N \times N$ matrices $(A_\mu)^c_d=(T^a)^c_d A^a_\mu$.
In the large-N limit we may neglect the tracelessness condition that characterizes $SU(N)$ matrices and  treat all $N^2$ components as independent. 
With this parametrization we replace the gluon line by two lines running in opposite directions and carrying the two conjugate color indices, c and d. The loop in Fig.~\ref{loop} (left) is then written as in Fig.~\ref{loop} (right). The factor N in the amplitude is now made evident by the closed circle which is realised in $N$ colors.

\begin{figure}[htb!]
 \begin{center}
   \includegraphics[width=9.0truecm]{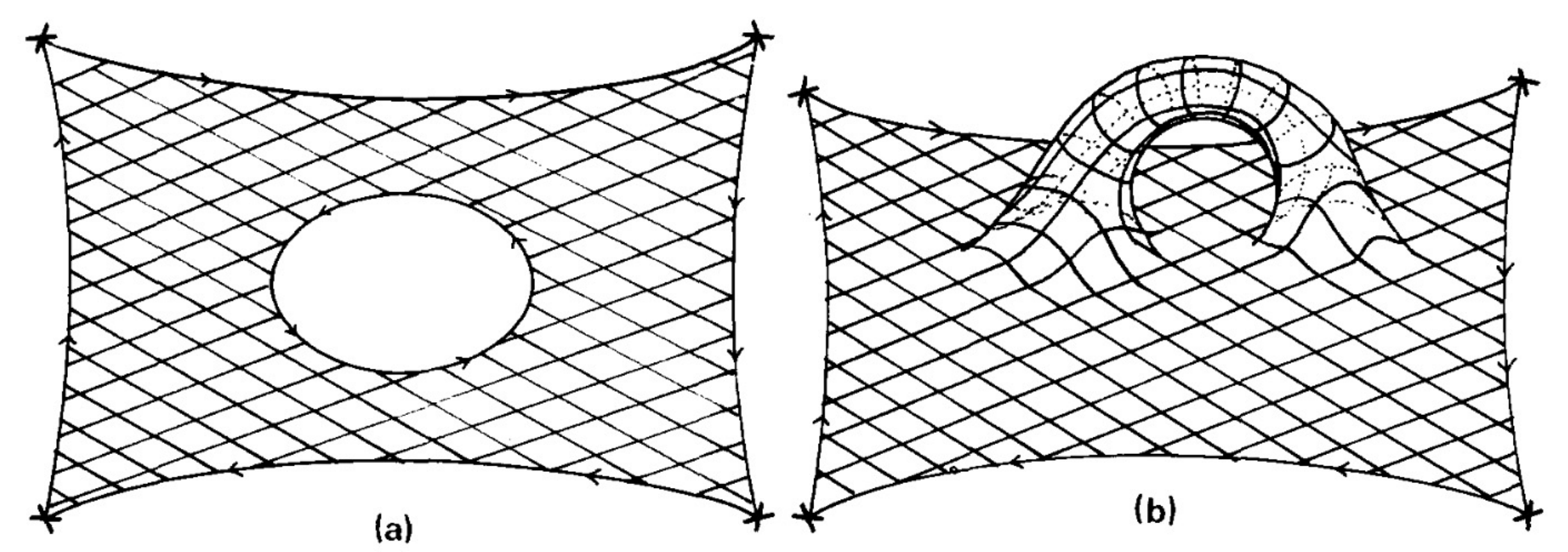}
 \caption{\footnotesize{.}}
\label{topo}
\end{center}
\end{figure}

\begin{figure}[htb!]
 \begin{center}
   \includegraphics[width=12.0truecm]{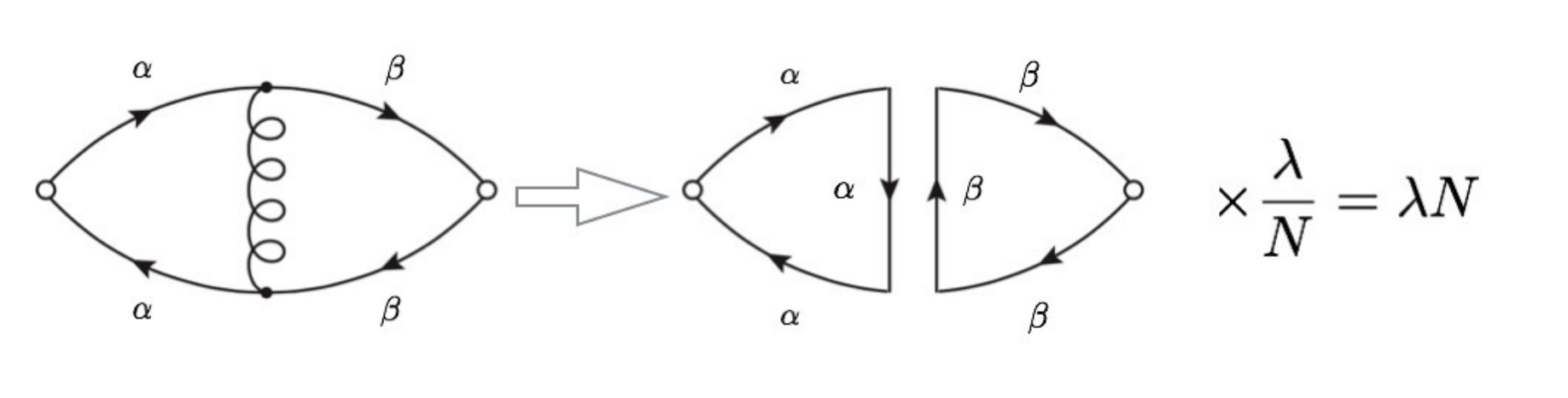}
 \caption{\footnotesize{.}}
\label{loop2}
\end{center}
\end{figure}

We shall be interested in correlation functions with a number of insertions of color singlet quark bilinears, corresponding to external mesons or tetraquarks. If $k$ is the exponent of the leading power of $N$,  't-Hooft has shown the general rule :
\be
k=2- L-H \label{rule}
\ee
where $L$ is the number of fermion loops and $H$ the number of handles, i.e. gluon lines that topologically are not in the same plane as the other lines, see Fig.~\ref{topo} (b). 
For a check, consider the diagram in Fig.~\ref{loop2}. Applying the gluon line writing just introduced to the diagram on the left, we obtain the diagram on the right which has two color loops, hence a factor $N^2$. 
Multiplying by $g_{QCD}^2$, we obtain:
\be
A= N^2 g_{QCD}^2= N^2\frac{\lambda}{N}=\lambda~N, ~{\rm i.e.} ~k=1
\ee
\begin{figure}[htb!]
 \begin{center}
   \includegraphics[width=6.0truecm]{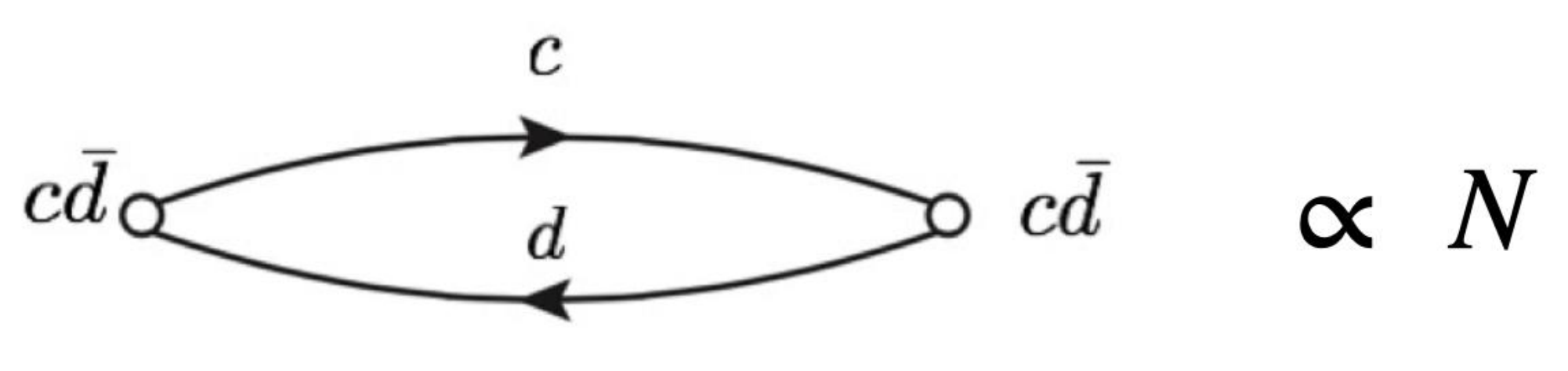}
 \caption{\footnotesize{.}}
\label{oneloop}
\end{center}
\end{figure}
which  agrees with 't-Hooft rule \eqref{rule} since the diagram in Fig.~\ref{loop2}(left) has $L=1, H=0$.

\emph{\bf{$(q\bar q)$ mesons : interactions and scattering.}}
External mesons are represented by the insertion color singlet $(q\bar q)$ operators in one fermion loop. 
The meson propagator is represented in Fig.~\ref{oneloop}, to be understood as representing the sum of all diagrams with one external fermion loop, dressed  with lines of interacting gluons in a {\it planar} topology (as illustrated in Fig.~\ref{topo}(a)).
To normalize the amplitude, we introduce one factor $1/\sqrt{N}$ for each insertion, so as to give a finite amplitude in the limit $N\to\infty$.

\begin{figure}[htb!]
 \begin{center}
\begin{minipage}[c]{6.6cm}
   \includegraphics[width=5.06truecm]{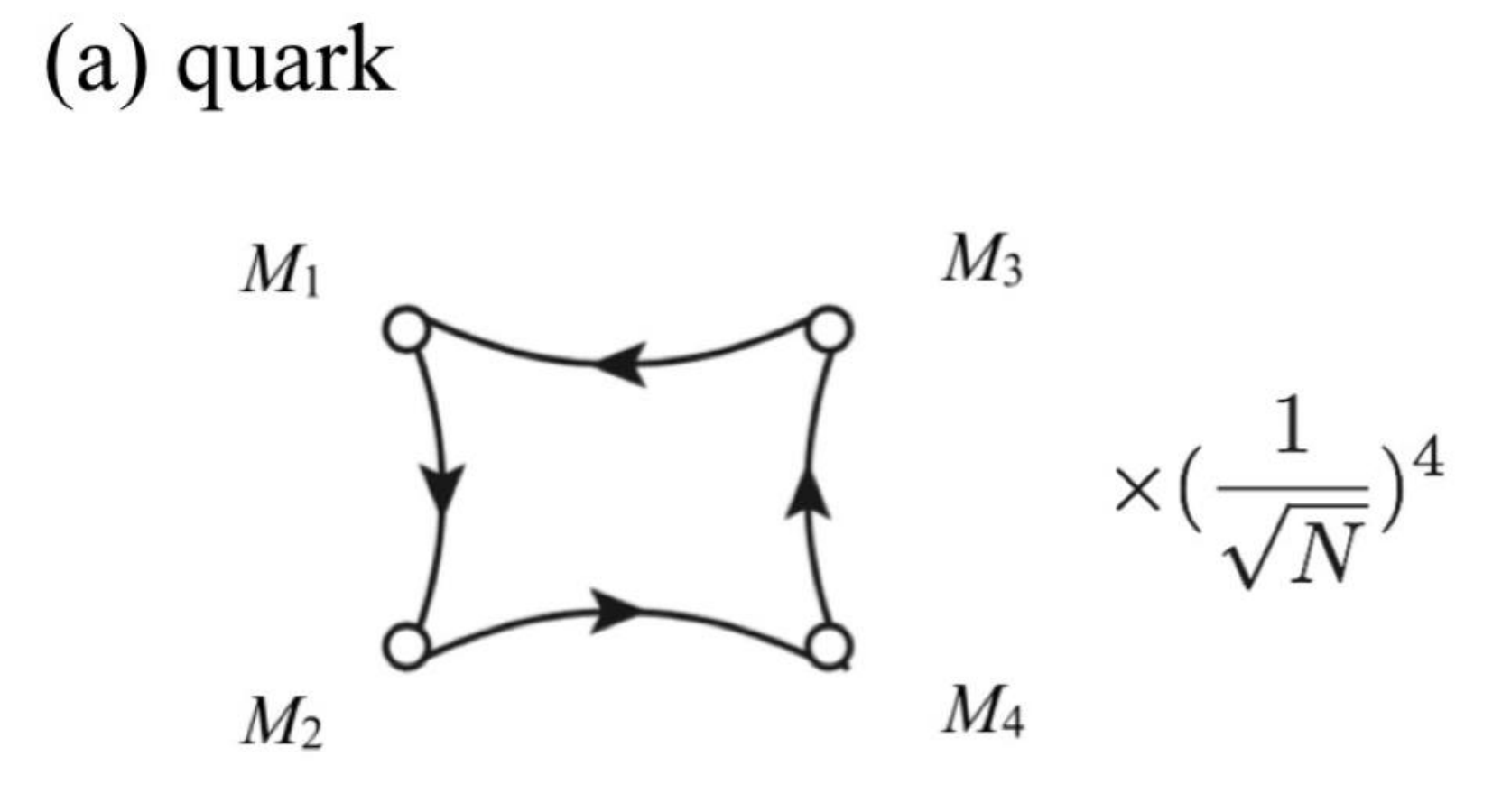}
  \end{minipage}%
\begin{minipage}[c]{6.6cm}
    \includegraphics[width=8.0truecm]{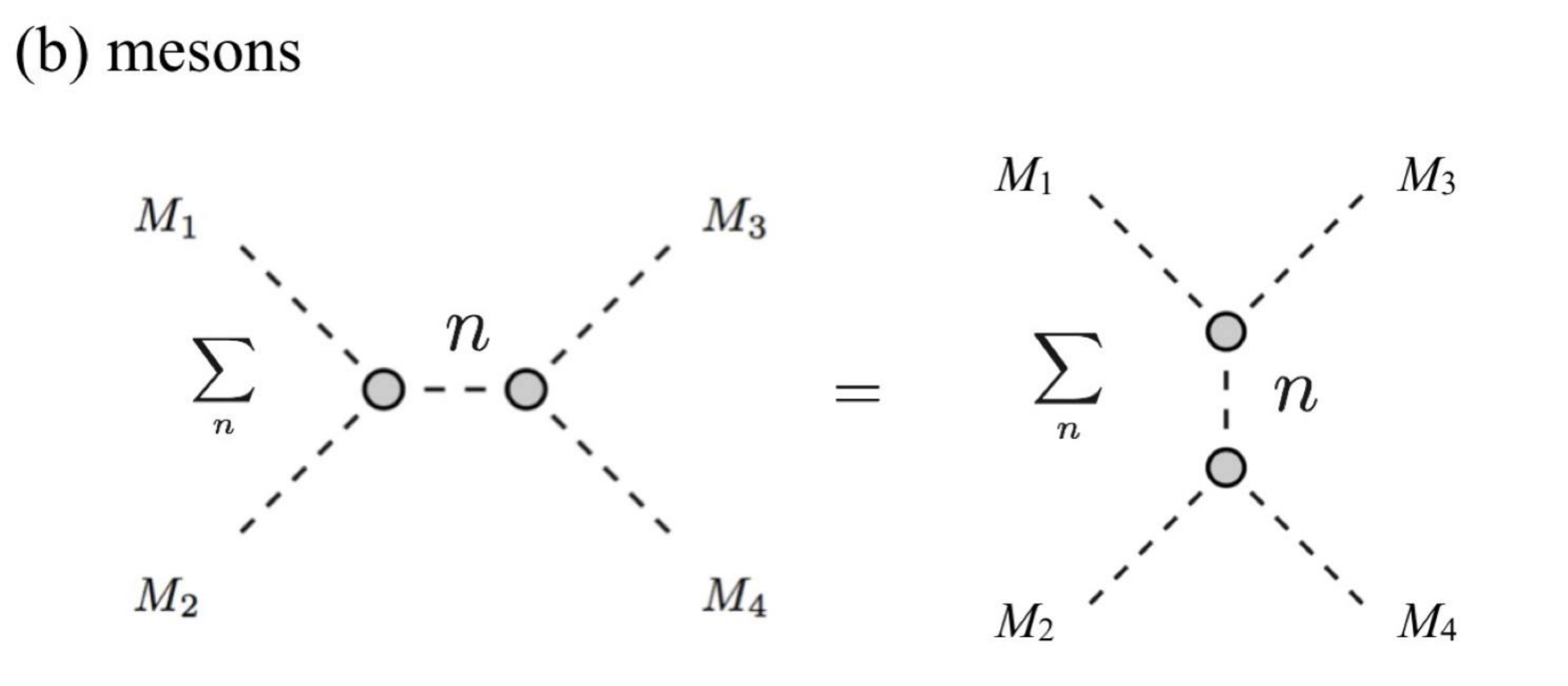}
   \end{minipage}%
 \caption{\footnotesize{ (left); (right) .}}
\label{mesint}
\end{center}
\end{figure}

With the same convention, the meson-meson scattering  amplitude, Fig.~\ref{mesint}(left), is 
\be
{\cal A}(M_1+M_2 \to M_3+M_4)\propto \frac{1}{N}
\ee
Cuts in the s-channel or in the t-channel give poles corresponding to stable mesons. Since the amplitudes at $q^2\to \infty$  increases logarithmically, there must be an infinite number of poles, corresponding to a tower of $(q\bar q)$ states with given flavour.

It is important to note that  in the leading $1/N$ expansion, {\it there is only one diagram} that represents at the same time $s-$ channel and $t$-channel poles. The equality in Fig.~\ref{mesint} (right) corresponds to the Dolen-Horn-Schmidt duality~\cite{Dolen:1967zz,Dolen:1967jr}: the sum of resonant amplitudes in $s$-channel reproduces the sum of  the poles exchanged in the $t$-channel. The planar approximation gives indeed a picture close to the Veneziano amplitude~\cite{Veneziano:1968yb}  (an illuminating discussion of Dolen-Horn-Schimdt duality, Veneziano model,  dual and string models is found in~\cite{Green:1987sp}).

In a way, we can say that each $s$-channel pole, e.g. the $\rho$ meson, is built by the forces represented by the exchange of $t$-channel, color singlet, resonances. The old-time bootstrap idea of Chew and Frautschi: ${\bf resonances=binding~forces}$, finds its realization in the leading order of the $1/N$ expansion.

Can we then  say that the $\rho$ meson is a $\pi-\pi$ hadron molecule? Well, yes and no. 

Yes, because the description of the $\rho$ meson as due to color singlet binding forces is correct. No, because the $\rho$ appears only if we allow, in an essential way, an {\it infinite number of exchanges}. Fur sure, not when we limit the exchange to the $\rho$ meson itself and few others, lowest lying color singlet resonances, as was done in the old-time bootstrap and as usually implied by the locution {\it hadron molecule}.

A pure dual model description of hadrons fails to account for deep inelastic processes, that reveal the point-like nature of constituents. To describe mesons and baryons, quarks are definitely better. 

\subsection{Tetraquarks in the large $N$ expansion}

\begin{figure}[htb!]
 \begin{center}
   \includegraphics[width=6.0truecm]{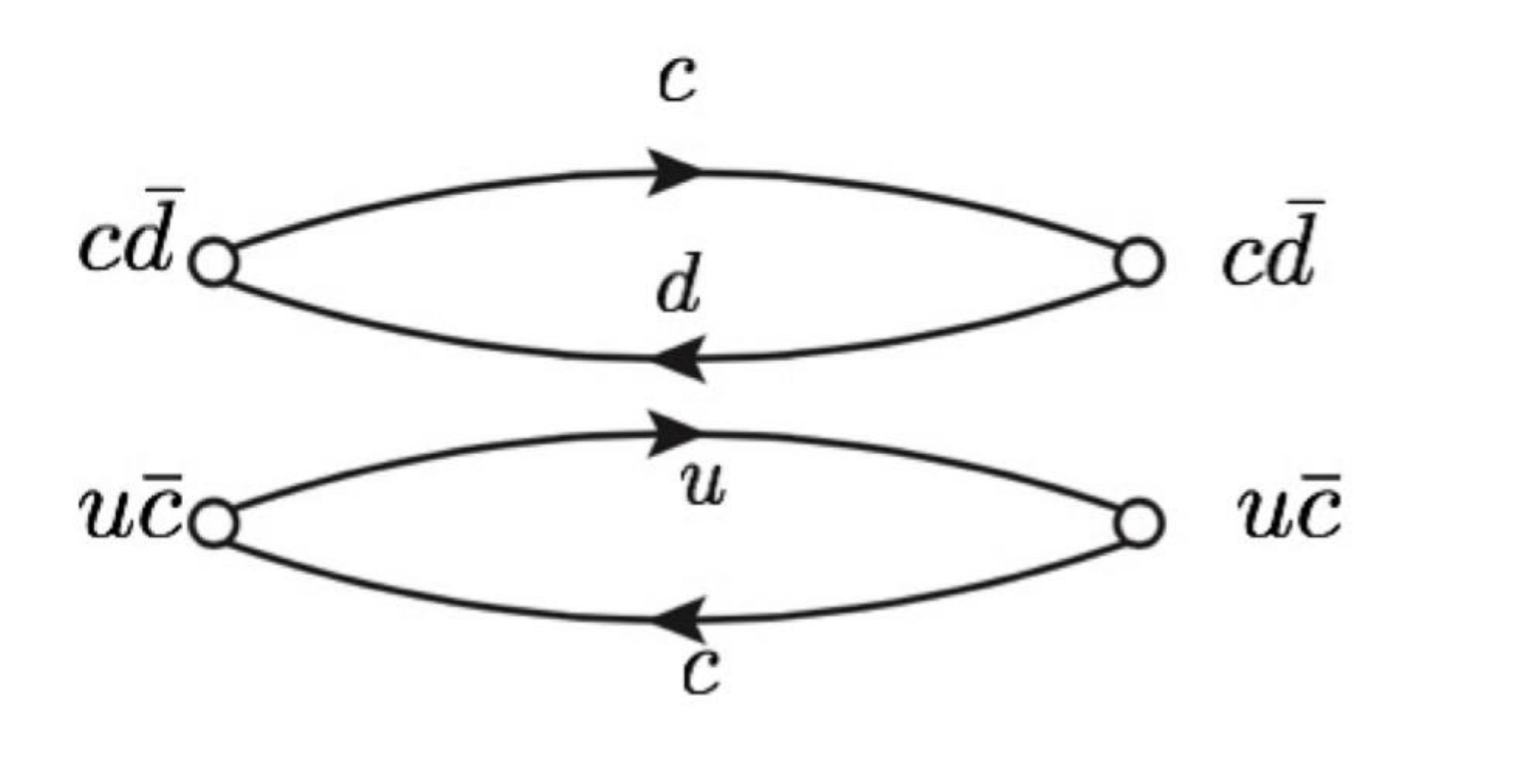}
 \caption{\footnotesize{.}}
\label{2loop}
\end{center}
\end{figure}

Respectability of tetraquarks was somehow tarnished by a theorem of S.~Coleman~\cite{Coleman:1985rnk}. The theorem is based on the fact that, by Fierz rearrangements, tetraquark operators can be reduced to a sum of products of color singlet bilinears. If so, the leading term in the $1/N$ expansion will be given by the disconnected diagrams of Fig.~\ref{2loop}, of order $N^2$. In Coleman's words:  {\it tetraquark operators, to leading order, create out of vacuum only pairs of mesons}.

The argument was reexamined by S. Weinberg~\cite{Weinberg:2013cfa} who argues that if the connected tetraquark correlator develops a pole, it will be irrelevant that the residue of the pole is of order 1/N with respect to the disconnected part: at the pole the connected part will dominate anyhow.

The real issue is the width of the tetraquark state: it may increase for large N, to the point of making the state unobservable.
Weinberg's conclusions was that the decay rate goes like $1/N$, making tetraquarks a respectable possibility.
Weinberg's discussion has been enlarged by Knecht and Peris~\cite{Knecht:2013yqa} and further considered by Cohen and Lebed~\cite{Cohen:2014via}.

\begin{figure}[htb!]
 \begin{center}
   \includegraphics[width=12.0truecm]{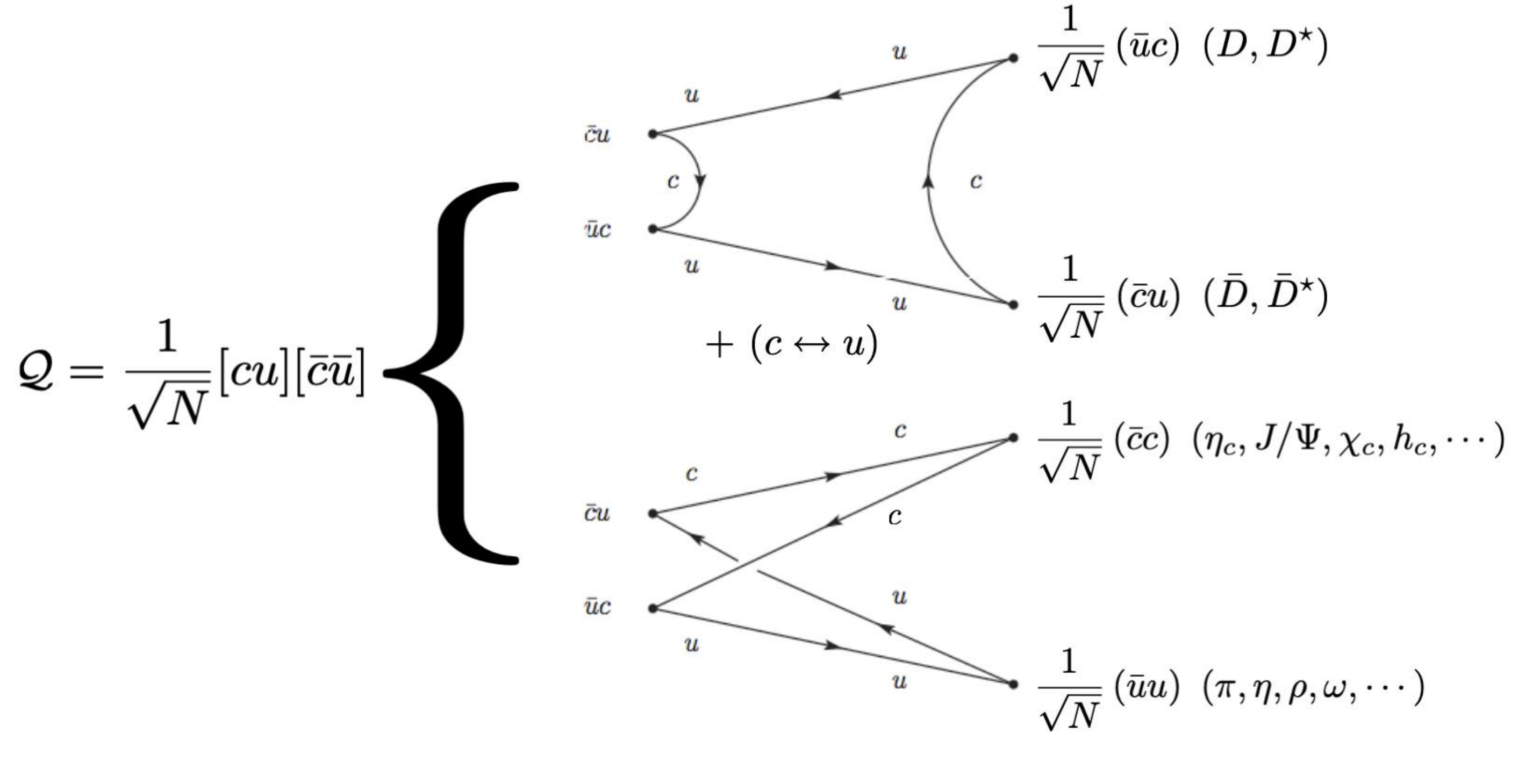}
 \caption{\footnotesize{.}}
\label{t1loop}
\end{center}
\end{figure}

\emph{\bf{Weinberg's result.}} Connected, one loop amplitudes with insertions of quark color singlet operators give a factor N, Fig~\ref{t1loop}. Interpolating field operators have to be multiplied by powers of N, such as to make the connected two-point correlators to be normalized to unity. Correspondingly, Weinberg finds decay amplitudes of order:
\be
{\cal A}_{conn.} \propto N \frac{1}{N\sqrt{N}}= \frac{1}{\sqrt{N}}
\ee
Rates do not diverge with $N$, tetraquarks are observable!


\emph{\bf{Beyond planar diagrams.}}
Typical connected diagrams of order $N$ show a 4 quarks cut. But~\cite{Cohen:2014tga,Maiani:2016hxw}: are these free or interacting quarks? The same question arises if we fill the quark loop by a multigluon, planar diagram {\it \'a la} 't-Hooft: do planar interactions resolve the problem?
\begin{figure}[htb!]
 \begin{center}
\begin{minipage}[c]{8.6cm}
   \includegraphics[width=8.0truecm]{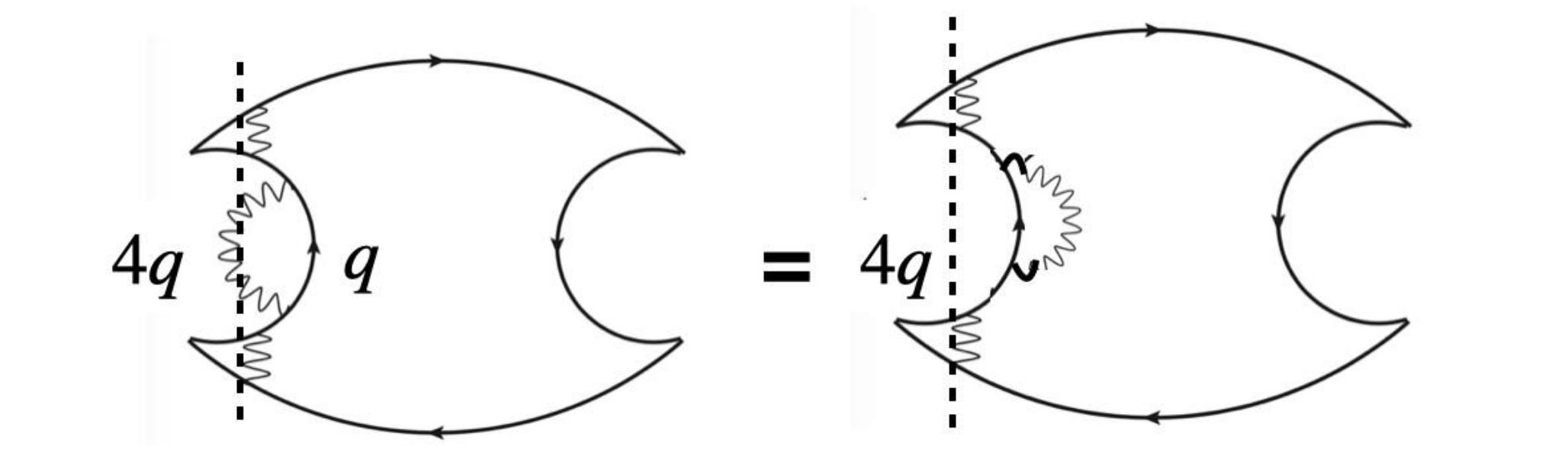}
  \end{minipage}
\begin{minipage}[c]{6.6cm}
    \includegraphics[width=7.0truecm]{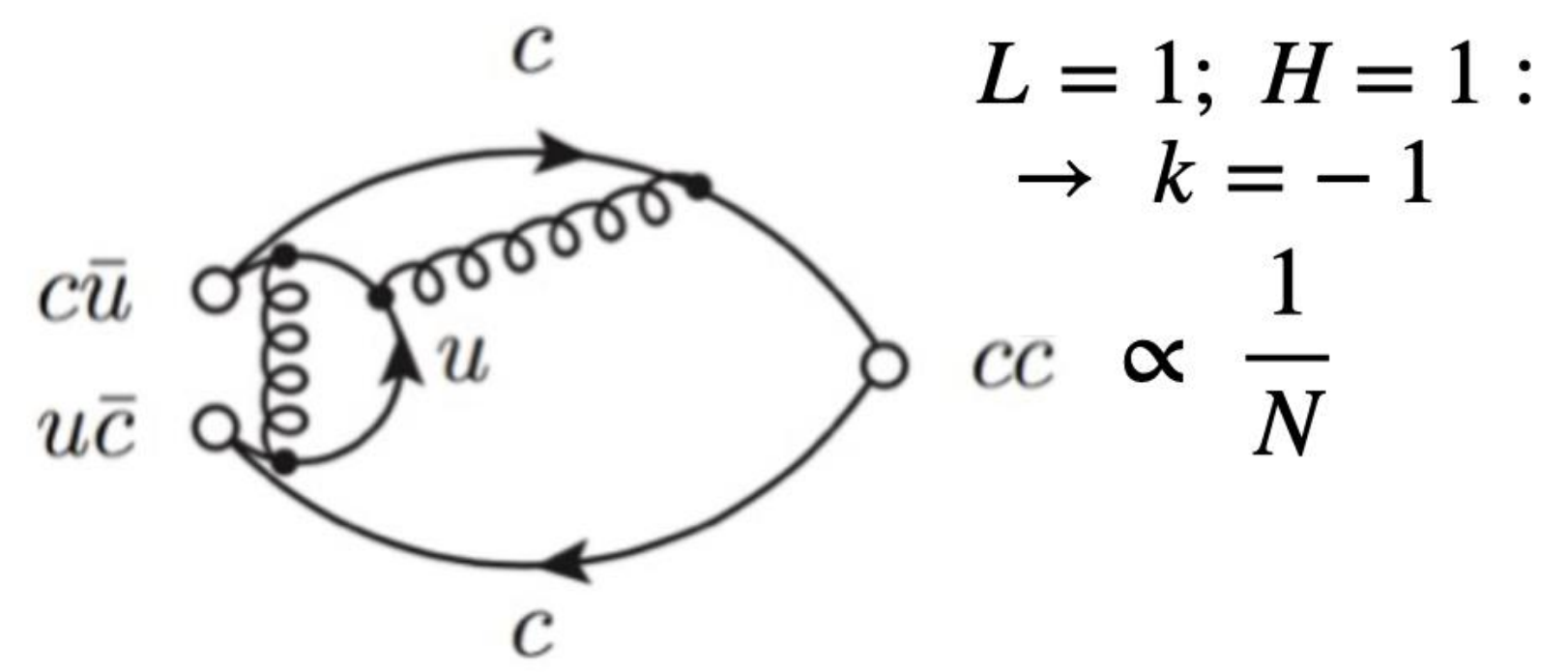}
   \end{minipage}%
 \caption{\footnotesize{ (left); (right) .}}
\label{4qcut}
\end{center}
\end{figure}

Consider Fig.~\ref{4qcut} (left). In a planar diagram all lines must be inside the quark loop. Then, to keep the result of order $N$, it must be possible to bring the gluon inside the  fermion loop as indicated by the equality in the Figure. But then, the four-quark  cut indicates that we are in presence of {\it{\bf {two non-interacting mesons}}}. It is difficult to imagine this kind of diagrams to be able to bind the mesons into a genuine tetraquark.

An internal gluon as on Fig.~\ref{4qcut}(right) would forbid moving the gluon inside the loop and it would leave us with a genuine interaction of the two quark pairs identified by the cut. The gluon external to the loop in Fig.~\ref{4qcut} (right), generates a topological handle, as indicated by the fact that the diagram is of order  $N^{-1}$: $L=1,~H=1\to k=-1$. 

\emph{\bf{Tetraquark diagrams are next-to next-to leading}}

\begin{figure}[htb!]
 \begin{center}
   \includegraphics[width=16.0truecm]{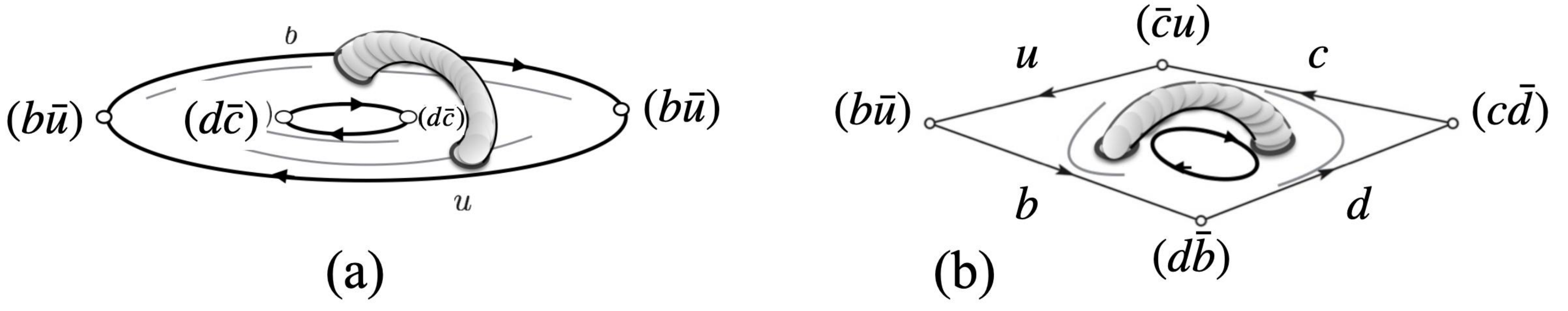}
 \caption{\footnotesize{.}}
\label{oursol}
\end{center}
\end{figure}

Investigations by several groups have addressed  the issue of the order in the $1/N$ expansion where bound tetraquark poles may appear. The question is still open. I will summarise here the conclusions of our group~\cite{Maiani:2018pef} illustrated in Fig.~\ref{oursol}.

Only one diagram is relevant, with meson insertions distributed along the two fermion loops, Fig.~\ref{oursol}.
\begin{itemize} 
\item for given flavours, color ${\bf{\bar 3}}$ tetraquarks suffice to obtain a consistent solution, in line with the fact that $(qq)_{\bf{\bar 3}}$ is attractive and $(qq)_{\bf6}$  is repulsive;
\item decay amplitudes are as follows:
\bea
&&{\cal A}(T\to M_1+M_2)\propto \frac{1}{N^2}\notag \\
&&{\cal A}(T^*\to T+M)\propto \frac{1}{\sqrt{N}} \notag\\
&&{\cal A}(T^*\to T+\gamma)\propto N^0; \notag \\
\eea
\item the decays:  $Y\to Z+\pi~{\rm and}~ Y\to X+\gamma$ are allowed transitions;
\item neutral hidden-charm tetraquarks mix with charmonia to order: $1/(N\sqrt{N})$; 
\item $Y$ states may be produce by and annihilate into $e^+ e^-$ via mixing.
\end{itemize}

\subsection{The alternative scheme, descending from Witten's baryons}

As we have just seen, tetraquarks of composition $[qq^\prime][\bar q^{\prime\prime}\bar q^{\prime\prime\prime}]$ can be generalized to N colours.
  
 Pentaquarks, however, are related to baryons, which in QCD with general $N$ require a completely different toolkit, since the three quark configuration in the baryon generalizes to the fully antisymmetric combination:
 \be
 B=q_{[1}q_2 . . . q_{N]}\label{baryon}
 \ee
first studied by Witten. 
 
The  {\it baryonium scheme}, a formulation {\it \'a la} Witten to generalize tetraquarks to any N, has been explored by Rossi and Veneziano in 1977 and reconsidered in 2016~\cite{Rossi:1977cy,Montanet:1980te,Rossi:2016szw}. Tetraquarks are defined by generalising  the antisymmetric quark and antiquark combinations:
\be
T=\epsilon_{a a_1\dots a_{N-1}}(q^{a_1}\dots q^{a_{N-1}})~\epsilon^{a b_1\dots b_{N-1}}(\bar q_{b_1}\dots \bar q_{b_{N-1}})
\ee
the generalised diquark $q^{[1}q^2 . . . q^{{N-1]} }$ transforms as an antiquark and can generalise the construction used in $N=3$ for pentaquarks and dibaryons.

\begin{figure}[htb!]
 \begin{center}
   \includegraphics[width=10.0truecm]{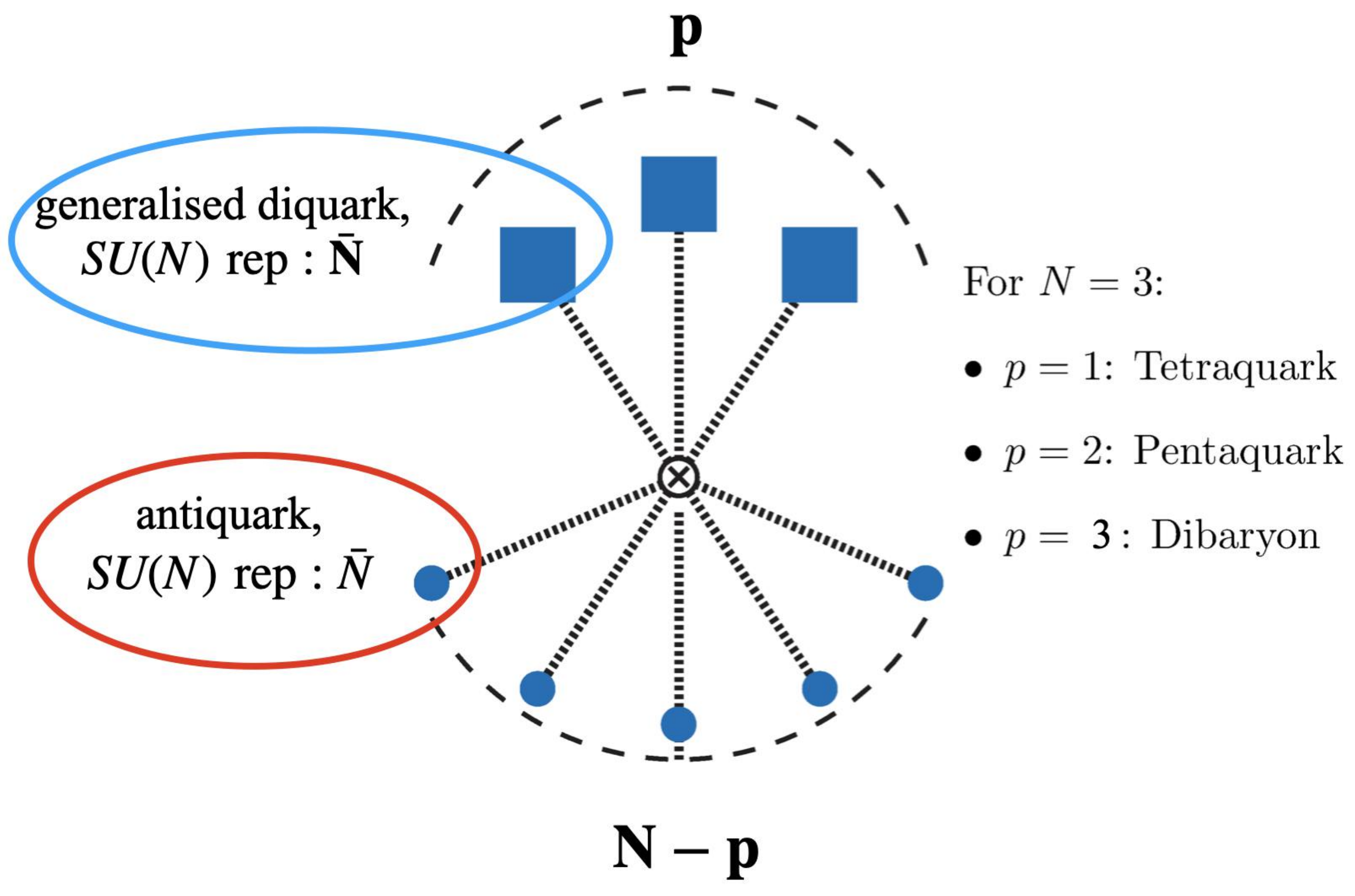}
 \caption{\footnotesize{.}}
\label{witt}
\end{center}
\end{figure}

 In a world of two colours, the new structures disappear:  $N=2$ QCD is made only of mesons, $\bar q q$, "baryons", $qq$, and molecules thereof. New  spectroscopic series start to appear at $N=3$, (our world!) and  can be  extended to $N$  colours~\cite{Maiani:2018tfe},  as shown in Fig.~\ref{witt}. 
 The generic multiquark structure is  characterized by an integer $0\leq p\leq N$ and consists of 
 \begin{itemize}
 \item $N-p$ antiquarks, each in the representation ${\bar{\bf N}}$ of $SU(N)$; 
 \item $p$  generalised diquarks, each in ${\bar{\bf N}}$ of $SU(N)$
\item  in all we have $N$ fundamental representations of $SU(N)$, which can be joined in a color singlet by the $\epsilon$ symbol in $N$ dimensions. 
 \end{itemize}
 For $N=3$: $p=0$, antibaryon; $p=1$, tetraquark; $p=2$, pentaquark; $p=3$, dibaryon. 
 
 By analogy, for generic $N$ we adopt the terminology: $p=0$, antibaryon; $p$=1, Rossi-Veneziano tetraquark; $2\leq p\leq N-1$, generalised pentaquarks, $p=N$, generalised dibaryons.

\emph{\bf{Baryon: structure and couplings to mesons.}} Witten assumes that each quark lives in Harthree-Fock wavefunctions $\phi_r(x)$, with $r=0, 1, ...$ ground and excited states. Given full color antisymmetry embodied in \eqref{baryon},  the space wave function of the baryon is  the symmetric wave function of N bosons.
\bea
&&{\bf Ground~state};\notag \\
&&\Psi_0^{1~2~\dots N}(x_1,x_2\dots x_N)=\epsilon^{1~2~\dots~N}\phi_0(x_1)\phi_0(x_2)\dots \phi_0(x_N)\label{ground}\\
&&{\bf Excited~state};\notag \\
&&\Psi_r^{1~2~\dots N}(x_1,x_2\dots x_N)=\epsilon^{1~2~\dots~N}\times\frac{1}{\sqrt{N}}\times \notag \\
&&\times [\phi_r(x_1)\phi_0(x_2)\dots \phi_0(x_N)+\phi_0(x_1)\phi_r(x_2)\dots \phi_0(x_N)+\dots +\phi_0(x_1)\phi_0(x_2)\dots \phi_r(x_N)]. \label{excd}
\eea
 
 We denote by $n_r$ the occupation number of $\phi_r(x)$. The energy of a generic baryon state is then
 \be
 M=N M_q+\sum_r n_r \epsilon_r;~~\sum_r n_r=N \notag
 \ee
 with $M_q$ the quark mass and $\epsilon_r$ the energies of the excited states, both assumed to be independent from $N$. Meson states are assumed to be normalised as
 \be
M=\frac{1}{\sqrt{N}} \sum_\ell \bar q_\ell q^\ell \label{meson}
\ee
\begin{figure}[htb!]
 \begin{center}
   \includegraphics[width=5.0truecm]{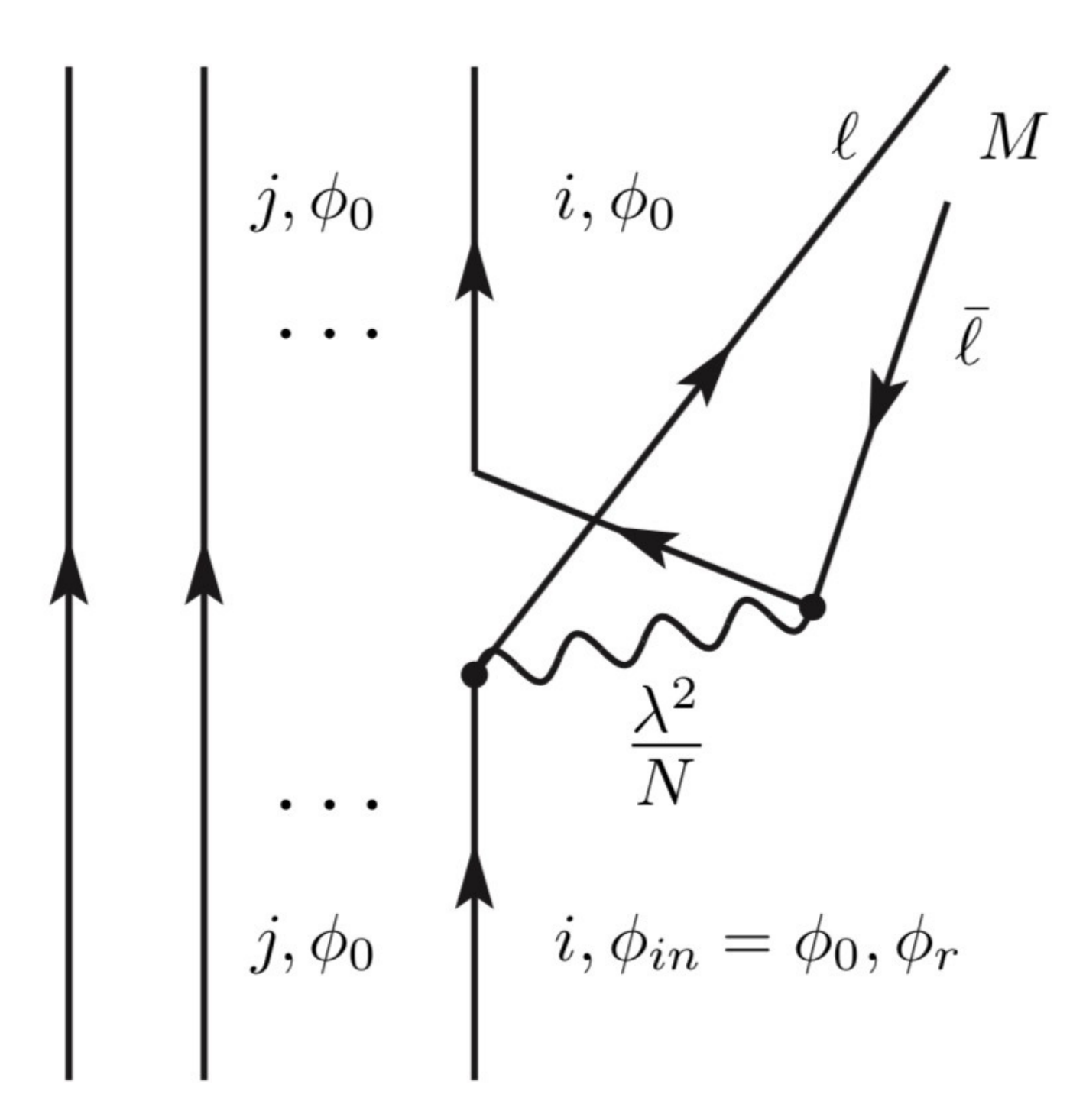}
 \caption{\footnotesize{.}}
\label{barexc}
\end{center}
\end{figure}
 The meson-baryon trilinear coupling is represented in Fig.~\ref{barexc}. The initial quark wave function is indicated by $\phi_{in}$, the final quark is in the ground state $\phi_0$. With $\phi_{in}=\phi_0$ or $\phi_r$, we obtain the ground state meson-baryon coupling, e.g. $g_{N\bar N \pi}$, or the transition amplitude of the excited state, e.g. $A(\Delta \to N \pi)$. 

The basic transition occurs via one gluon exchange:
\be q^i(\phi_{in})\to q^i (\phi_0)+q^\ell +{\bar q}^\ell~( \ell=1\dots N) \label{g-rad}
\ee 
Projecting over the colour singlet meson state
one finds the effective operator for the baryon to baryon transition
\be
{\it O}=\frac{\lambda^2}{N} \sqrt{N}~\bar q  {\cal O}(x) q. \notag
\ee
In the above, the ${\cal O}(x)$ is a $N$-independent operator acting on the single quark wave functions $\phi(x)$ and connecting $\phi_{in}$ to $\phi_0$.

The transition operator applied to  (\ref{ground}) gives $N$ equal terms and we obtain (see Witten 1979):
\be
g_{B\bar B M}\sim \frac{\lambda^2}{N} \sqrt{N} ~N\propto \sqrt{N}.\label{barcoup}
\ee

When applied to (\ref{excd}), the transition operator has to operate on $\phi_r$ only, to obtain a non-vanishing result when the scalar product with (\ref{ground}) is taken. We obtain $N$ equal factors, divided by the normalisation $\sqrt{N}$, so that
\be
A(B^* \to B+M)\sim \frac{\lambda^2}{N} \sqrt{N}~\frac{N}{ \sqrt{N}} \propto N^0\label{bardec}.
\ee

As noted by Witten, the tree-level meson-baryon low energy scattering amplitude obtained from (\ref{barcoup}) is ${\cal O}(1)$ since the baryon's  propagator brings in a factor $N^{-1}$ due to the baryon's mass.

\emph{\bf{Tetraquark decays: $ T\to B\bar B$ and $T^*\to T+M$.}}

We describe the tetraquark by the operator
\be
T=\frac{1}{\sqrt{N}} B_a\bar B^a\label{normatetra}, \notag
\ee
where $B_a=\frac{\partial}{\partial q^a }B$, is the operator $B$, \eqref{baryon}, with $q^a$ suppressed, and similarly for $\bar B^a$. Sum over $a=1,\cdots N$ is understood. 
\begin{figure}[htb!]
 \begin{center}
\begin{minipage}[c]{6.6cm}
   \includegraphics[width=5.60truecm]{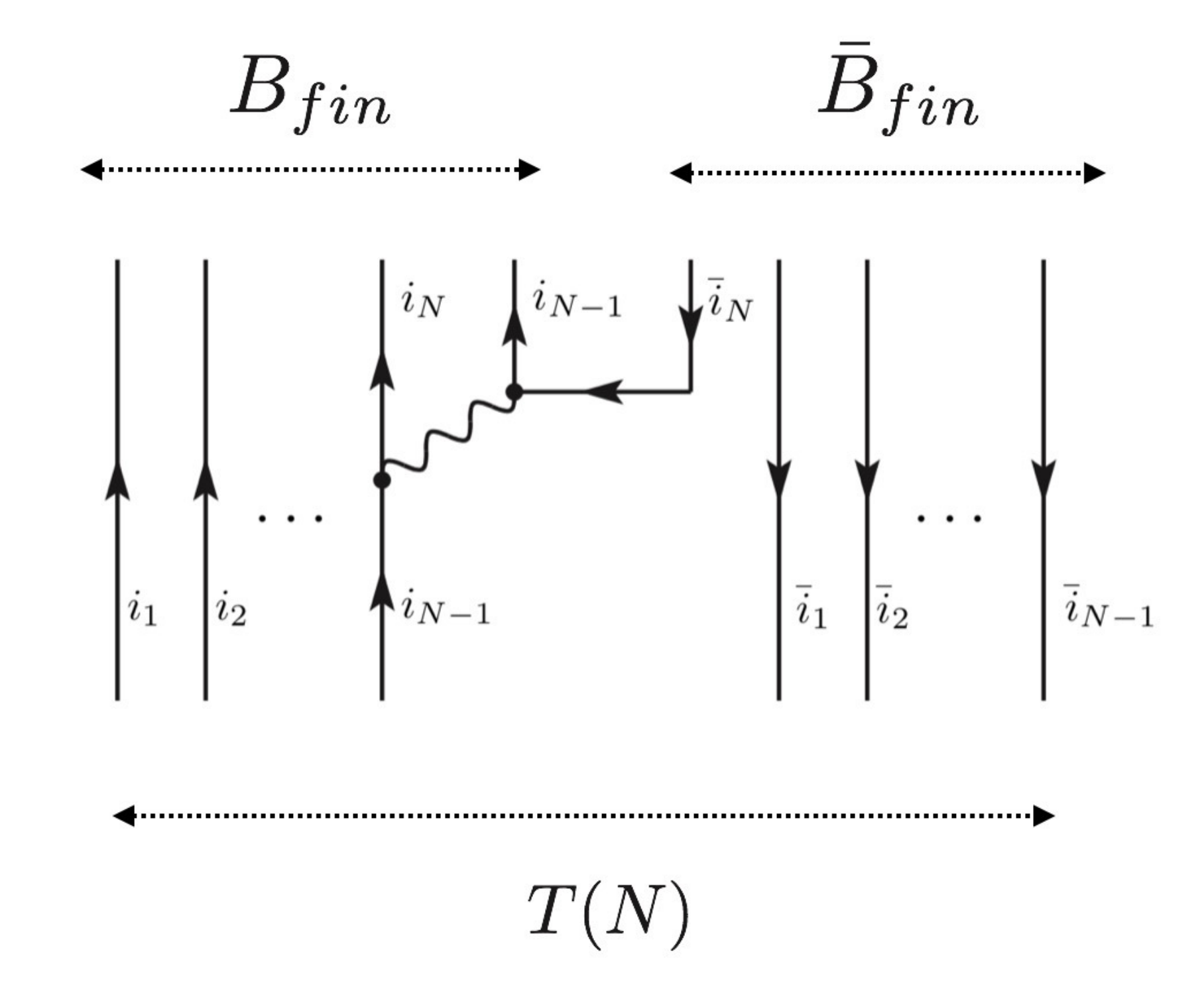}
  \end{minipage}
\begin{minipage}[c]{6.6cm}
    \includegraphics[width=6.2truecm]{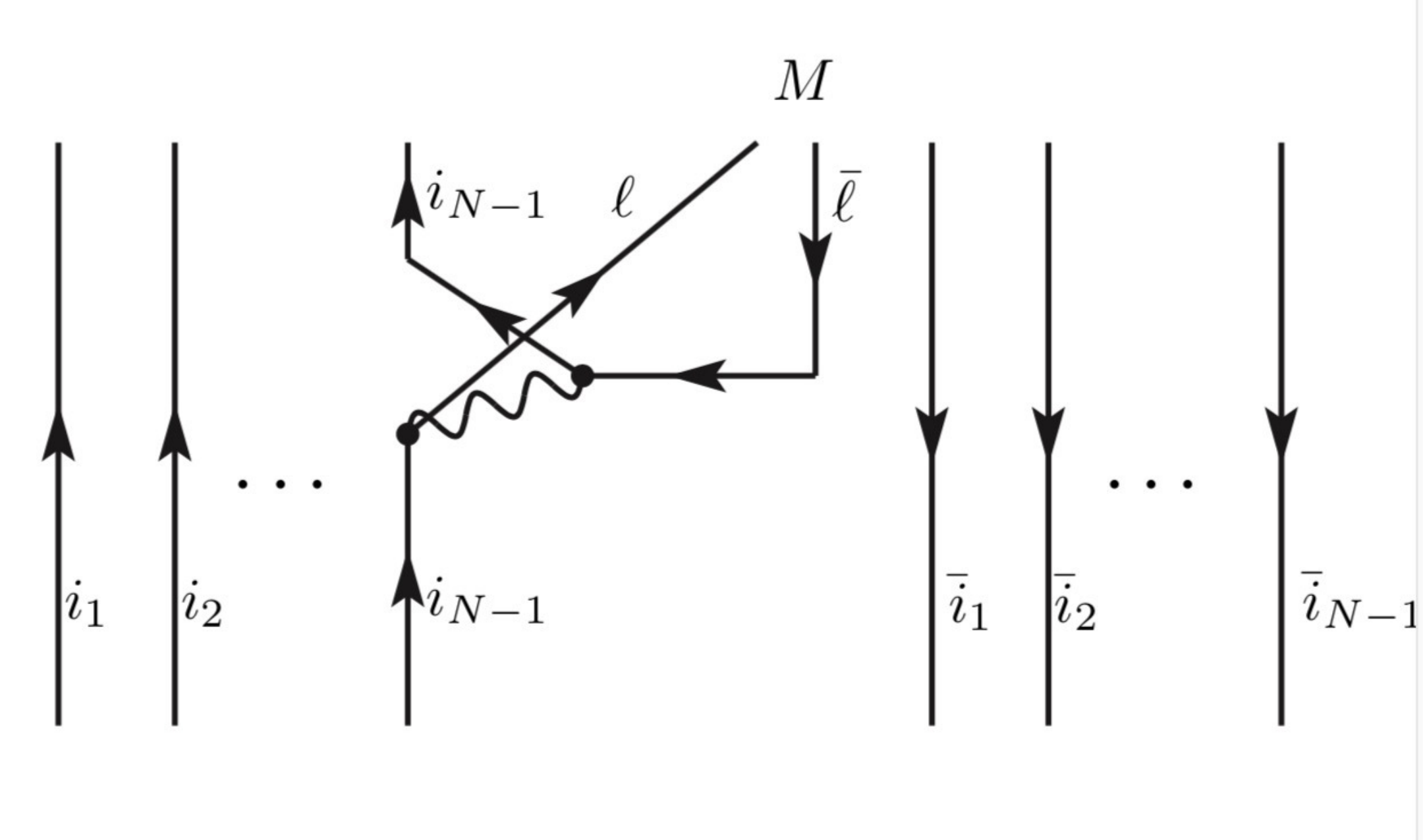}
   \end{minipage}%
 \caption{\footnotesize{ (left); (right) .}}
\label{tetradecs}
\end{center}
\end{figure}
  Note the identity
\be
q^a B_a=q^a\frac{\partial}{\partial q^a} B= B, {\rm for~any}~a \notag
\label{signfix}
\ee 
since $q^a\partial/\partial q^a$ is a bosonic operator.

We report in Fig.~\ref{tetradecs} the diagrams for the tetraquark decay in a baryon pair (left panel) and for the decay of an excited tetraquark state in the ground state plus a meson (right panel).

Amplitudes are obtained along similar lines to the baryon couplings. We refer to the original paper for details and report the results of calculation
\begin{itemize}
\item For $B\bar B$ decay:
 \be
 A(T\to B+ \bar B)=\frac{1}{\sqrt{N}}~\frac{\lambda^2}{N}~(N-1)~N  \propto{\sqrt{N}}. 
 \label{grdstbbar}
 \ee
For ground state tetraquark, the decay is unlikely to occur: the $B\bar B$ state has one extra pair of constituent quarks, compared to the initial state, and phase space is not enough . 
The decay in $B\bar B$ will occur for an excited (radial or orbital) state $T^*$, where the excitation energy $\epsilon_r-\epsilon_0$ can be used to create the quark pairl needed to transform the tetraquark into $B\bar B$. Only the excited quark emits the gluon in Fig.~\ref{tetradecs} (left) and we loose a factor of $\sqrt{N-1}$ to get:
\be
 A(T^*\to B+ \bar B) \propto N^0 \label{tetrastar1}.     
 \ee
The analogy with Eqs.~(\ref{bardec}) and (\ref{barcoup}) is evident.
\item Decay of an excited tetraquark to the ground  state by emitting a meson, which is shown in Fig.~\ref{tetradecs}(right). We find
\be
A(T^* \to T+M)\propto N^0. 
\ee
not growing with $N$.
\end{itemize}

\emph{\bf{Pentaquark decays: $ P \to T+B$ and $P\to 2B+\bar B$.}}

we define 
\be
B_{ab}=\frac{\partial}{\partial q^a}\frac{\partial}{\partial q^b} B=- B_{ba}. 
\label{signfix2}
\ee
with $B_{ab}$ antisymmetric for any value of $N$, and describe the pentaquark by the formula:
\begin{align}
P=\frac{1}{\sqrt{N(N-1)/2}}~\sum_{a< b} B_a B_b~ \bar B^{ab},\;\;a,b = 1\cdots, N
\label{pentaN2}. 
 \end{align}

 Decay into a baryon and a tetraquark goes through the process \eqref{g-rad}
where the gluon is radiated from either one or the other initial diquark. The additional quark $q^\ell$ changes the diquark into a baryon. The antiquark is absorbed by $\bar B^{ab}$ which becomes a (generalised) antidiquark to make a tetraquark with the other, spectator, diquark. 

The resulting amplitude is
\be
A(P\to B+ T)\propto \sqrt{N} \label{pentabar}. 
\ee
As before, decay from the ground state is forbidden by phase space. The decay amplitude from an excited pentaquark is reduced by a factor $\sqrt{N-1}$
\be
A(P^*\to B+ T)\propto N^0. 
\ee

For other decay modes, 
we  obtain similarly: 
\bea
&& A(P \to B+B+\bar B) = {\cal O}(N)\label{pentalead},  \\
&& A(P^* \to P+M)\propto N^0\label{pentadeexc},\\
\eea
\begin{figure}[htb!]
 \begin{center}
   \includegraphics[width=6.0truecm]{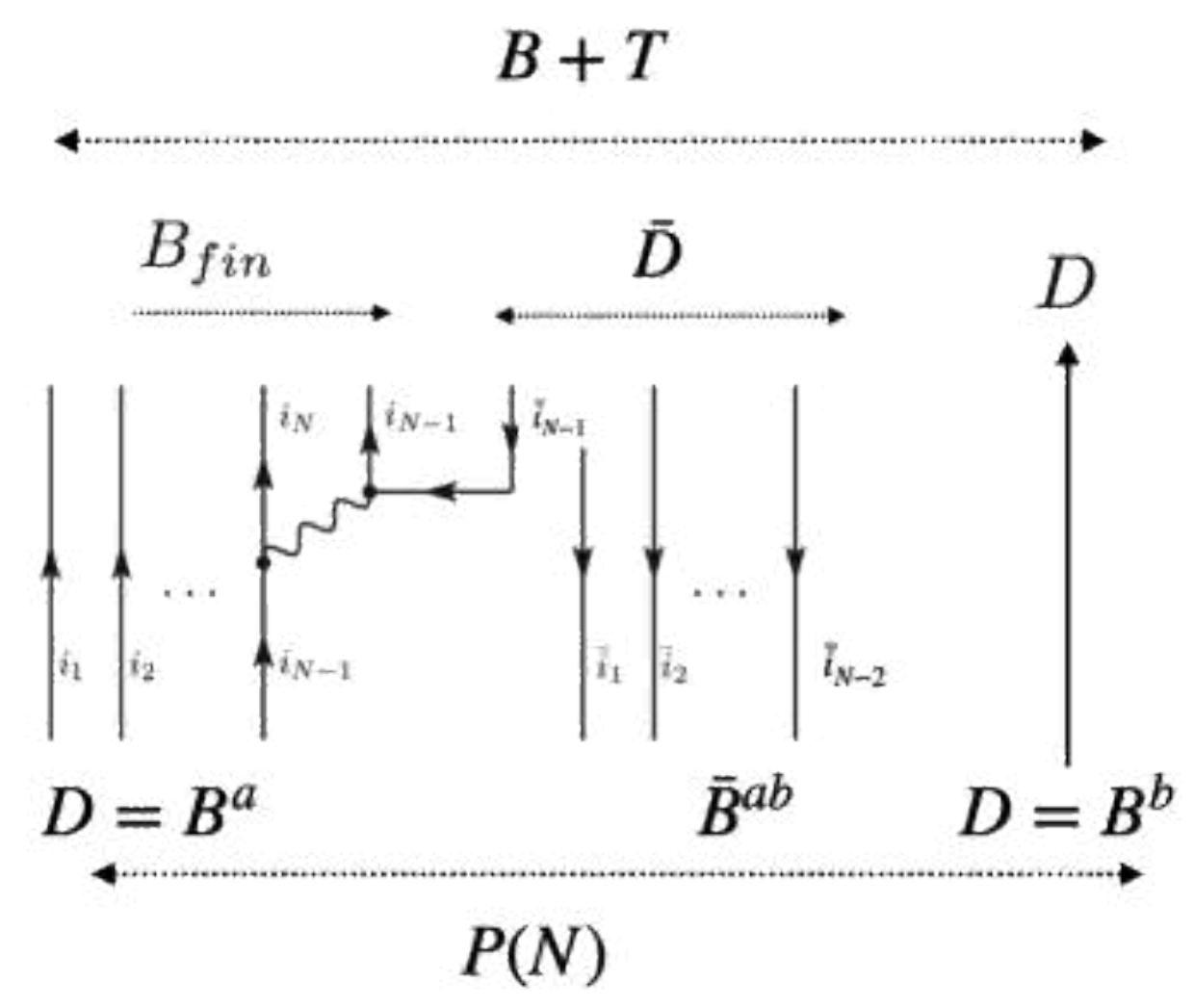}
 \caption{\footnotesize{.}}
\label{pentad}
\end{center}
\end{figure}

\subsection {Summary of Witten's multiquarks}

\begin{enumerate}
\item {\bf{Decay amplitudes from the ground states}} may diverge at large N. However such decays are generally  forbidden by phase space and the divergent amplitudes do not affect the observability for such particles. At $N=\infty$, bound states of multiquark hadrons are narrow or stable, particularly in the case of the dibaryon.
\item  {\bf {Decay amplitudes of excited states }}are summarised in the Table.
\begin{table}[htb!]
\centering
    \begin{tabular}{||c|c|c|c|c||}
     \hline
$T^*\to$& $B\bar B$ &  $  $ & $T+$ Meson & Mesons \\ \hline
$A*\propto$ & $N^0$ & $  $ & $N^0$ & $<e^{-\frac{N}{2}}$ \\ \hline \hline
$P^* \to$ & $B+T$ & $B+B+\bar B$ &$P+$~Mesons & $B+$~Mesons  \\ \hline
$A^* \propto$ & $N^0$  & $N^{1/2}$ & $N^0 $ & $<e^{-\frac{N}{2}}$  \\ \hline \hline
$D^* \to$ & $  $  & $N B \bar B$  & $D+$~Meson & $(N-1)B$\\ \hline
$A^* \to$ & $ $ &$>e^{+\frac{N}{2}}~ log N$  & $N^0$ & $>e^{+\frac{N}{2}~ log N}$\\ \hline \hline
\end{tabular}
 \caption{\footnotesize {Decay modes of excited Tetraquarks, Pentaquarks and Dibaryons and the corresponding amplitudes (denoted by $A^*$). Fourth column: decay of an excited into the ground state by meson emission. Last column refers to decays obtained by reorganising the quark-antiquark pairs of the initial state into a multi-meson state or redistribuiting the quarks of one diquark to the other diquarks, to form a set of N-1 baryons.}}
\label{witres}
\end{table}

\item {\bf{Excited tetraquarks}}: 
\begin{itemize}
\item the amplitudes for the decay of the excited states vanish or remain constant for $N\to \infty$: states are observable in this limit;
\item tetraquark de-excitation amplitudes are of the same order of $B\bar B$ decay amplitudes; 
\item For N=3 and flavour composition $[cu][\bar c\bar u]$ the threshold for  two-baryon decay is $2M(\Lambda_c)=4570$ MeV; Cotugno {\it et al}~\cite{Cotugno:2009ys} argued that X(4660) is a P-wave tetraquark decaying predominantly into $\Lambda_c \bar \Lambda_c$ in addition to the mode into $\psi(2S) \pi \pi$.
\item tetraquark-charmonium mixing is exponentially suppressed;
\item The divergence at large $N$ is not relevant for the width and the observability of the ground state, which is below threshold for the decay, 
\item Tetraquark ground state is dominant as intermediate state in elastic $B\bar B$ scattering. The $1/N$ behaviour we find for the latter amplitude is in agreement with the result given by Rossi and Veneziano (2016).
\end{itemize}
\item {\bf{Excited pentaquarks and dibaryon}}s:
\begin{itemize}
\item de-excitation amplitudes into the ground state and a meson remain limited for large N;
\item  at $N=\infty$ there are modes which give divergent amplitudes, namely $P^* \to B+B+\bar B ~{\rm and}~ D^*\to NB +\bar B ~{\rm or} ~(N-1)B$;\item  literally,   these results, imply sharp thresholds at $2B+\bar B$ and $(N-1)B$ respectively, below which we expect observable pentaquarks and dibaryons, and above which we expect large, unobservable widths, a situation similar to charmonia above and below the open charm-anticharm meson threshold.
\item  For N=3 and pentaquark with flavour composition:~$[cu][ud]\bar c$, corresponding to the states observed by LHCb, the threshold for ``non-observabilty" would be  $2M(\Lambda_c) + M(P)\sim 5510~{\rm MeV}$;
\item for a double charmed dibaryon with flavour [cu][cd][ud] the threshold would be at:  $2M(\Lambda_c)$.
\end{itemize}
\end{enumerate}

Finally, it is interesting to compare the results for tetraquarks with  the analysis based  on the large $N$ generalisation of tetraquarks following Weinberg. 
The results by the Roma group feature:
\begin{itemize}
\item a narrow ground state, with a suppressed decay amplitude into two mesons, of order $N^{-2}$. For large $N$ this is larger that the exponentially suppressed amplitude in Table~\ref{witres}, but it takes $N> 6 $ for the power suppression to  win over the exponential suppression;
\item amplitude of order$N^{-1/2}$ for the de-excitation into the ground state by meson emission;
\item tetraquark-charmonium mixing occurs to order $N^{-3/2}$; 
\item the decay of an excited tetraquark  into $B\bar B$ cannot be computed..
\end{itemize}

The similarities of two very different multiquark generalisations are indeed very intriguing.

%% file: chapts/WeinbergCrit.tex

\begin{figure}[htb!]
 \begin{center}
   \includegraphics[width=9.0truecm]{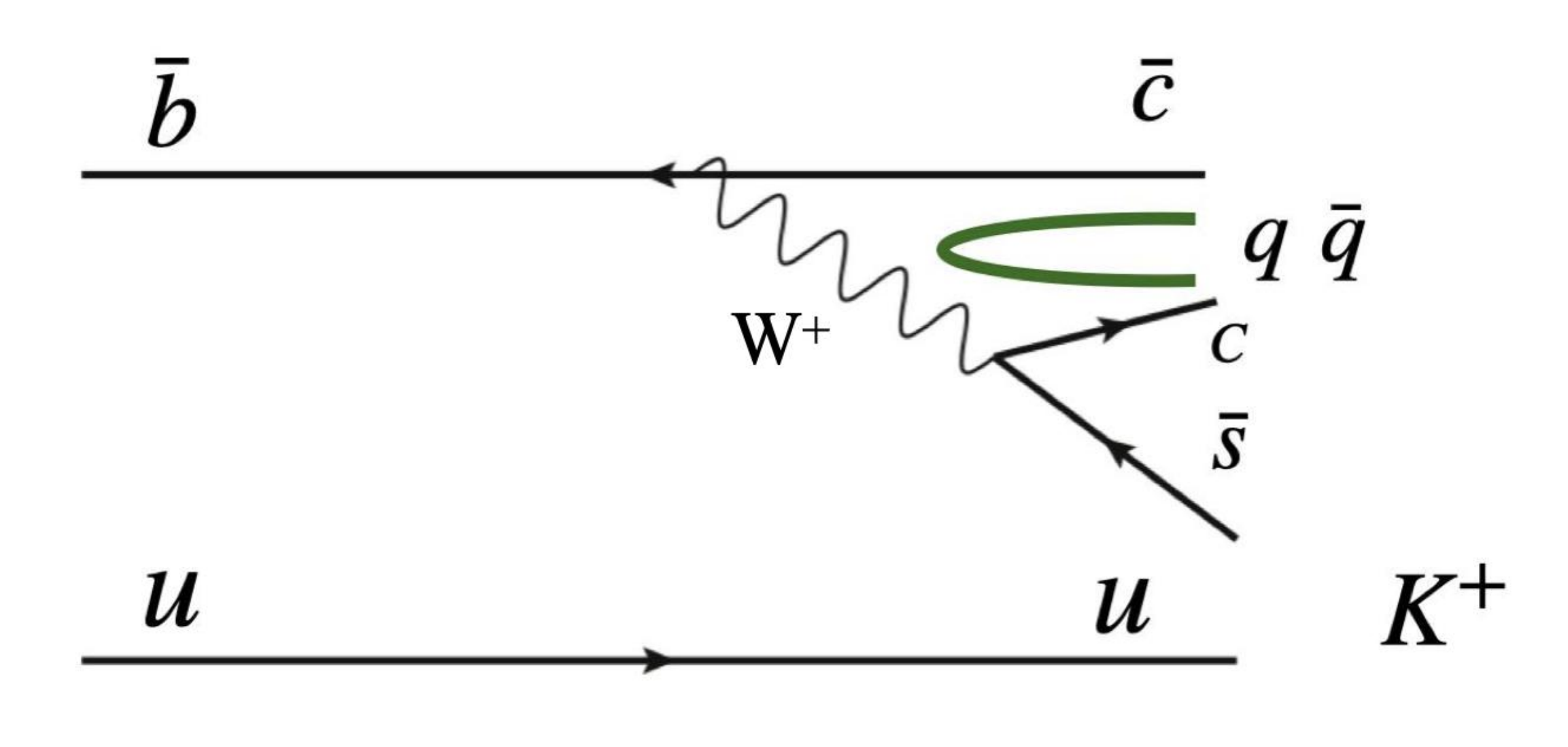}
 \caption{\footnotesize{  .}}
\label{bdec2}
\end{center}
\end{figure}

In Fig.~\ref{bdec2} we recall a figure given in Lect. 5, with the diagram of $B$ decaying into states with six quarks and focus on the upper right corner. There are two ways in which the $q\bar q$ pair may combine with the $\bar c c$ pair to form a color singlet, physical intermediate state $X$, which eventually will decay in the observed final state $f=D^*\bar D+\bar D^* D$:
\begin{enumerate}
\item molecule: $X=[\bar C q]^1[C\bar q]^1\to f$
\item tetraquark: $X=([Cq]^{\bar 3}~[\bar C \bar q]_3)_{\bf 1} \to f$ 
\end{enumerate}

In a different context, Weinberg  posed a similar question~\cite{Weinberg:1965zz}. Assume you produce a state with baryon number $B=2$ that materialises in a proton-neutron  pair, and call it "deuteron". Is  it the "deuteron" a proton-neutron bound state or is it an {\it elementary} dibaryon?\

\subsection{The QCD framework}

Let us rephrase Weinberg's question in a QCD framework.

We know that QCD produces confined $C=\pm 1$ mesons, and confined $C=0$  hidden charm hadrons: charmonia. We would like to ascertain  if QCD produces as well confined, hidden charm tetraquarks.

Consider the space of possible hidden charm states, $C=0,~J^{PC}=1^{++}$. Neglecting all interaction among them, e.g pion exchange, such a space is made by two components~\cite{Esposito:2021vhu}
\begin{itemize}
\item discrete energy states: charmonia and possibly tetraquarks:
\be
|C><C|+|T><T|
\ee
\item a continuum of charmed meson pairs: 
\be
|D^*\bar D(\alpha)><D^*\bar D(\alpha)|
\ee
(we neglect spin, $\alpha$ is the relative, c.o.m. momentum)
\item  Let $X$ denote the $X(3872)$ resonance. The completeness relation in the Hilbert space of states reads
\bea
&&|<X|X>|^2=1=Z+\int d\alpha~|<X|D^*\bar D(\alpha)>|^2; \label{complete} \\
&&Z=|<X|C>|^2+|<X|T>|^2 \notag
\eea
\end{itemize}

There are two regimes :
\begin{itemize}
\item $Z=0$:  corresponds to a pure molecular state: $X$  results from $D^*-\bar D$ interactions only (like a bound $n-p$ state)
\item  $Z\neq 0$: some compact, discrete state must exist. Is it charmonium or tetraquark?
\end{itemize}

The recent estimate of isospin violation in $X(3872)$ decay was given by LHCb~\cite{LHCb:2022bly}, with significantly better precision than previous data:
\be
\frac{g_{X(3872)\to \rho^0~\psi}}{g_{X(3872)\to \omega^0~\psi}}=0.29\pm 0.04\label{lhcbnew}
\ee
 The result indicates a considerable violation of Isospin symmetry, in particular compared to the analogous ratio, see PdG~\cite{pdg}
\be
\frac{g_{\psi(2S)\to \pi^0~\psi}}{g_{\psi(2S)\to \eta~\psi}}=0.045\pm 0.001
\ee
The LHCb results speaks against the association of $X(3872)$ with the charmonium family~\footnote{In this respect, the recent re-classification  by PdG of $X(3872)$ as $\chi_{c1}(3872)$ seems particularly unappropriate.} and it brings us to the "exotic" alternatives: (i) a pure $D^{*0}\bar D^0+D^{0}\bar D^{(*0)}$ molecule, for $Z=0$,  or (ii) a compact tetraquark, for $0<Z<1$.

\emph{\bf{Scattering amplitude.}} The key quantity is the $D^*\bar D$ scattering amplitude,$f$, normalised so that its square gives the differential cross section
\be
|f|^2=\frac{d\sigma}{d\Omega}
\ee

Near threshold  $f$ is parametrised as   
\be
f^{-1}= k \cot \delta(k) - i k=-\kappa_0 + \frac{1}{2} r_0 k^2 - i k \label{effek}
\ee
where $k\sim 0$ is the center of mass momentum and the real part of the denominator is the expansion of $ k \cot \delta(k)$, an even function of $k$,  to order $k^2$. 

If there is a shallow bound state or a resonance slightly below threshold, with mass:
\be
M(X)=M(D^*)+M(D)-B, ~(B={\rm binding~ energy})\label{binden}
\ee
the scattering amplitude, $f$, has a pole at imaginary momentum: $-i k=\kappa$ with
\be
 \kappa=\sqrt{2\mu B},~~(\mu={\rm reduced~mass})
 \ee
 The condition for the pole reads
 \be
 \kappa_0 + \frac{1}{2} r_0 \kappa^2-\kappa=0 \label{pole}
 \ee

Given $\kappa_0$ and $r_0$, $\kappa$ is the root of the second degree equation \eqref{pole}, which tends to $\kappa_0$ for $r_0\to 0$, that is:
\be
\kappa=\frac{1-\sqrt{1-2\kappa_0r_0}}{r_0}\notag
\ee
a relation we shall use later in the form:
\be
1-\kappa r_0=\sqrt{1-2\kappa_0r_0}\label{usrel}
\ee

It is convenient to introduce a representation of the scattering amplitude, which coincides with \eqref{effek} {\it at the pole}. We write
 \be
 f=-\frac{-\kappa_0+\frac{1}{2}r_0 k^2+ik}{(\kappa_0-\frac{1}{2}r_0 k^2)^2+k^2}=\frac{Num}{Den}
 \ee
 At the pole $-ik=\kappa$
 \be
 (Num)_{pole}=-2\kappa
 \ee
 while, using \eqref{pole}
 \bea
 &&(Den)_{pole}=[\kappa-\frac{1}{2}r_0(\kappa^2+k^2)]^2+k^2\sim \kappa^2-r_0 \kappa (\kappa^2 +k^2) +k^2=\notag \\
 &&=(\kappa^2+k^2)(1-r_0\kappa)=2\mu(1-r_0\kappa)~(B+T)
 \eea
 T is the kinetic energy in the center of mass $T= k^2/2\mu$, $\mu$ the reduced mass and we obtain
 \be
 f=\frac{\kappa}{\mu(1-r_0\kappa)}~\frac{1}{B+T}. \label{finf}
 \ee
 
\begin{figure}[htb!]
 \begin{center}
   \includegraphics[width=4.0truecm]{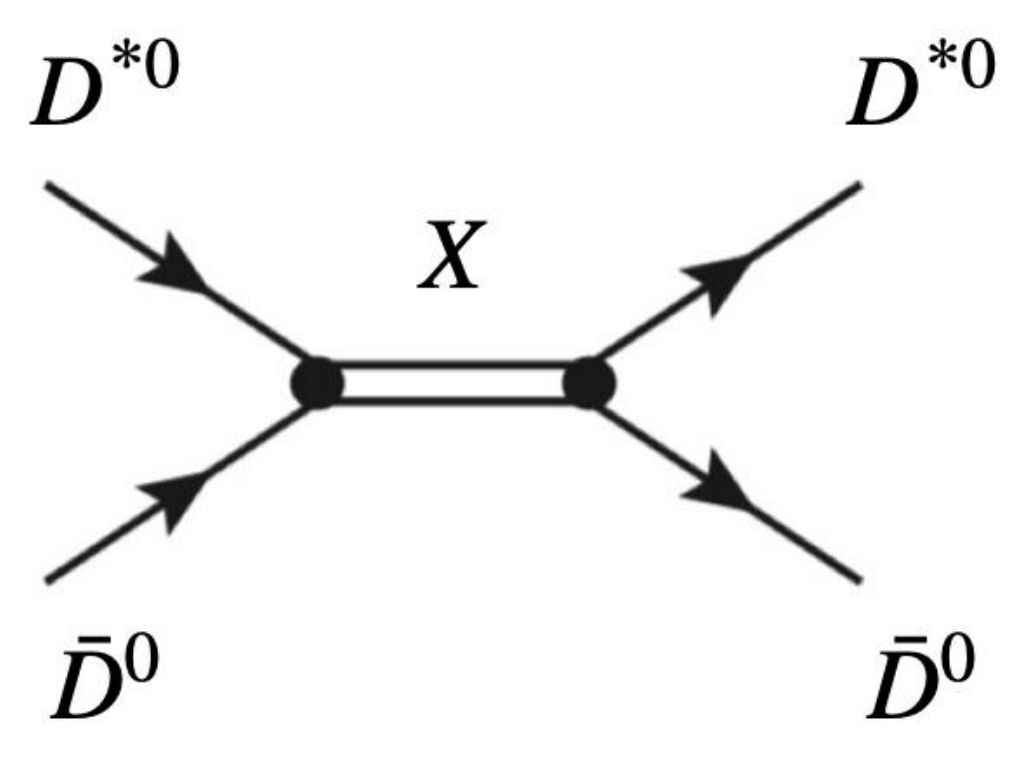}
 \caption{\footnotesize{  .}}
\label{loopexp}
\end{center}
\end{figure}

 \emph{\bf{Relation to the Feynman amplitude.}}
A more familiar representation of the scattering amplitude is given by the Feynman amplitude ${\cal M}_{fi}$, described by the diagram in Fig.~\ref{loopexp}. Define
\bea
&&S=1-i(2\pi)^4 \delta^{(4)}(p^\prime_a+p^\prime_b-p_a - p_b)~{\cal T}\notag \\
&& {\cal T}= \sqrt{N^\prime_a N^\prime_b N_a N_b} ~{\cal M}_{fi}\label{feynman}
\eea
where, for bosonic states,
\be 
N=\frac{1}{2E}
\ee
In the c.o.m., ${\bf p}_a=-{\bf p}_b={\bf k}$ and $v_{a,b}=p_{a,b}/E_{a,b}$. The relative velocity is
\be
v=|v_a-v_b|=k(\frac{1}{E_a}+\frac{1}{E_b})=k\frac{E_a+E_b}{E_a E_b}
\ee
and the cross-section is
\bea
&&d\sigma=\frac{1}{v}~(2\pi)^{4} \delta^{(4)}(p^\prime_a+p^\prime_b-p_a - p_b)\frac{d^3p^\prime_a}{(2\pi)^3}~\frac{d^3p^\prime_b}{(2\pi)^3}N^\prime_a N^\prime_b N_a N_b~|{\cal M}_{fi}|^2= \notag \\
&&=\frac{1}{(2\pi)^2}\frac{E_a E_b}{k(E_a+E_b)}(N_a N_b)^2~d\Omega~\int \delta(E^\prime_a+E^\prime_b-E_a-E_b)~k^2 dk~|{\cal M}_{fi}|^2
\eea
so that
\bea
&&\frac{d\sigma}{d\Omega}=\frac{1}{(2\pi)^2}(E_a E_b N_a N_b)^2 \frac{1}{(E_a+E_b)^2}~|{\cal M}_{fi}|^2=\notag \\
&&=\big[\frac{E_a N_a E_b N_b}{2\pi(E_a+E_b)}{\cal M}_{fi}\big]^2
\eea
In conclusion, up to an unessential phase~\cite{Weinbergqft}:
\be
f= \frac{E_a N_a E_b N_b}{2\pi(E_a+E_b)}{\cal M}_{fi}=\frac{\mu}{2\pi}\frac{1}{4 M_aM_b}{\cal M}_{fi}=\frac{\mu}{2\pi} {\cal T} \label{effe}
\ee
(non relativistic approximation).

\subsection{Weinberg's criteria  for $X(3872)$}

We call $H_0$ the Hamiltonian responsible for the formation of charmonia and, eventually, of tetraquarks and $V$ the additional interactions, including the color singlet exchange forces that may eventually lead to hadronic molecules. The total Hamiltonian is $H_0+V$. Computing enegies from the $D^* \bar D$ threshold, we have
\be
(H_0+V)|X>=-B|X>,~<D^* \bar D| H_0=<D^* \bar D(k)|T(k)
\ee
so that
\be
<D^* \bar D| (H_0+V)|X>=T<D^* \bar D|X>+<D^* \bar D|V|X>=-B<D^* \bar D|X>
\ee
 that is
\be
|<D^* \bar D|X>|^2=\frac{|<D^* \bar D|V|X>|^2}{(T+B)^2}=\frac{g^2}{(T+B)^2}
\ee
where, following Weinberg, we have defined
\be 
g=<D^* \bar D|V|X> \label{coupling}
\ee

We can insert this result in the completeness relation \eqref{complete} to find
\be
\int_0^\infty \frac{d^3k }{(2\pi)^3}\frac{g^2}{(B+\frac{k^2}{2\mu})^2}=1-Z
\ee 
Assuming $g$ to be essentially constant in the (very restricted) range of integration, ve obtain $g^2$ after an elementary integration\footnote{ Use the formula $\int_0^\infty dx \sqrt{x}/(1+x)^2=\frac{\pi}{2}$.}: 
\be
g^2=\frac{2\pi}{\mu^2 }\kappa(1-Z) \label{gvsz}
\ee
On the other hand, starting from \eqref{coupling} we obtain 
\be
{\cal T}=\frac{g^2}{B+T}
\ee
and the scattering amplitude
\be
f=\frac{\mu}{2\pi}    \frac{g^2}{B+T}=(1-Z) \frac{\kappa}{\mu(B+T)}=\frac{\kappa}{\mu(1-r_0\kappa)}~\frac{1}{B+T}
\ee
Comparing the last two terms we find
\be
1-Z=\frac{1}{1-\kappa r_0} \label{rzero}
\ee
This equation gives  $r_0$ as function $Z$. We can obtain the other parameter of the scattering amplitude, $\kappa_0$ by plugging the $r_0$ thus obtained into the pole equation \eqref{pole}. In conclusion, we find the results stated by Weinberg for the scattering parameters as function of $Z$ and of  the scale parameter $\kappa$:
\bea
&&\kappa_0^{-1}=2\frac{1-Z}{2-Z} \kappa^{-1}+ {\cal O}(m_\pi^{-1}) ;\notag\\
&& r_0=-\frac{Z}{1-Z}\kappa^{-1}+ {\cal O}(m_\pi^{-1}) \label{scattpar}
\eea
$Z$ dependent terms in Eqs.\eqref{scattpar} provide the leading contributions of order $\kappa^{-1}$, which, for a shallow state, is much larger than the typical range of hadronic forces. The latter, non-leading, contributions are indicated in \eqref{scattpar} as unspecified ${\cal O}(m_\pi^{-1})$ terms.

In the molecular case (Z=0) one has $r_0=O (1/m_\pi)$. A theorem due to Landau and Smorodinsky characterizes further the radius of an hadronic molecule by showing that {\it a $D^{*0}\bar D^0$ state bound by a fully attractive potential has necessarily $r_0 >0$}~\cite{smorodinsky,landau3,Bethe:1949yr}.

\emph{\bf{The Landau - Smorodinsky theorem.}}~Consider the Schr\"odinger's equation for the radial wave function of the molecular constituents
\be
u_k^{\prime\prime}(r) +\big[k^2-U(r)\big]u_k(r)=0\notag
\ee
$U(r) =2 \mu V(r), V(r) < 0$ is the potential, assumed to be of finite range, $a_0$ ($\simeq 1/m_\pi$), and attractive everywhere. We consider the wave function for two values of the momentum: $u_{k_{1,2}}\equiv u_{1,2}$. With simple manipulations we find the identity
\be
u_2u_1^\prime - u_2^\prime u_1 \Big |_0^R=(k_2^2- k_1^2)\int_0^R dr \, u_2 u_1 \label{uno}
\ee
$R >> a_0$.

Consider now the free equation: $\psi_k^{\prime\prime}(r) +k^2\psi_k(r)=0$, from which we also obtain
\be
\psi_2\psi_1^\prime - \psi_2^\prime \psi_1 \Big |_0^R=(k_2^2- k_1^2)\int_0^R dr \, \psi_2 \psi_1 \label{due}
\ee
Normalizing to unity at r=0, the general expression for $\psi_k$ is
\be
\psi_k(r) = \frac{\sin(kr+\delta(k))}{\sin\delta(k)},~\psi^\prime_k(0) = k \cot \delta(k) \notag
\ee
The radial wave function $u_k$ vanishes at r=0, and we normalize so that it tends exactly to the corresponding $\psi_k$ for large enough radii.
Now, subtract \eqref{uno} from \eqref{due} and let $R\to\infty$ (now the integral is convergent) to find
\be
k_2\cot \delta(k_2)-k_1\cot \delta(k_1) =(k_2^2- k_1^2)\int_0^\infty dr \, (\psi_2 \psi_1 - u_2 u_1) \label{tre}
\ee
We compare \eqref{tre} with the parameters of the scattering amplitude. First we set $k_1=0$. Since $\lim_{k_1\to0} k_1 \cot \delta(k_1) = -\kappa_0$
\be
k_2 \cot\delta(k_2) = - \kappa_0 + k_2^2 \int_0^\infty dr\, \left(\psi_2 \psi_0 - u_2 u_0\right) \notag
\ee
For small momenta:  $k_2 \cot \delta(k_2) = - \kappa_0 + \frac{1}{2}r_0 k_2^2$, so that:
\be
r_0=2\int_0^\infty dr\, (\psi _0^2-u_0^2) \notag
\ee
We know that $u_0(0) = 0,~\psi_0(0) = 1$. Defining $\Delta(r) = \psi_0(r) - u_0(r)$ we have:
\be
 \Delta(0) = +1, ~\Delta(\infty) = 0 \notag
 \ee
The equations of motion imply $\Delta^{\prime\prime}(r) = -U(r)u_0(r)$. In presence of a single bound state, where u(r) has no nodes, we get 
\be
\Delta''(r) > 0\
\ee
$\Delta$ goes from 1 to 0 monothonically, that is $ \psi_0(r) > u_0(r)$ and $r_0>0$.
Conversely, see \eqref{scattpar}, a negative value $r_0<0$ implies $Z>0$, from~\eqref{scattpar}.

Reassuringly: $r_0({\rm deuteron})=+ 1.75~{\rm fm}$.

\emph{\bf{The value of ${\bf Z}$.}}
From Eq.~\eqref{rzero}, we derive
\be
Z=\frac{-\kappa r_0}{1-\kappa r_0} \label{zvalue}
\ee
$Z$ is sometime identified with the admixture of the exotic hadron X with the compact (tetraquark) state and it is often stated that for small $Z$, X would be {\it essentially a molecule}.

The interpretation of Z as mixing coefficient, however, holds only in the free theory. With interaction, the compact state vector may be renormalized and the strength of Z loses its meaning.
A non-vanishing $Z$, indicates that there are, in the Hilbert space, states that cannot be made from the D D* continuum. Quoting Weinberg:
\begin{itemize}
\item {\it the true token that the deuteron is composite is an effective range $r_0$ small and positive rather than large and negative}.
\item {\it an elementary deuteron would have $0<Z<1$}.
\end{itemize}
 \emph{\bf{Compositeness indicator ${\bf X}$.}} The compositeness indicator $X$ is sometime introduced, defined as:
 \be
 X=\frac{1}{\sqrt{1-2\kappa_0 r_0}} \label{cindic}
 \ee
Note that $X$ is a function of the inverse scattering lenght, $\kappa_0$, while $Z$ depends upon the position of the pole, $\kappa$. The relation between $X$ and $Z$ is obtained from Eqs.~\eqref{usrel} and \eqref{zvalue}
\be
1-Z=\frac{1}{1-\kappa r_0}=\frac{1}{\sqrt{1-2\kappa_0r_0}}= X.  \label{compfac}
\ee
Thus, $X=1$ corresponds to the molecule while $0<X<1$ corresponds to $0<Z<1$, i.e. to Weinbeg's criterion for elementary X.

\subsection{X lineshape: from Breit-Wigner to scattering lengths}

Consider $D^{*0}\bar D^0$ scattering above threshold. If there is a resonance slight below, the amplitude takes the Breit-Wigner form
\be
f_{BW}=-\frac{\frac{1}{2}g^2_\text{BW}}{ E - m_\text{BW} + \frac{i}{2} g^2_\text{BW} k} \label{BW}
\ee
for $E=\frac{k^2}{2\mu}$, \eqref{BW} has the  same form of the scattering amplitude \eqref{effek}.
Thus, from the parameters of the line-shape we can determine the scattering parameters.

\begin{figure}[htb!]
 \begin{center}
   \includegraphics[width=9.0truecm]{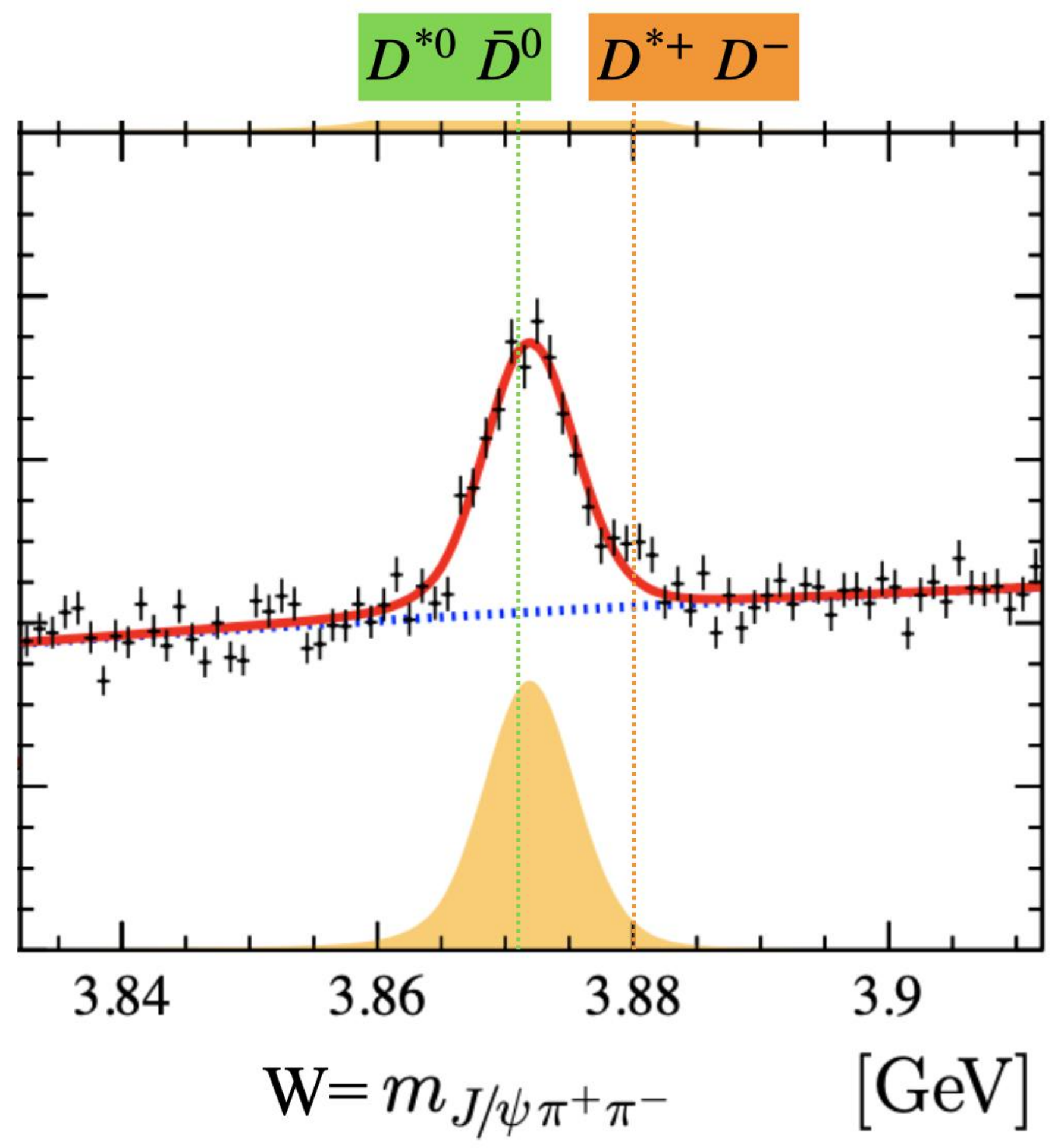}
 \caption{\footnotesize{The $X(3972)$ lineshape. Vertical lines mark the $D^{*0} \bar D^0$ and  $D^{*+} D^-$ thresholds.}}
\label{Xlsh}
\end{center}
\end{figure}

The line shape reported by LHCb~\cite{LHCb:2021auc} is given in Fig.~\ref{Xlsh} as function of the c.o.m energy, W, equal to the invariant mass of the observed final state $J/\Psi \pi^+\pi^-$. $D^{*0} \bar D^0$ and  $D^{*+} D^-$ thresholds are indicated.

From LHCb data, neglecting the experimental errors on the best fit parameters, we find~\cite{Esposito:2021vhu}
\be
{\bf X(3872)}:~ \kappa_0 \simeq 6.92~{\rm MeV};~r_0=-5.3~{\rm fm} \notag
\ee
We find $r_0$ well into the compact tetraquark region. Using the recent error analysis~\cite{Baru:2021ldu} we find the effective radius in the range;
\bea
&&-1.6~{\rm fm} >r_0> -5.3~{\rm fm};\notag\\
&& 0.14 > Z >0.052>0 \label{finalz}
\eea

\emph{\bf{Details.}}  The Flattt\'e function to fit the lineshape parameters is
\be
 f\left(X\to J/\psi \pi^+\pi^-\right) = -\frac{N}{E - m^0_X + \frac{i}{2} g_{\rm LHCb} \left(\sqrt{2 \mu E} + \sqrt{2 \mu_+ (E - \delta)}\right) + \frac{i}{2} \left(\Gamma^0_\rho(E) + \Gamma^0_\omega(E) + \Gamma^0_0\right)}\notag
\ee
where (particle symbols indicate the mass in MeV) and
\bea
&& E=W-(D^{*0}+\bar D^0)=T=\frac{k^2}{2\mu} ,~  \notag \\
&&\mu= \frac{D^{*0} \bar D^0}{D^{*0} + \bar D^0}=967~{\rm MeV},~\mu^+= \frac{D^{*+}  D^-}{D^{*+} +  D^-}=969~{\rm MeV} \notag \\
&& \delta=D^{*+}+D^- -D^{*0} -\bar D^0=8.3~{\rm MeV}>>E\notag
\eea
Details of fit and parameters are as follows.
\begin{itemize}
\item $\Gamma^0_\rho(E) + \Gamma^0_\omega(E) + \Gamma^0_0$ are fitted and subtracted from the denominator;
\item since $E=T<\delta$ we write
\be
 \frac{i}{2} g_{\rm LHCb} \sqrt{2 \mu_+ (T - \delta)}=\frac{g_{LHCb}}{2}\Big( - \sqrt{2\mu_+\delta} + T \sqrt{\frac{\mu_+}{2\delta}} \Big)
 \ee
\item Parametrization of the resulting denominator of $f$:
\bea  
&& Den=\frac{g_{LHCb}}{2}\cdot {\bar{D}},\notag \\
&&{\bar D}=\frac{2}{g_{\rm LHCb}}(T - m^0_X) - \sqrt{2\mu_+\delta} + T \sqrt{\frac{\mu_+}{2\delta}} + ik \notag
\eea
\item  Best fit: $g_{LHCb} = 0.108,~m^0_X=-7.18~{\rm MeV}$
\item Formulae for inverse scattering length and radius:
\be
\kappa_0 = -\frac{2m_X^0}{g_{\rm LHCb}} - \sqrt{2\mu_+\delta} \simeq 6.92~{\rm MeV};~r_0 = -\frac{2}{\mu g_{\rm LHCb}} - \sqrt{\frac{\mu_+}{2\mu^2\delta}} \simeq -5.34~{\rm fm} \notag
\ee
\item The Log Likelihood is very insensitive to the value of $g_{LHCb}$, leading to error estimate: $10>g_{LHCb}>0.108$; in correspondence, one obtains the range for $r_0$ and $Z$ reported in Eq.~\eqref{finalz}.
\end{itemize}

\subsection{The doubly charmed  Tetraquark, ${\bf T_{cc}^+(3875)}$}

The existence of doubly charmed tetraquarks, was considered in  2013 by Esposito {\it et al.}~\cite{Esposito:2013fma}.

Starting from the mass of the doubly charmed baryon, Karliner and Rosner~\cite{Karliner:2017qjm} estimated of the mass of  the lowest lying, I=0 state at $M(T_{cc}^+)=3882\pm 12~{\rm MeV}$, 7 MeV above the $D^0 D^{*+}$ threshold. A similar value was obtained by Eichten and Quigg~\cite{Eichten:2017ffp}.

A value close to the $D^0D^+$ threshold  was obtained in the Born-Oppenheimer approximation~\cite{Maiani:2019lpu}. The update considered in Lect. 6, brings the mass closer to the $D^0 D^{*+}$ threshold. 
Finally, the value $M(T_{cc}^+)-M(D^0D^{+})=-23\pm 11~{\rm MeV}$ is obtained in lattice QCD calculation by P.~Junnarkar {\it et al.}~\cite{Junnarkar:2018twb,Francis:2018jyb,Francis:2016hui,Leskovec:2019ioa}.
The closeness to the $D^0 D^{*+}$ threshold has nonetheless invited speculations about a molecular nature of $T_{cc}^+$.

\begin{figure}[htb!]
 \begin{center}
   \includegraphics[width=10.0truecm]{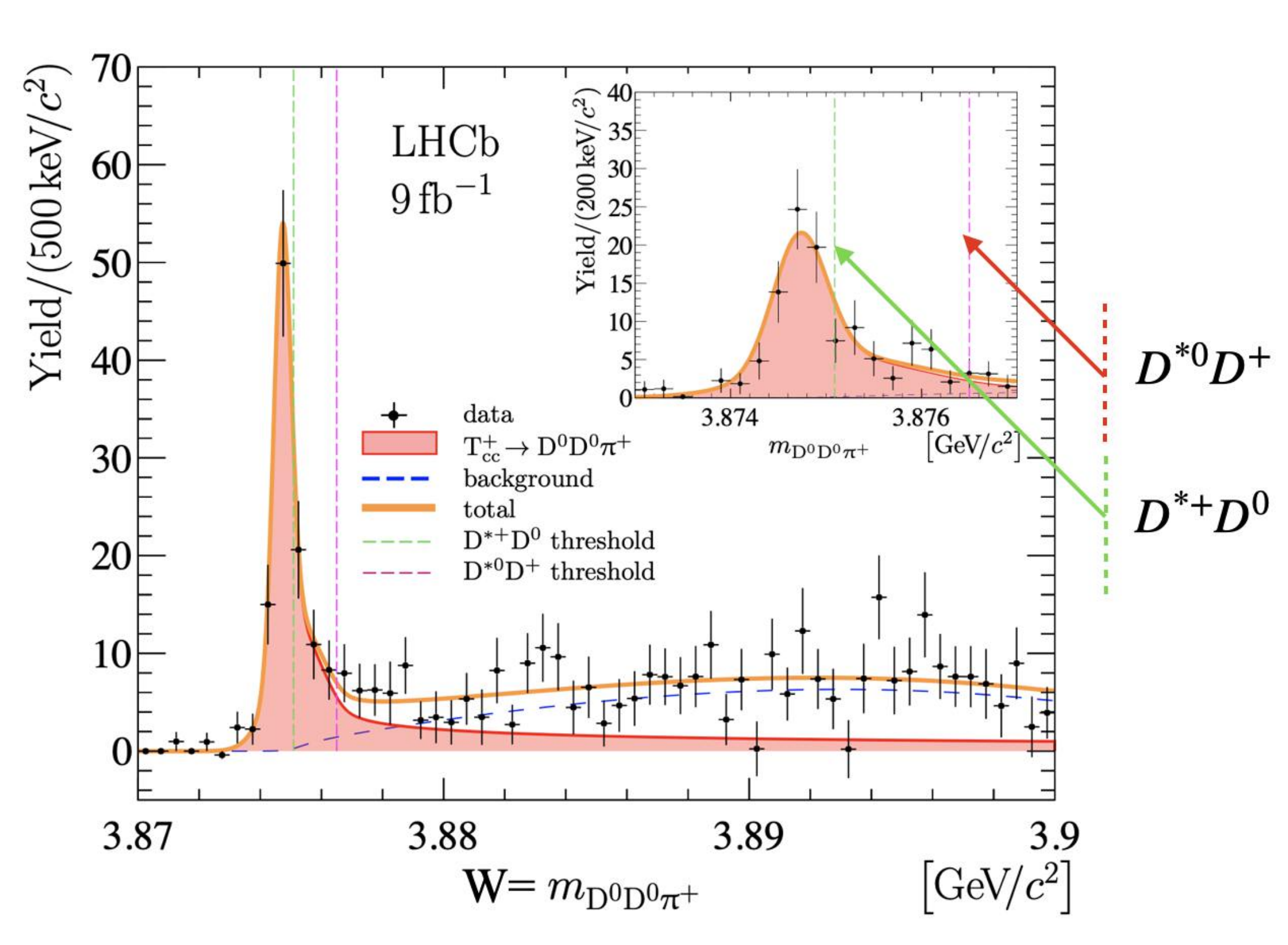}
 \caption{\footnotesize{The $$ lineshape for the decay ${\cal T}\to D^0 D^0 \pi^+$. Vertical lines mark the $D^{*0} \bar D^0$ and  $D^{*+} D^-$ thresholds.}}
\label{Tlsh}
\end{center}
\end{figure}

The line shape ot $T_{cc}^+$, in correspondence to the observed $D^0 D^0 \pi^+$ decay~\cite{LHCb:2021auc}
is reported in Fig.~\ref{Tlsh}.  The thresholds of the possible two-body decays: $D^0D^{*+}$ at $3875.1~{\rm MeV}$ and $D^{*0} D^+$ at $3876.5$~ MeV  are indicated.

We may  simply extend the analysis done for $X(3872)$ to estimate the effective radius of ${\cal T}(3875)$ .

 \emph{\bf{Details.}} We use the Flatt\'e formula
\be
 f({\cal T}\to D^0 D^0 \pi^+) = -\frac{N}{E - m^0_{\cal T} + \frac{i}{2} g_{\rm LHCb} \left(\sqrt{2 \mu E} + \sqrt{2 \mu_+ (E - \delta)}\right) }\notag
 \ee
with
\bea
 && E=W-(D^{*0}+ D^+)=T=\frac{k^2}{2\mu} ,~  \notag \\
&& \mu= \frac{D^{*+} D^0}{D^{*+} + \bar D^0}=967.5~{\rm MeV};~\mu^+= \frac{D^{*0}  D^+}{D^{*0} +  D^+}=968.0~{\rm MeV}\label{newparam1}\\
&&\delta=D^{*0}+D^+ -D^{*+} -\bar D^0=1.7~{\rm MeV}\label{newparam2}
 \eea

 Details of fit and parameters are as follows.
\begin{itemize}
\item Parametrization of the denominator of $f$:
\bea  
&& Den=\frac{g_{LHCb}}{2}\cdot {\bar{D}},\notag \\
&&{\bar D}=\frac{2}{g_{\rm LHCb}}(T - m^0_{\cal T}) - \sqrt{2m_+\delta} + T \sqrt{\frac{m_+}{2\delta}} + ik \notag
\eea
\item the effective radius is given by the same formula used for $X(3872)$, with the values of the parameters indicated in \eqref{newparam1} and \eqref{newparam2}:
\be
r_0 = -\frac{2}{\mu g_{\rm LHCb}} - \sqrt{\frac{\mu_+}{2\mu^2\delta}}\label{rzerot}
\ee
\item the constant $g_{LHCb}$ of the ${\cal T}$ lineshape is allowed by the fit to take large values, which may push the first term in \eqref{rzerot} to negligible values.
\end{itemize} 
 
\begin{figure}[htb!]
 \begin{center}
   \includegraphics[width=10.0truecm]{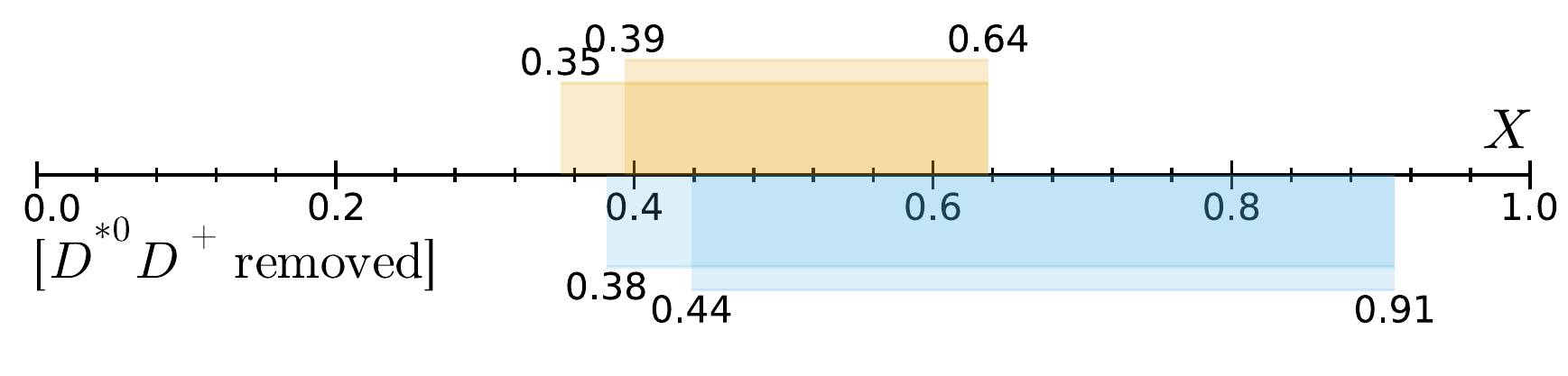}
 \caption{\footnotesize{${\cal T}(3875)$: intervals for the compositeness parameter $X$  of at 90(95) \% confidence
level. M. Mikhasenko, ArXiv:2203.04622.}}
\label{mikha}
\end{center}
\end{figure}

 With this simplifyed analysis, we may obtain the upper bound
\be
r_0 <- \sqrt{\frac{\mu_+}{2\mu^2\delta}}\simeq -3.4 ~{\rm fm}
\ee 
that can be combined with the lower bound $ r_0> - 11.9$~ fm given by LHCb to obtain
\be
{\bf {\cal T}(3875)}:~ -3.4>r_0 > - 11.9~ {\rm fm}
\ee

A  complete analysis of the scattering parameters of ${\cal T}$ has been done by Mikhasenko~\cite{Mikhasenko:2022rrl}  leading to
\be 
-4.3<r_0<-16.2
\ee

In Fig.~\ref{mikha} we report Mikhasenko's results for the intervals of the compositeness parameter $X$, Eq.~\eqref{cindic}, Using the relation: $X=1-Z$, Eq.~\eqref{compfac}, we derive the corresponding limitation for $Z$:
\bea
&&0.65>Z>0.36~({\rm brown~range})\notag \\
&&0.62> Z > 0.09~({\rm blue~range})
\eea

\emph{\bf{Our summary about ${\bf r_0}$.}} 

\begin{figure}[htb!]
 \begin{center}
   \includegraphics[width=9.0truecm]{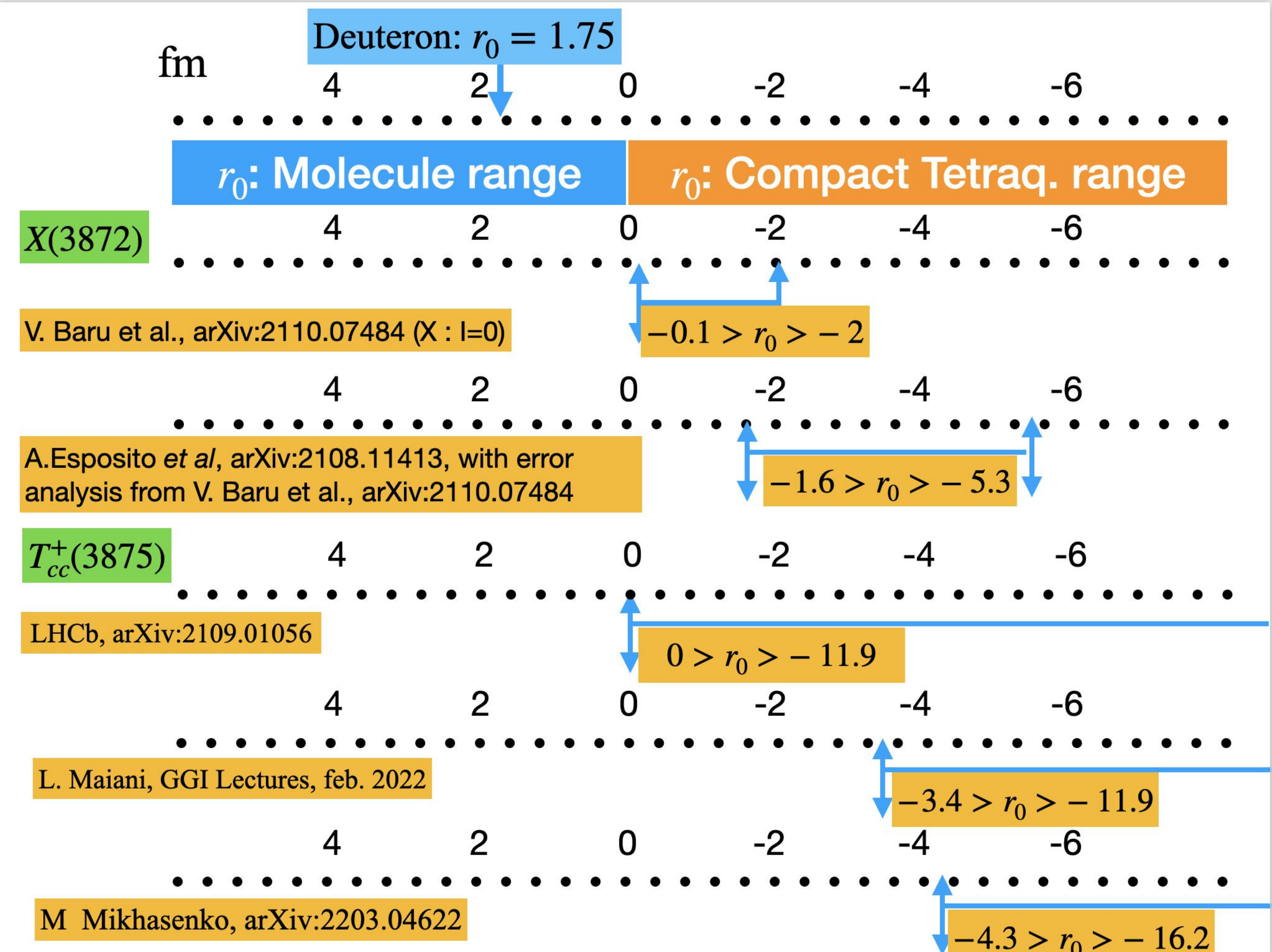}
 \caption{\footnotesize{ Compilation of recent reterminations of the effective radius, $r_0$, compared to the effective radius of the Deuteron.}}
\label{summ}
\end{center}
\end{figure}

Focusing on the determination of $r_0$ seems to lead to a convergent indication for a negative value, for both $X(3872)$ and ${\cal T}^+_{cc}(3875)$. The present situation about $r_0$ is reported in Fig.~\ref{summ}.
No consensus yet, but it seems that we are on a promising road. Stay tuned!
